\documentstyle[preprint,aps,eqsecnum,epsfig]{revtex}
\begin{document}
\tightenlines

\newcommand{\beq}{\begin{equation}}
\newcommand{\eeq}{\end{equation}}
\newcommand{\bea}{\begin{eqnarray}}
\newcommand{\eea}{\end{eqnarray}}
\newcommand{\cir}{{\buildrel \circ \over =}}

\title{Dynamical Body Frames, Orientation-Shape Variables and
Canonical Spin Bases  for the Non-Relativistic N-Body Problem.}

\author{David Alba}

\address
{Dipartimento di Fisica\\
Universita' di Firenze\\
L.go E.Fermi 2 (Arcetri)\\
50125 Firenze, Italy\\
E-mail: ALBA@FI.INFN.IT}

\author{and}

\author{Luca Lusanna}

\address
{Sezione INFN di Firenze\\
L.go E.Fermi 2 (Arcetri)\\
50125 Firenze, Italy\\
E-mail: LUSANNA@FI.INFN.IT}

\author{and}

\author{Massimo Pauri}

\address
{Dipartimento di Fisica\\
 Universita' di Parma\\
  Campus Universitario, Parco Area Scienze 7/A\\
  43100 Parma, Italy\\
   E-mail: PAURI@PR.INFN.IT}

\maketitle

\begin{abstract}

\noindent
After the separation of the center-of-mass motion, a new privileged
class of canonical Darboux bases is proposed for the non-relativistic
N-body problem by exploiting a geometrical and group theoretical
approach to the definition of {\it body frame} for deformable bodies.
This basis is adapted to the rotation group SO(3), whose canonical
realization is associated with a symmetry Hamiltonian {\it left
action}. The analysis of the SO(3) coadjoint orbits contained in the
N-body phase space implies the existence of a {\it spin frame} for the
N-body system. Then, the existence of appropriate non-symmetry
Hamiltonian {\it right actions} for non-rigid systems leads to the
construction of a N-dependent discrete number of {\it dynamical body
frames} for the N-body system, hence to the associated notions of {\it
dynamical} and {\it measurable} orientation and shape variables,
angular velocity, rotational and vibrational configurations. For N=3
the dynamical body frame turns out to be unique and our approach
reproduces the {\it xxzz gauge} of the gauge theory associated with
the {\it orientation-shape} SO(3) principal bundle approach of
Littlejohn and Reinsch. For $N \geq 4$ our description is different,
since the dynamical body frames turn out to be {\it momentum
dependent}. The resulting Darboux bases for $N\geq 4$ are connected to
the coupling of the {\it spins} of particle clusters rather than the
coupling of the {\it centers of mass} (based on Jacobi relative normal
coordinates). One  of the advantages of the spin coupling is that,
unlike the center-of-mass coupling, it admits a relativistic
generalization.

\vskip 1truecm

\today

\vskip 1truecm

\end{abstract}
\pacs{}
\vfill\eject

\vfill\eject

\section{Introduction.}

This paper deals with the construction of a specialized system of
coordinates for the non-relativistic N-body problem which could be
instrumental to nuclear, atomic and molecular physics, as well as to
celestial mechanics. In particular, we shall exploit the technique of
the canonical realizations of Lie symmetry groups
\cite{pauri2,pauri1,pauri3,pauri4,pauri5} within the framework of the
non-relativistic version of the {\it rest-frame Wigner covariant
instant form} of dynamics \cite{lu,india} to the effect of obtaining
coordinates adapted (locally in general) to the SO(3) group. In most
of the paper we consider only free particles, since the mutual
interactions are irrelevant to the definition of the kinematics in the
non-relativistic case.

Isolated systems of  N particles possess 3N degrees of freedom in
configuration space and 6N in phase space. The Abelian nature of the
overall translational invariance, with its associated three commuting
Noether constants of the motion, allows for the decoupling and,
therefore, for the elimination of either three configurational
variables or three pairs of canonical variables, respectively
(separation of the center-of-mass motion). In this way one is left
with either 3N-3 relative coordinates ${\vec \rho}_a$
 or 6N-6 relative phase space variables ${\vec \rho}_a$,
${\vec p}_a$, $a=1,.., N-1$ and the center-of-mass angular momentum or
spin is $\vec S= \sum_{a=1}^{N-1}{\vec \rho}_a \times {\vec p}_a$. In
the non-relativistic theory most of the calculations employ the sets
of $3N-3$ {\it Jacobi normal relative coordinates} ${\vec s}_a$ which
diagonalize the quadratic form associated with the  relative kinetic
energy [the spin becomes $\vec S=\sum_{a=1}^{N-1}{\vec s}_a \times
{\vec \pi}_{sa}$, with ${\vec \pi}_{sa}$  momenta conjugated to the
${\vec s}_a$'s]. Each set of relative Jacobi normal coordinates ${\vec
s}_a$, $a=1,..,N-1$, is associated with a {\it different clustering}
of the N particles, corresponding to  the centers of mass of the
various subclusters. In special relativity Jacobi normal coordinates
do not exist, as it will be shown in Ref.\cite{iten1}, and a different
strategy must be used.

On the other hand, the non-Abelian nature of the overall rotational
invariance entails the impossibility of an analogous intrinsic
separation of {\it rotational} (or {\it orientational})
configurational variables from others which could be called {\it
shape} or {\it vibrational}. As a matter of fact, this is one of the
main concerns of molecular physics and of advanced mechanics of
deformable bodies. Recently, a new approach inspired by the
geometrical techniques of fiber bundles has been proposed in these
fields of research:
 a self-contained and comprehensive exposition  of this
viewpoint and a rich bibliography can be found in Ref.\cite{little}.

In the theory of deformable bodies one looses any intrinsic notion of
{\it body frame}, which is a fundamental tool for the description of
rigid bodies and  their associated Euler equations. A priori, for a
given configuration of  a non-relativistic continuous body, and in
particular for a N-body system, any barycentric orthogonal frame could
be named {\it body frame} of the system with its associated notion of
{\it vibrations}.

This state of affairs suggested \cite{little} to replace the
kinematically accessible region of the non-singular configurations
\footnote{ See Refs.\cite{little,little1} for a discussion of the singular
(collinear and N-body collision) configurations.  Applying the SO(3)
operations to any given configuration of the 3N-3 relative variables
(a point in the relative configuration space) gives rise to 3
possibilities only: i) for {\it generic configurations} the orbit
containing all the rotated copies of the configuration is a
3-dimensional manifold [diffeomorphic to the group manifold of SO(3)];
ii) for {\it collinear configurations} the orbit is diffeomorphic to
the 2-sphere $S^2$; iii) for the {\it N-body collision configuration}
(in which all the particles coincide at a single point in space) the
orbit is a point.} in the (3N-3)-dimensional relative configuration
space by a SO(3) principal fiber bundle over a (3N-6)-dimensional base
manifold, called {\it shape} space
\footnote{It is the open set of all the orbits of {\it generic
non-singular configurations}.}. The SO(3) fiber on each shape
configuration carries the {\it orientational} variables (e.g. the
usual Euler angles) referred to the chosen {\it body frame}.  A local
cross section of the principal fiber bundle selects just one
orientation of a generic N-body configuration in each fiber (SO(3)
orbit) and this is equivalent to a {\it gauge convention}, namely to a
possible definition of a {\it body frame} ({\it reference
orientation}), to be adopted after a preliminary choice of the {\it
shape} variables. It turns out that this principal bundle is trivial
only for N=3, so that in this case only  global cross sections exist,
and in particular the identity cross section may be identified with
the {\it space frame}. In this case any global cross section is a copy
of the 3-body shape space and its coordinatization gives a description
of the {\it internal vibrational} motions associated with the chosen
gauge convention for the reference orientation. For $N\geq 4$,
however, global cross sections do not exist \footnote{This is due to
the topological complexity of the shape space generated by the
singular configurations \cite{little1}, which are dispersed among the
generic configurations for $N \geq 4$.} and the definition of the
reference orientation ({\it body frame}) can be given only locally.
This means that the  shape space  cannot be identified with a
(3N-6)-dimensional submanifold of the (3N-3)-dimensional relative
configuration space. The {\it gauge convention} about the reference
orientation and the consequent individuation of the internal
vibrational degrees of freedom requires the choice of a connection on
the SO(3) principal bundle (i.e. a concept of horizontality) and this
leads in turn to the introduction of a SO(3) gauge potential on the
base manifold. Obviously, physical quantities like the rotational or
vibrational kinetic energies and, in general, any observable feature
of the system must be gauge invariant, namely independent of the
chosen convention. Note that both the space frame and the body frame
components of the angular velocity are gauge quantities in the {\it
orientation-shape bundle} approach and their definition depends upon
the gauge convention.

While a natural gauge invariant concept of {\it purely rotational}
N-body configurations exists (when the N-body velocity vector field is
{\it vertical}, i.e. when the shape velocities vanish), a notion of
{\it horizontal} or {\it purely vibrational} configuration requires
the introduction a connection $\Gamma$ on the SO(3) principal bundle.
A gauge fixing is needed in addition in order to select a particular
$\Gamma$-horizontal cross section and the correlated  gauge potential
on the shape space. See Ref.\cite{little} for a review of the gauge
fixings used in molecular physics' literature  and, in particular, for
the virtues of a special connection C corresponding to the shape
configurations with vanishing center-of-mass angular momentum $\vec S$
\footnote{The C-horizontal cross sections are orthogonal to the fibers
with respect to the Riemannian metric dictated by the kinetic
energy.}.

This {\it orientation-shape} approach replaces the usual Euler
kinematics of rigid bodies and entails in general a coupling between
the internal {\it shape} variables and some of the {\it orientational}
degrees of freedom. In Ref.\cite{little} it is interestingly shown
that the non-triviality of the SO(3) principal bundle, when extended
to continuous deformable bodies, is at the heart of the explanation of
problems like the {\it falling cat} and the {\it diver}. A
characteristic role of SO(3) gauge potentials in this case is to {\it
generate rotations by changing the shape}.

In Ref.\cite{little}  the Hamiltonian formulation of this framework is
also given, but no explicit procedure for the construction of a
canonical Darboux basis for the orientational and shape variables is
worked out. See Refs.\cite{little,little1} for the existing sets of
shape variables for $N=3,4$ and for the determination of their
physical domain.

Independently of this SO(3) principal bundle framework and having in
mind the relativistic N-body problem where only Hamiltonian methods
are available, we have been induced to search for a {\it constructive
procedure} for building canonical Darboux bases in the
(6N-6)-dimensional relative phase space, suited to the non-Abelian
canonical reduction of the overall rotational symmetry. Our procedure
surfaced from the following independent pieces of information:

A) In recent years a systematic study of relativistic kinematics of
the N-body problem, in the framework of the {\it rest-frame Wigner
covariant instant form of dynamics} has been developed in
Ref.\cite{lus} and then applied to the isolated system composed  by N
scalar charged particles plus the electromagnetic field
\cite{albad,crater}.

These papers contain the construction of a special class of canonical
transformations, of the Shanmugadhasan type \cite{sha,lu}. These
transformations are simultaneously adapted to: i) the Dirac first
class constraints appearing in the  Hamiltonian formulation of
relativistic models; ii) the timelike Poincar\'e orbits associated
with most of their configurations. In the Darboux bases one of the
final canonical variables is the square root of the Poincar\'e
invariant $P^2$ ($P_{\mu}$  is the conserved timelike four-momentum of
the isolated system). Subsequently, by using the constructive theory
of the canonical realizations of Lie groups
\cite{pauri2,pauri1,pauri3,pauri4,pauri5} a new family of canonical
transformations was introduced in Ref.\cite{lucenti}. This latter
leads to the definition of the so-called {\it canonical spin bases},
in which also the Pauli-Lubanski Poincar\'e  invariant $W^2=-P^2 {\vec
S}^2_T$ for timelike Poincar\'e orbits \footnote{For the
configurations of the isolated system having a rest-frame Thomas
canonical spin ${\vec S}_T$ different from zero.} becomes one of the
final canonical variables. The construction of the spin bases exploits
the clustering of spins rather than the Jacobi clustering of centers
of mass.

In spite of its genesis in a relativistic context, the technique used
in the determination  of the spin bases, related to a {\it typical
form} \cite{pauri2} of the canonical realizations of the E(3) group,
can be easily adapted to the non-relativistic case, where $W^2$ is
replaced by the invariant ${\vec S}^2$ of the extended Galilei group.

B) These results provide the starting point for the construction of a
canonical Darboux basis adapted to the non-Abelian SO(3) symmetry. The
three non-Abelian Noether constants of motion $\vec S$ are arranged in
these canonical Darboux bases as an array containing the canonical
pair $S^3$, $\beta = tg^{-1}{{S^2}\over {S^1}}$ and the unpaired
variable $S=|\vec S|$ \footnote{In this context the configurations
with $\vec S
=0$ are singular and have to be treated separately.}
({\it scheme A} of the canonical realization of SO(3) \cite{pauri1}).
The angle canonically conjugated to $S$, say $\alpha$, is an {\it
orientational} variable, which, being coupled to the internal {\it
shape} degrees of freedom, cannot be a constant of motion. In
conclusion, in this non-Abelian case one has only two (instead of
three  as in the Abelian case) commuting constants of motion, namely
$S$ and $S^3$ (like in quantum mechanics).

This is also the outcome of the momentum map canonical reduction
\cite{mars,lib,con2} by means of adapted coordinates. Let us stress
that $\alpha$, $S^3$, $\beta$ are a local coordinatization of any
coadjoint orbit of SO(3) contained in the N-body phase space. Each
coadjoint orbit is a 3-dimensional embedded submanifold and is endowed
with a Poisson structure whose neutral element is $\alpha$. This
latter is also the essential coordinate for the definition of the {\it
flag} of spinors (see Refs. \cite{payne} and \cite{pauri1}, Section
V). On the other hand, the spinor flag is nothing else than a unit
vector $\hat R$ orthogonal to $\vec S$ \cite{pauu}, which is going to
be a fundamental tool in what follows.

By fixing non-zero values of the variables $S^3$, $\beta
= tg^{-1}{{S^2}\over {S^1}}$ through second class constraints, one can
define a (6N-8)-dimensional reduced phase space. However, the
canonical reduction cannot be furthered by eliminating $S$, since
$\alpha$ is not a constant of motion.

C) The group-theoretical treatment of rigid bodies \cite{lib} [Chapter
IV, Section 10] is based on the existence of the realization of the
(free and transitive) {\it left} and {\it right}  Hamiltonian actions
of the SO(3) rotation group on either the tangent or cotangent bundle
over their configuration space. Given a {\it laboratory or space
frame} ${\hat f}_r$, the generators of the {\it left} Hamiltonian
action\footnote{We follow the convention of Ref.\cite{little}; note
that this action is usually denoted as a {\it right} action in
mathematical texts. The $S^r_q$'s are the Hamiltonians, associated
with the momentum map from the symplectic manifold to $so(3)^{*}$ [the
dual of the Lie algebra $so(3)$],  which allow the implementation of
the symplectic action through Hamiltonian vector fields.} are the
non-Abelian constants of motion $S^1$, $S^2$, $S^3$, [$\{ S^r,S^s
\} = \epsilon^{rsu} S^u$], viz. the spin components in the {\it space frame}.
In the approach of Ref.\cite{little} the SO(3) principal bundle is
built starting from the {\it relative configuration space} and, upon
the choice of a body-frame convention, a gauge-dependent SO(3) {\it
right} action is introduced.

Similarly, taking into account the {\it relative phase space} of any
isolated system, one may investigate whether one or more SO(3) {\it
right} Hamiltonian actions could be implemented besides the global
canonical realization of the SO(3) {\it left} Hamiltonian action,
which is a symmetry action. In other words, one may look for solutions
${\check S}^r$, r=1,2,3, [with $\sum_r ({\check S}^r)^2
= \sum_r (S^r)^2 = S^2$], of the partial differential equations
$\{ S^r, {\check S}^s \} =0$, $\{ {\check S}^r, {\check S}^s \} =-
\epsilon^{rsu} {\check S}^u$ and then build corresponding {\it left} invariant
Hamiltonian vector fields. Alternatively, one may look for the
existence of a pair ${\check S}^3$, $\gamma = tg^{-1} {{{\check
S}^2}\over {{\check S}^1}}$, of canonical variables satisfying $\{
\gamma ,{\check S}^3 \} =-1$, $\{ \gamma ,S^r \} = \{ {\check S}^3,
S^r \} =0$ and also $\{ \gamma ,\alpha \} = \{ {\check S}^3, \alpha \}
=0$. Local theorems given in Refs.\cite{pauri2,pauri1} guarantee that
this is always possible provided $N \geq 3$. See Chapter IV of
Ref.\cite{lib} for what is known in general about the actions of Lie
groups on symplectic manifolds. Clearly, the functions ${\check S}^r$
do not generate symmetry actions, because they are not constants of
the motion.

The inputs coming from A), B), C) together with the technique of the
spin bases of  Ref.\cite{lucenti} suggest the following strategy for
the geometrical and group-theoretical identification of a privileged
class of canonical Darboux bases for the N-body problem:

1) Every such basis must be a {\it scheme B} (i.e. a canonical
completion of {\it scheme A})\cite{pauri2,pauri1} for the canonical
realization of the rotation group SO(3), viz. it must contain its
invariant $S$ and the canonical pair $S^3$, $\beta =
tg^{-1}{{S^2}\over {S^1}}$. Therefore, all the remaining variables in
the canonical basis except $\alpha$ are SO(3) scalars.

2) As said above, the existence of the angle $\alpha$  satisfying $\{
\alpha , S \} =1$ and $\{ \alpha ,S^3 \} = \{ \alpha ,\beta \} =0$
leads to the geometrical identification of a unit vector $\hat R$
orthogonal to $\vec S$ and, therefore, of a orthonormal frame $\hat
S$, $\hat R$, $\hat S\times \hat R$, which will be called {\it spin
frame}\footnote{The notation $\hat {}$ means unit vector.}.

3) The study of the equations ${\hat R}^2=1$ and $\{ \vec S\cdot {\hat
R}, {\hat R}^i \} =0$ entails the symplectic result $\{ {\hat R}^i,
{\hat R}^j \} =0$. As a byproduct we get a canonical realization of an
E(3) group with generators $\vec S$, $\hat R{}{}$ [$\{ {\hat
R}^i,{\hat R}^j \} =0$, $\{ {\hat R}^i, S^j \} = \epsilon^{ijk} {\hat
R}^k$] and fixed values of its invariants ${\hat R}^2=1$, $\vec S\cdot
\hat R=0$  ( non-irreducible type 3 realization according to
Ref.\cite{lucenti}).

4) In order to implement a SO(3) {\it Hamiltonian right action} in
analogy with  the rigid body theory\cite{lib}, we must construct an
orthonormal triad or {\it body frame} $\hat N$, $\hat
\chi$, $\hat N\times \hat \chi$. The decomposition

\beq
\vec S = {\check S}^1\hat \chi +{\check S}^2 \hat N\times \hat \chi +
{\check S}^3 \hat N \,\,\, {\buildrel {def}\over =}\,\,\, {\check S}^r
{\hat e}_r,
\label{I0}
\eeq

\noindent identifies the SO(3) scalar generators ${\check S}^r$ of the right
action provided they satisfy $\{ {\check S}^r,{\check S}^s
\} =
-\epsilon^{rsu} {\check S}^u$. This latter condition
together with the obvious requirement that $\hat N$, $\hat \chi$,
$\hat N\times \hat \chi$ be SO(3) vectors [$\{ {\hat N}^r,S^s
\} =\epsilon^{rsu}{\hat N}^u$, $\{ {\hat \chi}^r,S^s \}
=\epsilon^{rsu}{\hat \chi}^u$, $\{ {\hat N\times \hat \chi}^r,S^s \}
=\epsilon^{rsu}{\hat N\times \hat \chi}^u$] entails the equations
\footnote{With ${\check S}^r=\vec
S\cdot {\hat e}_r$, the conditions $\{ {\check S}^r,{\check S}^s \} =
-\epsilon^{rsu} {\check S}^u$ imply the equations $\vec S\cdot {\hat
e}_r\times {\hat e}_s +S^iS^j \{ {\hat e}^i_r,{\hat e}^j_s\} =
\epsilon_{rsu} S^k{\hat e}_k^u$, hence the quoted result.}

\beq
\{ {\hat N}^r,{\hat N}^s \} = \{ {\hat N}^r,{\hat \chi}^s \} = \{
{\hat \chi}^r,{\hat \chi}^s \} =0.
\label{I1}
\eeq

To each solution of these equations is associated a couple of
canonical realizations of the E(3) group (type 2, non-irreducible):
one with generators $\vec S$, $\vec N$ and non-fixed invariants
${\check S}^3=\vec S \cdot \hat N$ and $|{\vec N}|$; another with
generators $\vec S$, $\vec \chi$ and non-fixed invariants ${\check
S}^1=\vec S\cdot \hat \chi$ and $|{\vec \chi}|$. These latter contain
the relevant information for constructing the angle $\alpha$ and the
new canonical pair ${\check S}^3$, $\gamma =tg^{-1}{{{\check
S}^2}\over {{\check S}^1}}$ of SO(3) scalars. Since $\{  \alpha ,
{\check S}^3 \} = \{ \alpha , \gamma \} =0$ must hold, it follows that
the vector $\vec N$ necessarily belongs to the $\vec S$-$\hat  R$
plane. The three canonical pairs $S$, $\alpha$, $S^3$, $\beta$,
${\check S}^3$, $\gamma$ will describe the {\it orientational}
variables of our Darboux basis, while $|\vec N|$ and $|\vec \chi |$
will belong to the {\it shape} variables. Alternatively, an
anholonomic basis can be constructed by replacing the above six
variables by ${\check S}^r$ and three uniquely determined Euler angles
$\tilde \alpha$, $\tilde \beta$, $\tilde \gamma$.

Let us consider  the case N=3 as a first example. It turns out that a
solution of Eqs.(\ref{I1}) corresponding to a {\it body frame}
determined by the 3-body system configuration only, as in the {\it
rigid body} case, is completely individuated  once two orthonormal
vectors $\vec N$ and $\vec \chi$, functions of the relative
coordinates and independent of the momenta,  are found such that $\vec
N$ lies in the $\vec S$ - $\hat R$ plane\footnote{Let us remark that
any pair of orthonormal vectors $\vec N$, $\vec \chi$ function only of
the relative coordinates can be used to build a body frame. This
freedom is connected to the possibility of redefining a body frame by
using a configuration-dependent arbitrary rotation, which leaves $\vec
N$ in the $\vec S$-$\hat R$ plane.}. We do not known whether in the
case N=3 other solutions of Eqs.(\ref{I1}) exist leading to momentum
dependent body frames. Anyway, our constructive method necessarily
leads to momentum-dependent solutions of for Eqs.(\ref{I1}) for $N\geq
4$ and therefore to momentum-dependent or {\it dynamical body frames}.

We  can conclude that in the N-body problem there are {\it hidden
structures allowing the identification of special dynamical body
frames which, being independent of  gauge conditions, are endowed with
a physical meaning}.

The following particular results can be proven:

i) For $N=2$, a single E(3) group can be defined: it allows the
construction of an orthonormal {\it spin frame} $\hat S$, $\hat R$,
$\hat R\times \hat S$ in terms of the measurable relative coordinates
and momenta of the particles.

ii)  For $N=3$, $\vec S = {\vec S}_1+{\vec S}_2$, a  pair of E(3)
groups emerge associated with ${\vec S}_1$ and ${\vec S}_2$,
respectively. In this case, besides the orthonormal {\it spin frame},
an orthonormal {\it dynamical body frame} $\hat N$, $\hat
\chi$, $\hat N\times \hat \chi$ can be defined such that ${\check
S}^1=\vec S\cdot \hat \chi$, ${\check S}^2=\vec S\cdot \hat N\times
\hat \chi$, ${\check S}^3 = \vec S\cdot \hat N$ are the canonical
generators of a SO(3) Hamiltonian {\it right} action. The
non-conservation of ${\check S}^r$ entails that {\it the dynamical
body frame evolves in a way dictated by the equations of motion}, just
as it happens in the rigid body case.

It will be shown that for N=3 this definition of {\it dynamical body
frame} can be reinterpreted as a special global cross section ({\it
xxzz gauge}, where $x$ stays for $\hat \chi$ and $z$ for $\hat N$;
this outcome is independent from the particular choice made for $\vec
N$ and $\vec \chi$) of the trivial SO(3) principal bundle of
Ref.\cite{little}, namely a privileged choice of body frame. Actually,
the three canonical pairs of orientational variables $S^3$, $\beta =
tg^{-1} {{S^2}\over {S^1}}$; $S$, $\alpha$; ${\check S}^3$, $\gamma =
tg^{-1} {{{\check S}^2}\over {{\check S}^1}}$, can be replaced by the
anholonomic basis of three Euler angles $\tilde \alpha$, $\tilde
\beta$, $\tilde \gamma$ and by ${\check S}^1$, ${\check S}^2$, ${\check S}^3$
as it is done in Ref.\cite{little}. In our construction, however, the
Euler angles $\tilde \alpha$, $\tilde \beta$, $\tilde \gamma$  are
determined as the {\it unique set} of {\it dynamical orientation
variables}. Then, the remaining canonical pairs $q^{\mu}$, $p_{\mu}$,
$\mu =1,2,3$, of this spin-adapted Darboux basis describe the {\it
dynamical shape}  phase space.

While the above {\it dynamical body frame} can be identified with the
global cross section corresponding to the {\it xxzz gauge}, all other
global cross sections cannot be interpreted as {\it dynamical body
frames} (or {\it dynamical right} actions), because the SO(3)
principal bundle of Ref.\cite{little} is built starting from the
relative configuration space and, therefore, it is a {\it static},
velocity-independent, construction. As a matter of fact, after the
choice of the shape configuration variables $q^{\mu}$ and of a space
frame in which the relative variables have components $\rho^r_a$, the
approach of Ref.\cite{little} begins with the definition of the
body-frame components ${\check \rho}^r_a(q^{\mu})$ of the relative
coordinates, in the form $\rho^r_a=R^{rs}(\theta^{\alpha}) {\check
\rho}^s_a(q^{\mu})$  \footnote{$R$ is a rotation matrix, $\theta^{\alpha}$
are arbitrary gauge orientational parameters   and ${\check
\rho}^r_a(q^{\mu})$ is assumed to depend on the shape variables only
and not on their conjugate momenta.}, and then extends it in a {\it
velocity-independent} way to the relative velocities ${\dot
\rho}^r_a\, {\buildrel {def} \over =}\, R^{rs}(\theta^{\alpha})
{\check v}^s_a$. In our construction we get instead
$\rho^r_a=R^{rs}(\tilde \alpha ,\tilde \beta ,\tilde \gamma ) {\check
\rho}^s_a(q^{\mu})$ in the {\it xxzz gauge}, so that
in the present case (N=3)  all {\it dynamical} variables of our
construction coincide with the {\it static} variables in the xxzz
gauge. On  the other hand, in  the relative phase space, the
construction of the {\it evolving dynamical body frame} is based on
non-point canonical transformations

iii) For N=4, where $\vec S = {\vec S}_1+{\vec S}_2+{\vec S}_3$, it is
possible to construct {\it three} sets of {\it spin frames} and {\it
dynamical body frames}  corresponding to the hierarchy of clusterings
$((ab)c)$ [i.e. $((12)3)$, $((23)1)$, $((31)2)$] of the relative spins
${\vec S}_a$  \footnote{In a way analoguous to the angular momentum
composition in quantum mechanics; note that this spin clustering is
independent of the center-of-mass clustering associated with the
Jacobi coordinates: the existence of these two unrelated clusterings
might prove to be a useful and flexible tool in molecular physics.}.
The associated three canonical Darboux bases share the three variables
$S^3$, $\beta$, $S$ (viz. $\vec S$), while both the remaining three
orientational variables and the shape variables depend on the spin
clustering. This entails the existence of three different SO(3) {\it
right} actions with non-conserved canonical generators ${\check
S}^r_{(A)}$, A=1,2,3. Therefore, one can define three anholonomic
bases ${\tilde
\alpha}_{(A)}$, ${\tilde
\beta}_{(A)}$, ${\tilde \gamma}_{(A)}$, ${\check S}^r_{(A)}$ and
associated shape variables $q^{\mu}_{(A)}$, $p_{(A)\mu}$, $\mu
=1,..,6$, connected by canonical transformations leaving $S^r$
fixed. These anholonomic bases and the associated {\it evolving
dynamical body frames}, however, have no relations with the N=4 {\it
static} non-trivial SO(3) principal bundle of Ref.\cite{little}, which
admits only local cross sections. As a matter of fact, one gets
$\rho^r_a=R^{rs}({\tilde \alpha}_{(A)} ,{\tilde \beta}_{(A)} ,{\tilde
\gamma}_{(A)} ) {\check \rho}_{(A)a}(q^{\mu}_{(A)}, p_{(A)\mu}, {\check S}^r_{(A)})$
instead of  $\rho^r_a=R^{rs}(\tilde \alpha ,\tilde \beta ,\tilde
\gamma ) {\check \rho}^s_a(q^{\mu})$.

These results imply that, for N=4, the 18-dimensional relative phase
space  admits three operationally well defined {\it dynamical body
frames}, and associated {\it right} actions, and its coordinates are
naturally splitted in three different ways into 6 dynamical rotational
variables and 12 generalized dynamical shape variables. As a
consequence, we get three possible definitions of {\it dynamical
vibrations}. Each set of 12 generalized dynamical canonical shape
variables is obviously defined modulo canonical transformations so
that it should even be possible to find local canonical bases
corresponding to the local cross sections of the N=4 {\it static}
non-trivial SO(3) principal bundle of Ref.\cite{little}.

Our results can be extended to arbitrary N, with $\vec S =
\sum_{a=1}^{N-1}$ ${\vec S}_a$. There are as many independent ways (say $K$)
of spin clustering as in quantum mechanics. For instance for N=5,
$K=15$ : 12 spin clusterings correspond to the pattern $(((ab)c)d)$
and 3 to the pattern $((ab)(cd))$ [$a, b, c, d = 1,..,4$]. For N=6,
$K=105$: 60 spin clusterings correspond to the pattern
$((((ab)c)d)e)$, 15 to the pattern $(((ab)(cd)e)$ and 30 to the
pattern $(((ab)c)(de))$ [$a, b, c, d, e = 1,..,5$]. Each spin
clustering  is associated to: i) a related {\it spin frame}; ii) a
related {\it dynamical body frame}; iii) a related Darboux spin
canonical basis with orientational variables $S^3$, $\beta$, $S$,
$\alpha_{(A)}$, ${\check S}^3_{(A)}$, $\gamma_{(A)}=tg^{-1}\, {{
{\check S}^2_{(A)}}\over {{\check S}_{(A)}^1}}$, $A=1,.., K$ [their
anholonomic counterparts are ${\tilde \alpha}_{(A)}$, ${\tilde
\beta}_{(A)}$, ${\tilde \gamma}_{(A)}$, ${\check S}^r_{(A)}$ with
uniquely determined orientation angles] and  shape variables
$q^{\mu}_{(A)}$, $p_{\mu (A)}$, $\mu =1,.., 3N-6$. Furthermore, for $N
\geq 4$ we find the following relation between spin and  angular
velocity: ${\check S}^r = {\cal I}^{rs}(q^{\mu}_{(A)})\, {\check
\omega}^s_{(A)} + f^{\mu}(q^{\nu}_{(A)}) p_{(A)\mu}$.

Let us conclude this Introduction with some remarks.

The $\vec S=0$, {\it C-horizontal}, cross section of the {\it static}
SO(3) principal bundle corresponds to N-body configurations that
cannot be included in the previous Hamiltonian construction based on
the canonical realizations of SO(3): these configurations (which
include the singular ones) have to be analyzed independently since
they are related to the exceptional orbit of SO(3), whose little group
is the whole group.

While physical observables have to be obviously  independent  of the
gauge-dependent {\it static body frames}, they do depend on the {\it
dynamical body frame}, whose axes are operationally defined in terms
of the relative coordinates and momenta of the particles. In
particular, a {\it dynamical} definition of {\it vibration}, which
replaces the $\vec S=0$ {\it C-horizontal} cross section of the {\it
static} approach\footnote{Being connected to the Riemannian metric of
the non-relativistic Lagrangian, this concept does not survive the
transition to special relativity anyway.}, is based on the requirement
that the components of the {\it angular velocity} vanish. Actually,
the angular velocities with respect to the dynamical body frames
become now {\it measurable quantities}, in agreement with the
phenomenology of extended deformable bodies (e.g. the treatment of
spinning stars in astrophysics).

In Section II the rest-frame description of N free particles together
with the Jacobi normal relative coordinates is introduced and some
further informations are summarized about the orientation-shape SO(3)
principal bundle approach, both from the Lagrangian and the
Hamiltonian point of view.

In Section III the {\it canonical spin bases}, the {\it spin frame}
and the {\it dynamical body frames} are introduced and the cases
$N=2$, $N=3$ and $N\geq 4$ are analyzed separately.

In Section IV a short account is given of N particles interacting
through a potential.

Some final remarks are given in the Conclusions.

Appendix A contains the Lagrangian and Hamiltonian equations of motion
in the {\it static orientation-shape bundle} approach.

In Appendix  B the Lagrangian and Hamiltonian results of Subsection D
of Section II are reformulated in arbitrary (but not Jacobi normal)
relative coordinates.

In Appendix C detailed calculations are given for the case N=3.

In Appendix D some notions on Euler angles are reviewed.

In Appendix  E the gauge potential in the {\it xxzz gauge} is
evaluated.

In Appendix  F the construction of the {\it canonical spin bases} is
given for the $N=4$ case.

\vfill\eject

\section{The Center of Mass, the Jacobi Relative
Coordinates and  the Orientation-Shape  Bundle.}

In this  Section we first formulate the description of N
non-relativistic free particles in the rest frame and then we review
the theory of the orientation-shape principal SO(3) bundle of
Ref.\cite{little}.

Given the coordinates ${\vec \eta}_i(t)$, $i=1,..,N$, of N particles
of mass $m_i$, the standard Lagrangian is $L= \sum_{i=1}^N {{m_i}\over
2} {\dot {\vec \eta}}_i^2$. By introducing the center-of-mass
coordinates ${\vec q}_{nr}$ and a set of relative variables, the
Lagrangian can be rewritten as $L= {M\over 2}{\dot {\vec q}}_{nr} +$
(quadratic form in the relative velocities), $M=\sum_{i=1}^Nm_i$. The
canonical momenta are ${\vec p}_i=m_i {\dot {\vec \eta}}_i$, while the
total momentum conjugated to ${\vec q}_{nr}$ is $\vec P
=\sum_{i=1}^N{\vec p}_i$ . The Hamiltonian is $H=
\sum_{i=1}^N {{{\vec p}_i^2}\over {2m_i}} = {{{\vec P}^2}\over {2M}} +$
(quadratic form in the relative momenta).

The rest-frame description (equivalent to the decoupling of the center
of mass) is obtained by imposing the vanishing of the conserved total
momentum $\vec P =\sum_{i=1}^N {\vec p}_i =0$

\subsection{The non-relativistic rest-frame description.}

The rest-frame description of the relative motions can be obtained as
the non-relativistic $c\rightarrow \infty$ limit of the relativistic
rest-frame instant form of Ref.\cite{iten1}. Equivalently we can start
from the Lagrangian

\begin{equation}
L_{D}(t)=\sum_{i=1}^N{{m_i}\over 2} [{\dot {\vec \eta}}_i(t)+\vec
\lambda (t)]^2,\quad\quad S_{D}=\int dt L_{D}(t).
\label{II1}
\end{equation}

\noindent in which the Lagrange multipliers $\vec \lambda (t)$ are considered as
configurational variables.

The canonical momenta are

\begin{eqnarray}
 {\vec \kappa}_i(t)&=&{{\partial L_{D}(t)}\over {\partial {\dot
{\vec \eta}}_i(t)}}=m_i[{\dot {\vec \eta}}_i(t)+\vec
\lambda (t)],\nonumber \\
 {\vec \pi}_{\lambda}(t)&=&{{\partial
L_{D}(t)}\over {\partial {\dot {\vec
\lambda}}(t) }} =0.
\label{II2}
\end{eqnarray}

Therefore, ${\vec \pi}_{\lambda}(t) \approx 0$ is a primary
constraint. The canonical and Dirac Hamiltonians are [the variables
$\vec
\mu (t)$ being the Dirac multipliers in front of the primary constraints
${\vec \pi}_{\lambda}(t)\approx 0$]

\begin{eqnarray}
 H_{c}&=&{\vec \pi}_{\lambda}\cdot {\dot {\vec \lambda}}+\sum_{i=1}^N{\vec
\kappa}_i\cdot {\dot {\vec \eta}}_i-L_{D}=\sum_{i=1}^N {{ {\vec \kappa}
_i^{ 2}}\over {2m_i}}-\vec \lambda \cdot {\vec \kappa}_{+},\quad
{\vec \kappa}_{+}=\sum_{i=1}^N{\vec \kappa}_i,\nonumber \\
H_{D}&=&\sum_{i=1}^N {{ {\vec \kappa}_i^{ 2}}\over {2m_i}}-
\vec \lambda \cdot {\vec \kappa}_{+}+\vec \mu \cdot
{\vec \pi}_{\lambda}.
\label{II3}
\end{eqnarray}

The time constancy of the primary constraints implies the following
secondary constraints

\begin{equation}
 {\dot {\vec \pi}}_{\lambda}(t)\, {\buildrel \circ \over =}\,
\{ {\vec \pi}_{\lambda}(t), H_{D} \} = {\vec \kappa}_{+}\approx 0,
\label{II4}
\end{equation}

\noindent which is the non-relativistic rest frame condition.

There are two first class constraints ${\vec \pi}_{\lambda}\approx 0$, ${\vec
\kappa}_{+}\approx 0$ : $\vec \lambda (t)$ and a center-of-mass variable are
gauge variables. The Hamilton and Euler-Lagrange equations are
[${\buildrel \circ \over =}$ means evaluated on the trajectories which
minimize the action principle]

\begin{eqnarray}
{\dot {\vec \eta}}_i(t)\, &{\buildrel \circ \over =}\,& \{ {\vec
\eta}_i(t),H_{D} \}
={{ {\vec \kappa}_i(t)}\over {m_i}}-\vec \lambda (t),\nonumber \\
{\dot {\vec \lambda}}(t)\, &{\buildrel \circ \over =}\,& \vec \mu (t)
,\nonumber \\
 {\dot {\vec \kappa}}_i(t)\, &{\buildrel \circ \over
=}\,& 0,\nonumber \\
 {\dot {\vec \pi}}_{\lambda}(t)\, &{\buildrel \circ
\over =}\,& {\vec \kappa}_{+}\approx 0,\nonumber \\
 &&{}\nonumber \\
 m_i ({\ddot {\vec \eta}}_i+{\dot {\vec \lambda}})(t)\, &{\buildrel
 \circ \over =}\,& 0,\nonumber \\
\sum_{i=1}^Nm_i ({\dot {\vec \eta}}_i+\vec \lambda )(t)\, &{\buildrel \circ
\over =}\,& 0.
\label{II5}
\end{eqnarray}

This is the non-relativistic limit of the relativistic rest-frame
instant form of dynamics: Minkowski spacetime is replaced by Galilei
spacetime and the Wigner hyperplanes are replaced by the inertial
observers seeing the isolated system is istantaneously at rest in the
$t=const.$ hyperplanes.

Defining the non-relativistic center of mass

\begin{equation}
{\vec q}_{nr}=\sum_{i=1}^N{{m_i}\over m}\, {\vec \eta}_i,
\label{II6}
\end{equation}

\noindent with $m=\sum_{i=1}^N
m_i$, the gauge fixing ${\vec q}_{nr}\approx 0$ implies $\vec
\lambda (\tau )\approx 0$  and the decoupling of the center of mass,
see Eq.(\ref{II9}). Instead, the gauge fixing ${\vec
\eta}_{+}= {1\over \sqrt{N}}\sum_{i=1}^N{\vec \eta}_i\approx 0$ does not imply $\vec
\lambda (\tau )\approx 0$ and the decoupling, just as it happens in the
relativistic case\cite{iten1}.

In analogy with the relativistic case of Ref.\cite{lus}, let us
introduce the following family of non-relativistic point canonical
transformations \footnote{They can be used inside the Lagrangian
$L_{D}$; the first one is the non-relativistic analogue of that used
in Ref.\cite{lus}.}

\begin{equation}
\begin{minipage}[t]{4cm}
\begin{tabular}{|l|} \hline
${\vec \eta}_i$ \\  \hline
 ${\vec \kappa}_i$ \\ \hline
\end{tabular}
\end{minipage} \ {\longrightarrow \hspace{.2cm}} \
\begin{minipage}[t]{4cm}
\begin{tabular}{|l|l|}\hline
${\vec \eta}_{+}$& ${\vec \rho}_a$ \\ \hline ${\vec \kappa}_{+}$&
${\vec \pi}_a$ \\ \hline
\end{tabular}
\end{minipage} \ {\longrightarrow \hspace{.2cm}} \
\begin{minipage}[t]{4 cm}
\begin{tabular}{|l|l|} \hline
${\vec q}_{nr}$   & ${\vec \rho}_{qa}={\vec \rho}_a$   \\ \hline
 ${\vec \kappa}_{+}$&${\vec \pi}_{qa}$ \\ \hline
\end{tabular}
\end{minipage}
\label{II7}
\end{equation}

\noindent defined by \footnote{The total angular momentum of the N-body system is
$\vec J=\sum_{i=1}^N {\vec \eta}_i \times {\vec \kappa}_i = {\vec
q}_{nr} \times {\vec \kappa}_{+} + {\vec S}_q = {\vec \eta}_{+}
\times {\vec \kappa}_{+} + \vec S$; $\vec S$ is the barycentric angular momentum or spin.}:

\begin{eqnarray}
{\vec \eta}_i&=&{\vec \eta}_{+} + {1\over {\sqrt{N}}} \sum_{a=1}^{N-1} \gamma
_{ai} {\vec \rho}_a=
 {\vec q}_{nr} + {1\over {\sqrt{N}}} \sum_{a=1}^{N-1} \Gamma_{ai} {\vec
\rho}_{qa},\nonumber \\
 {\vec \kappa}_i&=&{1\over N}{\vec
\kappa}_{+} + \sqrt{N} \sum_{a=1}^{N-1} \gamma_{ai} {\vec \pi}_a=
 {{m_i}\over m} {\vec \kappa}_{+} + \sqrt{N} \sum_{a=1}^{N-1}
\gamma_{ai} {\vec \pi}_{qa},\nonumber \\
 &&{}\nonumber \\
 &&{}\nonumber \\
 {\vec \eta}_{+} &=& {1\over \sqrt{N}}\sum_{i=1}^N{\vec \eta}_i,\nonumber \\
  {\vec \kappa}_{+}&=& \sum_{i=1}^N {\vec \kappa}_i,\nonumber \\
  {\vec \rho}_a&=&\sqrt{N} \sum_{i=1}^N
\gamma_{ai} {\vec \eta}_i,\nonumber \\
 {\vec \pi}_a&=&{1\over {\sqrt{N}}} \sum_{i=1}^N\gamma_{ai}{\vec \kappa}_i
={\vec \pi}_{qa}+{1\over {\sqrt{N}}} (\sum_{k=1}^N{{m_k}\over m}\gamma_{ak})
{\vec \kappa}_{+},\nonumber \\
 &&{}\nonumber \\
 &&{}\nonumber \\
 {\vec q}_{nr}&=&\sum_{i=1}^N{{m_i}\over m} {\vec
\eta}_i = {\vec \eta}_{+}+ {1\over
{\sqrt{N}}}\sum_{a=1}^{N-1}(\sum_{i=1}^N{{m_i}\over m}\gamma_{ai})
{\vec \rho}_a,\nonumber \\
  {\vec \kappa}_{+}&=& \sum_{i=1}^N {\vec \kappa}_i,\nonumber \\
 {\vec \rho}_{qa} &=& {\vec \rho}_a,\nonumber \\
  {\vec \pi}_{qa}&=&{1\over {\sqrt{N}}} \sum_{i=1}^N \Gamma_{ai} {\vec
\kappa}_i
={\vec \pi}_a-{1\over {\sqrt{N}}} (\sum_{k=1}^N{{m_k}\over m}\gamma_{ak})
{\vec \kappa}_{+}\approx {\vec \pi}_{qa},\nonumber \\
 &&{}\nonumber \\
 &&{}\nonumber \\
\vec S&=&\sum_{a=1}^{N-1} {\vec \rho}_a\times {\vec \pi}_a=\sum_{a=1}^{N-1}
{\vec \rho}_a\times {\vec \pi}_{qa}+{1\over
{\sqrt{N}}}\sum_{a=1}^{N-1} (\sum_{k=1}^N{{m_k}\over m}\gamma_{ak})
{\vec \rho}_a\times {\vec \kappa}_{+} \approx \nonumber \\
 &\approx& {\vec S}_q = \sum_{a=1}^{N-1}{\vec \rho}_a\times {\vec \pi}_{qa},\nonumber \\
 &&{}\nonumber \\
 &&{}\nonumber \\
 \Gamma_{ai}&=& \gamma_{ai} - \sum_{k=1}^N {{m_k}\over m} \gamma_{ak},\quad\quad
\gamma_{ai}=\Gamma_{ai}-{1\over N} \sum_{k=1}^N \Gamma_{ak},\nonumber \\
&&\sum_{i=1}^N\gamma_{ai}=0,\quad\quad \sum_{i=1}^N{{m_i}\over
m}\Gamma_{ai}=0,\nonumber \\
&&\sum_{i=1}^N\gamma_{ai}\gamma_{bi}=\delta_{ab},\quad\quad
\sum_{i=1}^N\gamma_{ai} \Gamma_{bi}=\delta_{ab},\nonumber \\
&&\sum_{a=1}^{N-1}\gamma_{ai}\gamma_{aj}=\delta_{ij}-{1\over N},\quad\quad
\sum_{a=1}^{N-1}\gamma_{ai}\Gamma_{aj}=\delta_{ij}-{{m_i}\over m}.
\label{II8}
\end{eqnarray}

Here, the $\gamma_{ai}$'s [and the $\Gamma_{ai}$'s] are numerical
parameters depending on ${1\over 2}(N-1)(N-2)$ free
parameters\cite{lus,lu1}. From now on we shall use the notation ${\vec
\rho}_{qa}$ for ${\vec \rho}_a$.

Then, by using the equations of motion $m [{\dot {\vec
q}}_{nr}(t)+\vec \lambda (t)]\, {\buildrel \circ \over =}\, 0$, we get
the Lagrangian $L_{rel}$ and the Hamiltonian $H_{rel}$ describing the
relative motions after the separation of the center-of-mass motion

\begin{eqnarray}
L_{D}(t)&=&\sum_{i=1}^N{{m_i}\over 2} \Big[ {\dot {\vec q}}_{nr}(t
)+\vec \lambda (t)+{1\over {\sqrt{N}}} \sum_{a=1}^{N-1}
\Gamma_{ai} {\dot {\vec \rho}}_{qa}(t) \Big] {}^2 \,
{\buildrel \circ \over =}\nonumber \\
 &{\buildrel \circ \over =}\,& L_{rel}(t)={1\over 2}
\sum_{a,b}^{1..N-1} k_{ab}[m_i,\Gamma_{ai}]\, {\dot {\vec \rho}}_{qa}(t)\cdot {\dot {\vec
\rho}}_{qb}(t),\nonumber \\
 &&{}\nonumber \\
 &&k_{ab}[m_i,\Gamma_{ci}]=k_{ba}[m_i,\Gamma_{ci}]={1\over N}\sum_{i=1}^N
 m_i\Gamma_{ai}\Gamma_{bi}=\nonumber \\
 &&={1\over N}\Big[ \sum_{i=1}^Nm_i\gamma_{ai}\gamma_{bi}-{{\sum_{h=1}^Nm_h
 \gamma_{ah}\sum_{k=1}^Nm_k\gamma_{bk}}\over m}\Big],
 \nonumber \\
 &&k^{-1}_{ab}[m_i,\Gamma_{ci}] = N \sum_{i=1}^N {{\gamma_{ai}\gamma_{bi}}\over {m_i}}=
 N \sum_{i=1}^N{{\Gamma_{ai}\Gamma_{bi}}\over {m_i}}-\nonumber \\
 &-&\sum_{k=1}^N\big[ \sum_{i=1}^N{{\Gamma_{ai}}\over {m_i}}\sum_{k=1}^N\Gamma_{bk}+
 \sum_{k=1}^N\Gamma_{ak} \sum_{i=1}^N{{\Gamma_{bi}}\over {m_i}}\Big] +{1\over N}
 (\sum_{i=1}^N{1\over {m_i}})\sum_{h=1}^N\Gamma_{ah}\sum_{k=1}^N\Gamma_{bk},\nonumber \\
 &&{}\nonumber \\
 &&\Downarrow \nonumber \\
 {\vec \pi}_{qa}(t)&=&\sum_{b=1}^{N-1} k_{ab}[m_i,\Gamma_{ci}]\,
  {\dot {\vec \rho}}_{qb}(t),\nonumber \\
 &&{}\nonumber \\
\Rightarrow&& H_{rel}={1\over 2} \sum_{ab}^{1..N-1} k^{-1}_{ab}[m_i,\Gamma_{ai}]\,\,
 {\vec \pi}_{qa}(t)\cdot {\vec \pi}_{qb}(t).
\label{II9}
\end{eqnarray}

\noindent The same result can be obtained by adding the gauge fixings
${\vec q}_{nr}\approx 0$ which imply $\vec \lambda (t)=0$, and by
going to Dirac brackets with respect to the second class constraints
${\vec \kappa}_{+}\approx 0$, ${\vec q}_{nr}\approx 0$. The
(6N-6)-dimensional reduced phase space is now spanned by ${\vec
\rho}_{qa}$, ${\vec \pi}_{qa}\equiv {\vec \pi}_a$ and from Eq.(\ref{II8})
we have $\vec S\equiv {\vec S}_q=\sum_{a=1}^{N-1} {\vec \rho}_a\times
{\vec \pi}_{qa}$.

 At the non-relativistic level \cite{little,pack} the next problem for each N
is to diagonalize the matrix $k_{ab}[m_i,\Gamma_{ai}]$. The
off-diagonal terms of the matrix $k_{ab}[m_i,\Gamma_{ai}]$ are called
{\it mass polarization terms}, while its eigenvalues are the {\it
reduced masses}.

\subsection{Jacobi normal relative coordinates.}

There is a discrete set of point transformations

\begin{equation}
{\vec \rho}_a\, \mapsto {\vec s}^{(k)}_a
=\sum_{b=1}^{N-1}\zeta^{(k)}_{ab} {\vec \rho}_{qb},
\label{II10}
\end{equation}

\noindent which defines the relative Jacobi normal coordinates
\footnote{$k=1,..,k_N={1\over 2}(N-1)(N-2)$; for two values $k_1$ and $k_2$ one
has ${\vec s}^{(k_1)}_a=\sum_{b=1}^{N-1} D^{(k_1k_2)}_{ab} {\vec
s}^{(k_2)}_b$, $D^{(k_1k_2)\, T}=D^{(k_1k_2)\, -1}$, with the set of
matrices $D^{(k_1k_2)}$ (democracy transformations or kinematical
rotations) forming the {\it democracy group}, which is a subgroup of
O(N-1) \cite{little,aqui}} and diagonalizes the kinetic energy term of
the Lagrangian

\begin{eqnarray}
k_{ab}&=& k_{ab}[m_i,\Gamma_{ai}]=\zeta^{(k)\, T}_{ac} \Big(
\mu^{(k)}_c
\delta_{cd}\Big)
\zeta^{(k)}
_{db},\nonumber \\
 &&\mu^{(k)}_c\delta_{cd}=\zeta^{(k)\, -1T}_{ca}k_{ab}\zeta
^{(k)\, -1}_{bd},\nonumber \\
&\Downarrow&\nonumber \\
 L_{rel,nr}(t)&=&{1\over 2} \sum_{a=1}^{N-1}
\mu^{(k)}_a {\dot {\vec s}}^{(k)\, 2}
_a(t) = {1\over 2} \sum_{a=1}^{N-1} {\dot {\vec {\tilde s}}}^{(k)\, 2}_a(t),
\nonumber \\
&&{\vec {\tilde s}}^{(k)}_a=\sqrt{\mu^{(k)}_a} {\vec s}^{(k)}_a,
\label{II11}
\end{eqnarray}

\noindent where $\mu^{(k)}_c$ are the reduced masses of the clusters and the
${\vec {\tilde s}}^{(k)}_a$ are called {\it mass-weighted Jacobi
coordinates}. This form of the Lagrangian defines an Euclidean metric
on the relative configuration space.

The general Jacobi coordinates or vectors ${\vec s}^{(k)}_a$ organize
the particles into a ``hierarchy of clusters", in which each cluster,
of mass $\mu^{(k)}_a$, consists of one or more particles and where
each Jacobi vector ${\vec s}^{(k)}
_a$ joins the centers of mass of two clusters, thereby creating a larger
cluster; the discrete set ($k=1,..,k_N$) of choices of Jacobi vectors
corresponds to the possible different clusterings of N particles.
Usually, by ``standard Jacobi coordinates" ${\vec s}_a$ one means the
special set \footnote{$M_i=m_i+M_{i-1}$, $M_N=m=\sum_{i=1}^Nm_i$; for
the reduced masses we have $\mu_{(12)}={{m_1m_2}\over {M_2}}$,
$\mu_{((12)3)}={{ M_2m_3}\over {M_3}}$,...}

\begin{eqnarray}
{\vec s}_1&=&\sqrt{ {{m_1m_2}\over {mM_2}}} \Big[ {\vec \eta}_1-{\vec
\eta}_2\Big]=\sqrt{ {{\mu_{(12)}}\over m} }\Big[ {\vec \eta}_1-{\vec \eta}_2\Big]=\nonumber \\
 &=& \sqrt{ {{\mu_{(12)}}\over {Nm}} } \sum_{a=1}^{N-1}( \Gamma_{a1}-\Gamma_{a2})
 {\vec \rho}_{qa},\nonumber \\
 {\vec s}_2&=&\sqrt{ {{m_3}\over {mM_2M_3}}}\Big[ m_1{\vec
\eta}_1+m_2{\vec \eta}_2-M_2{\vec \eta}_3\Big] =,\nonumber \\
 &=&\sqrt{ {{\mu_{((12)3)}}\over m} } \Big[ {{m_1{\vec \eta}_1+m_2{\vec \eta}_2}\over {M_2}}-
 {\vec \eta}_3\Big] =\nonumber \\
 &=& \sqrt{ {{\mu_{((12)3)}}\over {Nm}} } \sum_{a=1}^{N-1}\Big[ {{m_1\Gamma_{a1}+
 m_2\Gamma_{a2}}\over {M_2}}- \Gamma_{a3}\Big] {\vec \rho}_{qa},\nonumber \\
 &&{}\nonumber \\
  &&......,\nonumber \\
  &&{}\nonumber \\
  {\vec s}_{N-1}&=&\sqrt{
{{m_N}\over {mM_{N-1}M_N}}} \Big[ \sum_{i=1}^{N-1}m_i{\vec \eta}_i
-M_{N-1}{\vec \eta}_N\Big] =\nonumber \\
&=& \sqrt{ {{\mu_{(...(12)3)...)N-1)}}\over m} } \Big[
{{\sum_{i=1}^{N-1}m_i{\vec \eta}_i}\over {M_{N-1}}} -{\vec
\eta}_N\Big]=\nonumber \\
 &=& \sqrt{ {{\mu_{(...(12)3)...)N-1)}}\over {Nm}} } \sum_{a=1}^{N-1}\Big[
{{\sum_{i=1}^{N-1}m_i\Gamma_{ai}}\over {M_{N-1}}} -\Gamma_{aN}\Big]
{\vec \rho}_{qa},
\label{II12}
\end{eqnarray}

\noindent in which ${\vec s}_1$ joins particles 1 and 2 and is directed towards 1,
while ${\vec s}_a$ is directed from the (a+1)th particle to the center
of mass of the first {\it a} particles \footnote{See Ref.\cite{aqui}
for more details on the general Jacobi coordinates and on special
classes of them (like the Radau ones) treating the particles either in
a more symmetric way or according to a more complex patterns of
clustering; they are connected to the standard ones by {\it kinematic
rotations} belonging to the {\it democracy group}.}.

Let us remark that our family [there are ${1\over 2}(N-1)(N-2)$  free
parameters inside the $\gamma_{ai}$'s] of point canonical
transformations of Eqs. (\ref{II7})  contains as a special case the
transition to the normal Jacobi coordinates of Eqs.(\ref{II13}).

An induced set of canonical transformations from the canonical basis
${\vec \rho}_{qa}$, ${\vec \pi}_{qa}$ to the Jacobi bases is the
following

\begin{eqnarray}
{\vec s}^{(k)}_a&=&\sum_{b=1}^{N-1} \zeta^{(k)}_{ab} {\vec \rho}_{qb}
=\sqrt{N} \sum_{i=1}^N \sum_{b=1}^{N-1}
\zeta^{(k)}_{ab} \gamma_{bi} {\vec \eta}_i,\nonumber \\
{\vec \pi}^{(k)}_{sa}&=&\sum_{b=1}^{N-1} {\vec \pi}_{qb} \zeta^{(k)\, -1}_{ba}
={1\over {\sqrt{N}}} \sum_{i=1}^N\sum_{b=1}^{N-1} \Gamma_{bi}\zeta^{(k)\, -1}
_{ba} {\vec \kappa}_i,\nonumber \\
 &&{}\nonumber \\
 &&\delta^{rs}\delta_{ab}=\{ \rho^r_{qa},\pi^s_{qb} \} = \{ s^{(k)\,
r}_a,\pi^{(k)\, s}_{sb} \} ,\nonumber \\
 &&{}\nonumber \\
  {\vec S}_q&=& \sum_{a=1}^{N-1} {\vec \rho}_{qa}\times {\vec \pi}_{qa} = \sum
_{a=1}^{N-1} {\vec s}^{(k)}_a\times {\vec \pi}^{(k)}_{sa},\nonumber \\
 &&{}\nonumber \\
\Rightarrow&& H_{rel}=\sum_{a=1}^{N-1} {{ {\vec \pi}_{sa}^{(k)\, 2}(\tau )}\over
{2 \mu^{(k)}_a}},\nonumber \\
 &&{}\nonumber \\
  {\vec \eta}_i&=&{\vec
q}_{nr}+{1\over {\sqrt{N}}} \sum_{ab}^{1..N-1} \Gamma
_{ai} \zeta^{(k)\, -1}_{ab} {\vec s}^{(k)}_b,\nonumber \\
{\vec \kappa}_i&=&{{m_i}\over M}{\vec \kappa}_{+}+\sqrt{N} \sum_{ab}^{1..N-1}
\gamma_{ai} \zeta^{(k)}_{ab} {\vec \pi}^{(k)}_{sb}.
\label{II13}
\end{eqnarray}

\subsection{More about the {\it static} orientation-shape SO(3) principal bundle approach.}

As said in the Introduction, the attempt of decoupling {\it absolute
global rotations} from {\it vibrational degrees of freedom} led to the
development of the {\it static} theory of the orientation-shape SO(3)
principal bundle \cite{little}, which generalizes the traditional
concept of {\it body frame} of rigid bodies.  This bundle is a
non-trivial (for $N \geq 4$) principal SO(3)-bundle with the
(3N-6)-dimensional shape manifold (with coordinates $q^{\mu}$) as base
and standard fiber SO(3) (parametrized e.g. by the Euler angles
$\theta^{\alpha}$).  For each given shape we need:  i) the assignement
of an arbitrary {it reference frame}; ii) the assignement of a {\it
body frame}, identified by the value of the orientational variables
$\theta^{\alpha}$ with respect to the reference frame. Recall that the
$\theta^{\alpha}$'s are {\it gauge} variables in this approach.
.

A convention  about which is the {\it body frame} for generic
configurations of the N-body system, namely a local cross section of
the non-trivial orientation-shape bundle, is equivalent to two
independent statements: i) a choice of a set of SO(3)-scalar shape
variables $q^{\mu}$, $\mu=1,..,3N-6$; ii) a choice of the explicit
form of the components ${\check \rho}^r_a(q^{\mu})$ [${\check
s}^{(k)r}_a(q^{(k)\mu})$] of the relative coordinate vectors with
respect to the chosen body frame axes \footnote{This is the choice of
a gauge for the orientation variables, independent of the shape
coordinates $q^{\mu}$; for each shape, one gives the positions of the
N particles relative to the body frame axes ${\hat e}_r$. The
orientation variables, for example the Euler angles $\theta^{\alpha}$,
identify an SO(3)-element $R(\theta^{\alpha})$ in the fiber over the
given shape $q^{\mu}$; the {\it reference orientation} for each shape
is such that $\rho^r_{qa}={\check \rho}^r_{qa}$.}. These components
are connected to the coordinates $\rho^r_a$ in the {\it space frame}
by

 \begin{equation}
 \rho^r_{qa} = R^{rs}(\theta^{\alpha}) {\check
\rho}^s_{qa}(q^{\mu})\quad\quad [s^{(k)r}_a=R^{rs}(\theta^{\alpha}){\check
s}^{(k)s}_a(q^{(k)\mu})].
\label{II14}
\end{equation}

\noindent The same relation holds for the components of every vector,
like $S_q^r=R^{rs}(\theta^{\alpha}) {\check S}^s_q$. If ${\hat f}_r$
are the space frame axes and  ${\hat e}_r$ the axes of the chosen body
frame, we have  ${\vec \rho}_{qa} = \rho^r_{qa}{\hat f}_r
={\check \rho}^r_{qa}{\hat e}_r$ and ${\vec S}_q=S^r_q {\hat f}_r=
{\check S}^r_q {\hat e}_r$ for each possible shape of the N-body
system. In particular the SO(3)-scalars have the same functional form
in both space and body frames: $ A^r B^r={\check A}^r {\check B}^r$.
See Refs.\cite{con2,con1,con3,con4,con5} for the mathematical and
physical aspects of the {\it orientation-shape principal bundle}
approach.

The main result from the theory of the orientation-shape bundle is
that the transitions among different body frame conventions are
interpreted as gauge transformations among the local cross sections of
the principal bundle. Therefore, a gauge transformation is a
shape-dependent proper rotation $S(q)\in SO(3)$ that maps the body
frame with axes ${\hat e}_r$ into the body frame with axes ${\hat
e}_r^{'}$ \footnote{We have ${\hat e}_r=S_{rs}(q){\hat e}_s^{'}$,
$R=R^{'}S^T(q)$, $\vec A=A^r{\hat f}_r={\check A}^r{\hat e}_r={\check
A}^{{'}r}{\hat e}_r^{'}$, $A^r=R^{rs}{\check A}^s=R^{{'}\, rs}{\check
A}^{{'}\, s}$. The new orientation angles $\theta^{{'}\, \alpha}$
depend on the old ones $\theta^{\alpha}$ and on the shape variables
$q^{\mu}$ too.}. Instead of this passive change of coordinates on the
fibers, one can consider an active (gauge dependent) {\it right
action} of SO(3): $(R,q^{\mu}) \mapsto (RQ,q^{\mu})$, $Q\in SO(3)$.
The corresponding symplectic right action in phase space, associated
with the left-invariant vector fields on SO(3), is generated by the
non-conserved body frame spin components ${\check S}^r_q$. On the
other hand, the {\it left action} of SO(3) $\rho^r_{qa}
\mapsto Q^{rs}\rho^s_{qa}$, $(R,q^{\mu})\mapsto (QR,q^{\mu})$, $Q\in
SO(3)$ \footnote{I.e. the action of the structure group on the bundle,
which is independent  of the choice of any cross section and  is
called a `gauge-invariant action'.} is generated by the space frame
spin components $S^r_q$ (Noether constants of motion), associated to
the right-invariant vector fields on SO(3). As already said, it holds
$\{ S_q^r,S_q^s \}
=\epsilon^{rsu}S_q^u$, $\{ {\check S}_q^r,{\check S}_q^s \}
=-\epsilon^{rsu}{\check S}_q^u$, $\{ S_q^r,{\check S}_q^s \} =0$.

In conclusion, within the static orientation-shape bundle approach the
$6N-6$ shape variables $q^{\mu}$ are gauge invariant quantities,
because their definition does not depend on the body frame convention,
while the body frame components of any vector are gauge quantities. On
the other hand, let us  stress that only the vectors independent of
the body frame convention have their space frame components as
observable physical quantities. The angular velocity $\vec \omega$ is
a clear instantiation of a body frame dependent vector, for which both
the body frame ${\check \omega}^r$ and the space frame $\omega^r$
components are gauge quantities. Note that they are not even {\it
gauge covariant} \footnote{A gauge covariant quantity ${\check A}^r$
transforms as ${\check A}^r=S^{rs}(q){\check A}^{{'} s}$.}, because
under a change of body frame $R(\theta^{\alpha}) \mapsto
R^{'}(\theta^{\alpha})=R(\theta^{\alpha}) S(q^{\mu})$, it holds
${\check \omega}^r=S^{rs} [{\check \omega}^{{'} s}-\gamma^s_{\mu}{\dot
q}^{\mu}]$ with $\epsilon^{rsu}\gamma^u_{\mu}{\dot q}^{\mu}=[S^T\dot
S]^{rs}$, so that $\omega^r\not= \omega^{{'} r}$.

These results suggest to consider local point canonical
transformations of the form [$\mu =1,..,3N-6$]

\begin{equation}
\begin{minipage}[t]{4cm}
\begin{tabular}{|l|} \hline
${\vec \rho}_{qa}$ \\  \hline
 ${\vec \pi}_{qa}$ \\ \hline
\end{tabular}
\end{minipage} \ {\longrightarrow \hspace{.2cm}} \
\begin{minipage}[t]{4cm}
\begin{tabular}{|l|}\hline
${\vec s}^{(k)}_a$ \\ \hline
 ${\vec \pi}^{(k)}_{qa}$ \\ \hline
\end{tabular}
\end{minipage} \ {\longrightarrow \hspace{.2cm}} \
\begin{minipage}[t]{4 cm}
\begin{tabular}{|l|l|} \hline
$\theta^{\alpha}$   & $q^{\mu}$   \\ \hline
 $p_{\alpha}$&$p_{\mu}$ \\ \hline
\end{tabular}
\end{minipage}
\label{II15}
\end{equation}

\noindent They define a canonical basis, in which the local ``orientation"
coordinates $\theta^{\alpha}$, $\alpha =1,2,3$ are either the Euler
angles or any other parametrization of the group manifold of SO(3).

\subsection{Non-Relativistic Rotational Kinematics in Jacobi Coordinates}

In this Subsection, using Jacobi coordinates, we shall elucidate the
Lagrangian and Hamiltonian treatment of the orientation-shape bundle
approach. In Appendix B a reformulation of these results is given in
terms of arbitrary (non-Jacobi normal) relative coordinates.

Given a set of Jacobi coordinates $s^{(k)r}_a$, let us introduce the
associated body frame coordinates ${\check s}_a^{(k)r}$ and velocities
${\check v}^{(k)r}_a$

\begin{eqnarray}
s^{(k)r}_a&=& R^{rs}(\theta^{\alpha})\, {\check
s}^{(k)s}_a(q^{(k)}),\qquad {\vec s}^{(k)}_a=s^{(k)r}_a{\hat
f}_r={\check s}^{(k)r}_a{\hat e}_r,\nonumber \\
 &&{}\nonumber \\
 {\dot s}^{(k)r}_a&=&{\dot R}^{rs}(\theta^{\alpha})\, {\check
s}^{(k)s}_a(q^{(k)})+ R^{rs}(\theta^{\alpha})\, {{\partial {\check
s}^{(k)s}_a(q^{(k)})}\over {\partial q^{(k)\mu}}} {\dot q}^{(k)\mu}\,
{\buildrel {def} \over =}\, R^{rs}(\theta^{\alpha})\, {\check v}
^{(k)s}_{sa},\nonumber \\
 &&{}\nonumber \\
 {\check v}^{(k)r}_{sa} &{\buildrel {def} \over =}& R^{T\, rs}(\theta^{\alpha})\, {\dot
s}^{(k)s}_a=[R^T(\theta^{\alpha})
\dot R(\theta^{\alpha})]^{rs}\, {\check s}^{(k)s}_a+{{\partial {\check s}^{(k)r}_a(q
^{(k)})}\over {\partial q^{(k)\mu}}}{\dot q}^{(k)\mu}\, {\buildrel {def}
\over =}\, \nonumber \\
&{\buildrel {def} \over =}\,& \epsilon^{rus} {\check
\omega}^u(\theta^{\alpha},\dot \theta^{\alpha})\,
{\check s}_a^{(k)s}(q^{(k)})+{{\partial {\check s}^{(k)r}_a(q
^{(k)})}\over {\partial q^{(k)\mu}}}{\dot q}^{(k)\mu}.
\label{II16}
\end{eqnarray}

The body frame components of the angular velocity $\vec \omega$ are

\begin{equation}
 {\check \omega}^r(\theta^{\alpha},\dot \theta^{\alpha})=-
 {1\over 2}\epsilon^{ruv}[R^T\dot R]^{uv}(\theta^{\alpha},\dot \theta^{\alpha})
 ={1\over 2}\epsilon^{ruv}[{\dot R}^TR]^{uv}(\theta^{\alpha},\dot \theta^{\alpha}).
 \label{II17}
 \end{equation}

As said before, also the space frame angular velocity components
$\omega^r(\theta^{\alpha},\dot
\theta^{\alpha})=R^{rs}(\theta^{\alpha})\, {\check \omega}^s(\theta^{\alpha},
\dot \theta^{\alpha})$ are  gauge dependent.

The Jacobi momenta are

\begin{eqnarray}
\pi^{(k)r}_{sa}&=&\mu^{(k)}_a {\dot s}^{(k)r}_a=R^{rs}(\theta^{\alpha}){\check \pi}
^{(k)s}_{sa},\nonumber \\
{\check \pi}^{(k)r}_{sa}&=&\mu^{(k)}_a \Big[ ({\vec
\omega}\times {\vec s}_a)^r +{{\partial {\check s}^r_a}\over
{\partial q^{(k)\mu}}} {\dot q}^{(k)\mu} \Big],
\label{II18}
\end{eqnarray}

For the spin we have

\begin{eqnarray}
S^r_q&=&R^{rs}(\theta^{\alpha}) {\check S}_q^r =
\sum_{a=1}^{N-1}({\vec s}^{(k)}_a\times {\vec
\pi}^{(k)}_{sa})^r,\nonumber \\
 {\vec S}_q&=& \sum_{a=1}^{N-1}{\vec s}^{(k)}_a\times {\vec
\pi}^{(k)}_{sa} =\sum_{a=1}^{N-1}
\mu^{(k)}_a {\vec s}
^{(k)}_a\times {\vec v}^{(k)}_{sa}=\nonumber \\
&=&\sum_{a=1}^{N-1}\mu^{(k)}_a \Big[ {\vec s}_a^{(k)\, 2}{\vec
\omega}-{\vec s}^{(k)}_a\cdot {\vec
\omega} {\vec s}^{(k)}_a+{\vec s}^{(k)}_a\times
{{\partial {\vec s}
^{(k)}_a}\over {\partial q^{(k)\mu}}}{\dot q}^{(k)\mu} \Big] .
\label{II19}
\end{eqnarray}

By introducing the Euclidean tensors\footnote{In the relativistic
case\cite{iten1}, where the tensor of inertia does not exist, only the
tensors (\ref{II20}) can be defined.}

\begin{equation}
{\check I}^{uv}_{(aa)}(q^{(k)})={\vec s}^{(k) 2}_a
\delta^{uv}-{\check s}^{(k) u}_a  {\check s}^{(k) v}_a,
\label{II20}
\end{equation}

\noindent and the body frame barycentric inertia tensor
\footnote{This tensor is gauge covariant, while
${\check I}^{(k)uv}=S^{ur}(q) {\check I}^{(k){'} rs} (S^T)^{sv}(q)$;
the space frame inertia tensor $I^{uv}=R^{ur}R^{vs}{\check I}^{rs}$ is
gauge invariant.}

\begin{equation}
{\check I}^{(k)uv}(q^{(k)},\mu^{(k)})=\sum_{a=1}^{N-1}\mu_a^{(k)}\,\,
{\check I}^{uv}_{(aa)}(q^{(k)}),
\label{II21}
\end{equation}

\noindent we get the following expression of the body frame spin
components\footnote{For ${\dot q}^{\mu}=0$ we get the rigid body
result ${\check S}^r_q={\check I}^{(k)rs} {\check \omega}^s$. Let us
remark that also these relations  no longer hold in the relativistic
case \cite{iten1}.}

\begin{eqnarray}
{\check S}_q^u&=& \sum_{a=1}^{N-1} \mu^{(k)}_a \Big[ {\check
I}^{uv}_{(aa)}(q^{(k)})\, {\check
\omega}^v+{\check a}^u_{(aa)\mu}(q^{(k)})\, {\dot q}^{(k)\mu} \Big] =\nonumber
\\ &=& \sum_{a=1}^{N-1}\mu^{(k)}_a {\check I}^{uv}_{(aa)}(q^{(k)})
\Big[ {\check \omega}^v +{\check A}^v_{(aa)\mu}(q^{(k)})\, {\dot q}^{(k)\mu}
\Big] =\nonumber \\
 &=& {\check I}^{(k) uv}(q^{(k)},\mu^{(k)})\, {\check \omega}^v
+{\check a}^{(k)u}_{\mu}(q^{(k)},\mu^{(k)})\, {\dot
q}^{(k)\mu}=\nonumber
\\ &=&{\check I}^{(k)uv}(q^{(k)},\mu^{(k)}) \Big[ {\check \omega}
^v+{\check A}^{(k)v}_{\mu}(q^{(k)},\mu^{(k)})\, {\dot q}^{(k)\mu}\Big],\nonumber \\
 &&{}\nonumber \\
 \Rightarrow&& S^u_q = I^{(k)uv}(q^{(k)},\mu^{(k)})\Big[ \omega^v
 +A^{(k)v}_{\mu}(q^{(k)},\mu^{(k)})\, {\dot q}^{(k)\mu}\Big],\quad
 A^{(k)v}_{\mu}=R^{vu}(\theta^{\alpha}){\check A}^{(k)u}_{\mu},\nonumber \\
  &&{}
\label{II22}
\end{eqnarray}

\noindent where

\begin{eqnarray}
&&{\check a}^u_{(aa)\mu}(q^{(k)})=\Big( {\vec s}^{(k)}_a\times
{{\partial {\vec s}^{(k)}_a}\over {\partial q^{(k)\mu}}}\Big)^r\,
{\buildrel {def}\over
=}\, {\check I}^{uv}_{(aa)}(q^{(k)})\, {\check A}^v_{(aa)\mu}(q^{(k)}),\nonumber \\
 &&{\check a}^{(k)u}_{\mu}(q^{(k)},\mu^{(k)})=\sum_{a=1}^{N-1}\mu^{(k)}_a
  {\check a}^u_{(aa)\mu}(q^{(k)})=\sum_{a=1}^{N-1}\mu^{(k)}_a
  {\check I}^{uv}_{(aa)}(q^{(k)})\, {\check A}^v_{(aa)
\mu}(q^{(k)})=\nonumber \\
&=&{\check I}^{(k)uv}(q^{(k)},\mu^{(k)})\, {\check
A}^{(k)v}_{\mu}(q^{(k)},\mu^{(k)}).
\label{II23}
\end{eqnarray}

The quantity ${\vec A}^{(k)}_{\mu}(q^{(k)},\mu^{(k)})$ is the SO(3)
gauge potential of the orientation-shape bundle formulation
\footnote{See Ref.\cite{little} for the monopole-like singularities of the gauge
potential at the N-body collision configuration.}. Note that it is not
gauge covariant: ${\check A}^{(k)r}_{\mu}=S^{rs}(q)[{\check
A}_{\mu}^{(k){'} s}+\gamma_{\mu}^s]$. Its field strength (curvature
form), called the Coriolis tensor, is the gauge covariant quantity
[${\check B}^{(k)r}_{\mu\nu}=S^{rs}(q) {\check B}^{(k){'}
s}_{\mu\nu}$]

\begin{equation}
{\vec B}^{(k)}_{\mu\nu}={{\partial {\vec A}^{(k)}_{\nu}}\over
{\partial q^{(k)\mu}}}-{{\partial {\vec A}^{(k)}_{\mu}}\over {\partial
q^{(k)\nu}}}- {\vec A}^{(k)}_{\mu}\times {\vec A}^{(k)}_{\nu}.
\label{II24}
\end{equation}

Let us stress that Eq.(\ref{II22}) does not provide an effective
separation of rotational and internal (vibrational) contributions to
the angular momentum, since the separation is {\it gauge} dependent
within this approach .

Eq.(\ref{II22}) can be inverted to express the body frame angular
velocity components in terms of the body frame spin components and of
the gauge potential

\begin{equation}
 {\check \omega}^u=[{\check I}^{(k)\,-1}(q^{(k)},\mu^{(k)})]^{uv}{\check S}_q^v-
{\check A}^{(k)u}_{\mu}(q^{(k)},\mu^{(k)})\, {\dot q}^{(k)
\mu}.
\label{II25}
\end{equation}

The non-relativistic Lagrangian for relative motions can then be
rewritten in the following forms

\begin{eqnarray}
 L_{rel}&=&{1\over 2}\sum_{ab}^{1..N-1}k_{ab}{\dot {\vec
\rho}}_a\cdot {\dot {\vec \rho}}_b={1\over 2} \sum_{a=1}^{N-1}
\mu^{(k)}_a {\dot {\vec s}}^{(k)}_a= {1\over 2} \sum_{a=1}^{N-1}
\mu^{(k)}_a {\check {\vec v}}^{(k) 2}_{sa}=
\nonumber \\
&=&{1\over 2} \Big( {\check I}^{(k)uv}{\check \omega}^u{\check
\omega}^v+2{\check a}^{(k)u}
_{\mu}{\check \omega}^u{\dot q}^{(k)\mu}+h^{(k)}_{\mu\nu}{\dot q}^{(k)\mu}
{\dot q}^{(k)\nu} \Big)=\nonumber \\
 &=&{1\over 2} \Big[ {\check
I}^{(k)uv}({\check \omega}^u+{\check A}^{(k)u}_{\mu}{\dot q}
^{(k)\mu})({\check \omega}^v+{\check A}^{(k)v}_{\nu}{\dot q}^{(k)\nu})+g^{(k)}_{\mu\nu}
{\dot q}^{(k)\mu}{\dot q}^{(k)\nu} \Big] ,\nonumber \\
 &=&{1\over 2}
\Big[ ({\check I}^{(k) -1})^{uv} {\check S}_q^u{\check S}_q^v+g^{(k)}_{\mu\nu}
{\dot q}^{(k)\mu} {\dot q}^{(k)\nu} \Big] {\buildrel {def}
\over =}\,
  L_{rel} [\mu^{(k)}, {\check {\vec \omega}}(\theta^{\alpha},{\dot \theta}^{\alpha}),
 q^{\mu}, {\dot q}^{\mu}],
 \label{II26}
 \end{eqnarray}

\noindent where

\begin{equation}
h^{(k)}_{\mu\nu}(q^{(k)},\mu^{(k)})=\sum_{a=1}^{N-1}\mu^{(k)}_a{{\partial
{\check {\vec s}}_a}\over {\partial q^{(k)\mu}}}\cdot {{\partial
{\check {\vec s}}_a}\over {\partial q^{(k)\nu}}}{\dot q}^{(k)\mu}{\dot
q}^{(k)\nu},
\label{II27}
\end{equation}

\noindent is a pseudo-metric on shape space \footnote{It is neither gauge invariant
nor gauge covariant:
$h^{(k)}_{\mu\nu}=h^{(k){'}}_{\mu\nu}+\gamma^r_{\mu}{\check I}^{(k){'}
rs}{\check A}^{(k){'} s}_{\nu}+\gamma^r_{\nu}{\check I}^{(k){'}
rs}{\check A}^{(k){'} s}_{\mu}+{\check A}^{(k){'} r}_{\mu}{\check
I}^{(k){'} rs} {\check A}^{(k){'} s}_{\nu}$.}, while

\begin{equation}
g^{(k)}_{\mu\nu}(q^{(k)},\mu^{(k)})=\Big[ h^{(k)}_{\mu\nu}-{\check
A}^{(k)u}_{\mu} {\check I}^{(k)uv} {\check A}^{(k)v}_{\nu}\Big]
(q^{(k)},\mu^{(k)}),
\label{II28}
\end{equation}

\noindent is a true gauge invariant  metric on shape space\cite{little}
[$g^{(k)}_{\mu\nu}= g^{(k){'}}_{\mu\nu}$] \footnote{It can be shown
\cite{little} that the inverse metric is $g^{\mu\nu}=\sum_{a=1}^{N-1}
{{\partial q^{\mu}}\over {\partial {\vec
\rho}_a}} \cdot {{\partial q^{\nu}}\over {\partial {\vec \rho}_a}}$.}.

A manifestly gauge invariant separation between rotational and
vibrational kinetic energies is exhibited only in the last two lines
of Eq.(\ref{II26}) \footnote{For ${\dot q}^{\mu}=0$ we get the rigid
body result $L={1\over 2} {\check I}^{(k)rs} {\check
\omega}^r{\check \omega}^r={1\over 2} I^{(k)rs} \omega^r\omega^s$.}.
In order to clarify this point, a velocity multivector $({\dot
\theta}^{\alpha}, {\dot q}^{\mu})$ was introduced in Ref.\cite{little}
together with its anholonomic version $({\check \omega}^r,{\dot
q}^{\mu})$ in which Euler angle velocities are replaced by the body
frame angular velocity components. A metric tensor for these
multivectors is naturally induced by the Euclidean metric of the
kinetic energy. The intrinsic notion of {\it vertical vector fields}
of the SO(3) principal orientation-shape bundle corresponds to the
{\it purely rotational} velocity multivectors $({\check \omega}^r, 0)$
defined by the gauge invariant condition ${\dot q}^{\mu}=0$. On the
other hand, there is no gauge invariant definition of {\it purely
vibrational} velocity multivectors\footnote{For instance the simplest
choice ${\check \omega}^r=0$ is clearly not gauge invariant.}, since
any such definition is connected to a {\it horizontal cross section}
of the principal bundle and, therefore, to the assignement of a
connection form. Each connection gives a definition of {\it
horizontality} and the possibility, through a gauge fixing, to choose
a certain {\it horizontal cross section} as a {\it
connection-dependent definition of vibration}.

In Ref.\cite{little} it is shown that a special connection C can be
defined by requiring that the C-horizontal vector fields are
orthogonal (in the sense of the  multivector metric) to the vertical
ones and that the associated C-horizontal cross sections (defined only
locally for $N\geq 4$) are identified by the vanishing of the body
frame spin components ${\check S}^r=0$ \footnote{A system velocity is
C-horizontal if and only if the associated spin vanishes and
horizontal vector fields describe purely vibrational effects in a
gauge invariant way; this also implies that the C-horizontal cross
sections cannot be interpreted as $(6N-6)$-dimensional submanifolds of
the configuration space.}. By privileging the connection C, we get the
following splitting of an arbitrary velocity multivector into vertical
and C-horizontal parts

\begin{eqnarray}
({\check \omega}^r, {\dot q}^{\mu})&=&({\check \omega}^r, {\dot
q}^{\mu})_v+({\check \omega}^r, {\dot q}^{\mu})_{Ch},\nonumber \\
 &&{}\nonumber \\
 ({\check \omega}^r, {\dot q}^{\mu})_v&=&({\check \omega}^r+{\check
 A}^{(k)r}_{\mu}{\dot q}^{\mu},0),\nonumber \\
 ({\check \omega}^r, {\dot q}^{\mu})_{Ch}&=&(-{\check A}^{(k)r}_{\mu}{\dot
 q}^{\mu}, {\dot q}^{\mu}).
\label{II29}
\end{eqnarray}

\noindent It is just this C-splitting which identifies the metric (\ref{II28})
on shape space\cite{little} and the manifestly gauge invariant
separation of the kinetic energy given in the last lines of
Eq.(\ref{II26}).

The standard fiber of the orientation-shape bundle is SO(3). Its group
manifold admits many parametrizations. If one uses the local
parametrization given by the Euler angles $\theta^{\alpha}$
\footnote{See Ref.\cite{deaz} for the parametrization  of the SO(3) group manifold
with a 3-vector $\vec \epsilon$ determining the rotation axis and the
rotation angle $\psi$ by $|\vec \epsilon | = 2sin\, {{\psi}\over 2}$.
.} the form of the right-invariant vector fields and 1-forms on
SO(3) \footnote{Recall that they are the generators of the
infinitesimal left translations on SO(3) and are thought as body frame
quantities.} is

\begin{eqnarray}
{\check X}^{(R)}_{(\alpha )}&=&{\check X}^{(R)}{}^{\beta}_{(\alpha
)}(\theta^{\gamma})
 {{\partial}\over {\partial \theta^{\beta}}},\quad\quad
 [ {\check X}^{(R)}_{(\alpha )},{\check X}^{(R)}_{(\beta )} ]=
 -\epsilon_{\alpha\beta\gamma} {\check X}^{(R)}_{(\gamma )},\nonumber \\
 {\check \Lambda}^{(R)}{}^{(\alpha )}&=&{\check \Lambda}^{(R)}{}^{(\alpha )}_{\beta}
 (\theta^{\gamma}) d\theta^{\beta},\nonumber \\
 &&{}\nonumber \\
 &&{\check X}^{(R)}{}^{\beta}_{(\alpha )}(\theta^{\gamma})=
 [{\check \Lambda}^{(R)-1}]{}^{\beta}_{(\alpha )}(\theta^{\gamma}),\nonumber \\
 {\check \Lambda}^{(R)}{}^{(\alpha )}_{\beta}(\theta^{\gamma})&=&-{1\over 2} \epsilon
 _{\alpha\gamma\delta}R_{\zeta\gamma}(\theta^{\rho}) {{\partial R_{\zeta\delta}(\theta^{\rho})}
 \over {\partial \theta^{\beta}}},\nonumber \\
 &&{{\partial R_{\alpha\beta}(\theta^{\rho})}\over {\partial \theta^{\gamma}}}=
 \epsilon_{\beta\delta\zeta} R_{\alpha\delta}(\theta^{\rho})
 {\check \Lambda}^{(R)}{}^{(\zeta )}_{\gamma};
 \label{II30}
 \end{eqnarray}

\noindent If the Euler angles are defined by the convention
$R(\theta^{\alpha})=R_z(\theta^1) R_y(\theta^2) R_z(\theta^3)$, one
has\hfill\break
\hfill\break
$\left( {\check \Lambda}^{(R)}{}^{(\alpha )}_{\beta} \right) =
\left(  \begin{array}{ccc}
-sin\, \theta^2 cos\, \theta^3 & sin\, \theta^3 & 0\\
sin\, \theta^2 sin\, \theta^3 & cos\, \theta^3 & 0\\
 cos\, \theta^2 & 0 & 1 \\
 \end{array} \right) $
 and $det\, {\check \Lambda}^{(R)}= -sin\, \theta^2$.

The left-invariant vector fields on SO(3) \footnote{Recall that they
are the generators of the infinitesimal right translations on SO(3)
and interpreted as space frame quantities.} are  ($[ {\check
X}^{(R)}_{(\alpha )},X^{(L)}_{(\beta )} ] =0$)

 \begin{eqnarray}
X^{(L)}_{(\alpha )}&=&X^{(L)}{}^{\beta}_{(\alpha )}(\theta^{\gamma})
 {{\partial}\over {\partial \theta^{\beta}}},\quad\quad
 [ X^{(L)}_{(\alpha )},X^{(L)}_{(\beta )} ]=
 \epsilon_{\alpha\beta\gamma} X^{(L)}_{(\gamma )},\nonumber \\
 \Lambda^{(L)}{}^{(\alpha )}&=&\Lambda^{(L)}{}^{(\alpha )}_{\beta}
 (\theta^{\gamma}) d\theta^{\beta}.
 \label{II31}
 \end{eqnarray}

The linear relation between the body frame angular velocity and the
velocities ${\dot \theta}^{\alpha}$ is

\begin{equation}
{\check \omega}^r= {\check
\Lambda}^{(R)}{}^{(\alpha =r)}_{\beta}(\theta^{\gamma})\,  {\dot \theta}^{\beta}.
\label{II32}
\end{equation}

Using the anholonomic components $({\check \omega}^r,{\dot q}^{\mu})$
\footnote{Using the dreibein given by the right-invariant vector
fields on SO(3) and regarding the Lagrangian as function of
$\theta^{\alpha}$, ${\check \omega}^r$, $q^{\mu}$, ${\dot q}^{\mu}$
[see Eq.(\ref{II26})].} of the velocities instead of the holonomic basis $({\dot
\theta}^{\alpha}, {\dot q}^{\mu})$, one gets the following canonical
anholonomic momenta \footnote{The body frame spin components ${\check
S}^r_q$ replace the momenta $\pi_{\theta \,\alpha}={\check
\Lambda}^{(R)}{}^{(\beta =r)}_{\alpha} {\check S}_{qr}$ conjugate to $\theta^{\alpha}$.}
\footnote{Let us remark that one could also use anholonomic gauge invariant
shape momenta ${\tilde p}_{\mu}=p_{\mu}- {\vec S}_q\cdot {\vec
A}^{(k)}_{\mu}$ with Poisson brackets: $\{
\theta^{\alpha},{\tilde p}_{\mu} \} =-{\check X}^{(R)}{}^{\alpha}_{(\beta =r)}
{\check A}^{(k)r}_{\mu}$, $\{ {\vec S}_q, {\tilde p}_{\mu} \}
={\vec A}^{(k)}_{\mu}\times {\vec S}_q$, $\{ q^{\mu}, {\tilde p}_{\mu}
\} =\delta^{\mu}_{\nu}$, $\{ {\tilde p}_{\mu},{\tilde p}_{\nu} \} =
{\vec S}_q \cdot {\vec B}^{(k)}_{\mu\nu}$.}

\begin{eqnarray}
{\check S}_q^r&=&{{\partial L_{rel}}\over {\partial {\check
\omega}^r}},\nonumber \\
 p^{(k)}_{\mu}&=&{{\partial L_{rel}}\over {\partial
{\dot q}^{(k)\mu}}}= g^{(k)}_{\mu\nu}{\dot q}^{(k)\nu}+{\vec S}_q\cdot
{\vec A}^{(k)}_{\mu},\nonumber \\
 &&{}\nonumber \\
 &&\{ \theta^{\alpha},\theta^{\beta} \} = 0,\quad\quad \{ \theta^{\alpha},
 {\check S}_{qr} \} ={\check X}^{(R)}{}^{\alpha}_{(\beta =r)},\quad\quad
 \{{\check S}^r_q,{\check S}^s_q \} =-\epsilon^{rsu}{\check S}_q^u,\nonumber \\
 &&\{ f,g \} ={\check X}^{(R)}{}^{\alpha }_{(\beta =r)} \Big( {{\partial f}\over
 {\partial \theta^{\alpha}}} {{\partial g}\over {\partial {\check S}_{qr}}}-
 {{\partial f}\over {\partial {\check S}_{qr}}}{{\partial g}\over
 {\partial \theta^{\alpha}}}\Big) -{\vec S}_q\cdot \Big( {{\partial f}\over
 {\partial {\vec S}_q }} \times {{\partial g}\over {\partial
 {\vec S}_q }}\Big) +\nonumber \\
 &+& \Big( {{\partial f}\over {q^{\mu}}}{{\partial g}\over {\partial p_{\mu}}}-
 {{\partial f}\over {\partial p_{\mu}}}{{\partial g}
 \over {\partial q^{\mu}}}\Big),\nonumber \\
 &&\Downarrow \nonumber \\
{\check \omega}^u&=&[{\check I}^{(k)
-1}(q^{(k)},\mu^{(k)})]^{uv}{\check S}_q^v
-\Big( {\check A}^{(k)u}_{\mu} g^{(k)\mu\nu} [p^{(k)}_{\nu}-{\vec
S}_q\cdot {\vec A}^{(k)}_{\nu}]\Big) (q^{(k)},\mu^{(k)}),\nonumber
\\
 {\dot q}^{(k)\mu}&=&g^{(k)\mu\nu}(q^{(k)},\mu^{(k)}) [p^{(k)}_{\nu}-
{\vec S}_q\cdot {\vec A}^{(k)}_{\nu}(q^{(k)},\mu^{(k)})],
\nonumber \\
 &&{}\nonumber \\
 ({\check S}^r_q,\quad p_{\mu}) &=& ({\check S}^r_q,\quad p_{\mu})_v +
 ({\check S}^r_q,\quad p_{\mu})_{Ch},\nonumber \\
 &&({\check S}^r_q,\quad p_{\mu})_v=
 ({\check S}^r_q,\quad {\vec S}_q\cdot {\vec
 A}^{(k)}_{\mu}(q^{(k)},\mu^{(k)}) ),\nonumber \\
 &&({\check S}^r_q,\quad p_{\mu})_{Ch}= (0,\quad p_{\mu}-{\vec S}_q\cdot {\vec
 A}^{(k)}_{\mu}(q^{(k)},\mu^{(k)}) ).
\label{II33}
\end{eqnarray}

The last lines  show the decomposition of the momenta into vertical
and C-horizontal parts.

Finally, the Hamiltonian becomes

\begin{eqnarray}
H_{rel}&=&{1\over 2}\sum_{a=1}^{N-1}{{ {\vec \pi}^{(k) 2}_{sa}}\over
{\mu_a^{(k)}}}= {\vec \omega}\cdot {\vec S}_q+p^{(k)}_{\mu}{\dot q}
^{(k)\mu}-L_{rel}=\nonumber \\
 &=&{1\over 2} \Big[ {\check S}_q^u ({\check I}^{(k)-1}(q^{(k)},\mu
^{(k)}))^{uv} {\check S}^v_q+\nonumber \\
&+&\Big( g^{(k)\mu\nu}(p^{(k)}_{\mu}-{\vec S}_q
\cdot {\vec A}^{(k)}_{\mu})(p^{(k)}_{\nu}-{\vec S}_q\cdot
{\vec A}^{(k)}_{\nu})\Big) (q^{(k)},\mu^{(k)}) \Big] .
\label{II34}
\end{eqnarray}

For ${\dot q}^{\mu}=0$, namely $p_{\mu}= {\vec S}_q\cdot {\vec
A}^{(k)}_{\mu}$, one gets the rigid body  Hamiltonian for pure
rotations without vibrations $H={1\over 2} ({\check I}^{-1})^{(k)rs}
{\check S}^r_q{\check S}^s_q = {1\over 2} (I^{-1})^{(k)rs}
S^r_qS^s_q$.

See Appendix A for the form of the Lagrangian and Hamiltonian
equations of motion in the orientation-shape bundle approach and
Appendix B for the reformulation of the results of this Subsection
with arbitrary coordinates.

\vfill\eject

\section { Canonical Spin Bases.}

The {\it static} theory of the orientation-shape SO(3) principal
bundle is based on point canonical transformations of the type of
Eq.(\ref{II15}).

Following the preliminary work of Ref.\cite{lucenti}, we look for a
set of non-point canonical transformations from the relative canonical
variables ${\vec \rho}_{qa}={\vec \rho}_a=\sqrt{N}
\sum_{i=1}^N\gamma_{ai} {\vec \eta}_i$, ${\vec \pi}_{qa}= {1\over {\sqrt{N}}}
\sum_{i=1}^N \Gamma_{ai} {\vec \kappa}_i$ of Eq.(\ref{II8})
 to a canonical basis adapted to the SO(3) subgroup
\cite{pauri2,pauri1} of the extended Galilei group and containing one
of its invariants, namely the modulus of the spin. Jacobi coordinates
will not be used in this Section and the comparison with the
orientation-shape formalism has to be done by using Appendix B.

Again, the configurations with $\vec S \not= 0$ and with $\vec S
=0$ have to be treated separately. The special connection C of
the static orientation-shape SO(3) principal bundle is not included in
our description, which is valid only for  the $\vec S
\not= 0$ configurations.  Accordingly, after the exceptional case N=2,
we shall find that in the  case N=3 the results of the static trivial
orientation-shape SO(3) principal bundle are recovered in a gauge of
the {\it xxz} type. On the other hand, our results for $N \geq 4$ will
differ substantially from the static non-trivial SO(3) principal
bundle approach.

\subsection{2-Body Systems.}

The relative variables are ${\vec \rho}_q=\vec \rho$, ${\vec
\pi}_q$ and the Hamiltonian is $H_{rel}={{{\vec \pi}_q^2}\over
{2\mu}}$, where $\mu={{m_1m_2}\over {m_1+m_2}}$ is the reduced mass.
The spin  is ${\vec S}_q={\vec \rho}_q\times {\vec \pi}_q$
[$S_q=\sqrt{{\vec S}_q^2}$].

Let us define the following decomposition \footnote{The notation $\hat
R$ for the unit vector ${\hat \rho}_q$ is used for comparison with
Ref.\cite{lucenti}.}

\begin{eqnarray}
{\vec \rho}_q&=& \rho_q \hat R,\quad\quad \rho_q=\sqrt{{\vec \rho}_q^2},\quad
\quad \hat R={{{\vec \rho}_q}\over {\rho_q}}={\hat \rho}_q,\quad\quad {\hat R}^2=1,
\nonumber \\
&&{}\nonumber \\ {\vec \pi}_q&=& {\tilde \pi}_q \hat R -{{S_q}\over
{\rho_q}} \hat R\times {\hat S}_q={\tilde \pi}_q{\hat
\rho}_q-{{S_q}\over {\rho_q}}{\hat \rho}_q\times {\hat S}_q,\nonumber \\
&& {\tilde \pi}_q={\vec \pi}_q\cdot \hat R={\vec \pi}_q\cdot {\hat
\rho}_q,\quad\quad {\hat S}_q={{{\vec S}_q}\over {S_q}},\quad\quad
{\hat S}_q \cdot \hat R =0.
\label{III1}
\end{eqnarray}

Therefore, besides the standard {\it space or laboratory frame} with
unit vectors ${\hat f}_r$, we can  build a {\it spin frame}, whose
basis unit vectors ${\hat S}_q$, $\hat R$, ${\hat S}_q\times \hat R$
are identified by the 2-body system itself. Since $\{ S_q^i,S_q^j \} =
\epsilon^{ijk} S^k_q$, $\{ {\hat R}^i, {\hat R}^j
\} =0$, $\{ {\hat R}^i, S^j_q \} = \epsilon^{ijk} {\hat R}^k$,
${\vec S}_q$ and $\hat R$ are the generators of an E(3) group
containing SO(3) as a subgroup. The E(3) invariants turn out to have
the fixed values ${\hat R}^2=1$ and ${\vec S}_q\cdot \hat R=0$.

Usually one considers the following local point canonical
transformation [like in Eqs.(\ref{II15})] to polar coordinates
\cite{pauri1}

\begin{equation}
\begin{minipage}[t]{1cm}
\begin{tabular}{|l|} \hline
${\vec \rho}_q$ \\ \hline
${\vec \pi}_q$ \\ \hline
\end{tabular}
\end{minipage} \ {\longrightarrow \hspace{.2cm}} \
\begin{minipage}[t]{2 cm}
\begin{tabular}{|ll|l|} \hline $\theta$ & $\varphi$ & $\rho_q$ \\
\hline
   $\pi_{q\theta}$ & $\pi_{q\varphi}$ & $\pi_{q\rho}={\tilde \pi}_q$\\ \hline
\end{tabular}
\end{minipage}
\label{III2}
\end{equation}

\begin{eqnarray}
\rho_q^1&=&\rho_q sin\, \theta cos\, \varphi ,\nonumber \\
\rho^2_q&=&\rho_q sin\, \theta sin\, \varphi ,\nonumber \\
\rho^3_q&=&\rho_q cos\, \theta,\nonumber \\
&&{}\nonumber \\
\pi^1_q&=&sin\, \theta cos\, \varphi \pi_{q\rho} -{{sin\, \varphi}\over {\rho_q
sin\, \theta}} \pi_{q\varphi} +{{cos\, \varphi cos\, \theta}\over {\rho_q}}
\pi_{q\theta},\nonumber \\
\pi^2_q&=&sin\, \theta sin\, \varphi \pi_{q\rho} +{{cos\, \varphi}\over {\rho_q
sin\, \theta}} \pi_{q\varphi} +{{sin\, \varphi cos\, \theta}\over {\rho_q}}
\pi_{q\theta},\nonumber \\
\pi^3_q&=&cos\, \theta \pi_{q\rho} -{{sin\, \theta}\over {\rho_q}}
\pi_{q\theta},\nonumber \\
 &&{}\nonumber \\
 &&{\vec \pi}_q^2 = \pi^2_{q\rho} +{1\over {\rho_q^2}} \Big[ \pi^2_{q\theta}
 + {{\pi^2_{q\varphi}}\over {sin^2\, \theta}}\Big] = \pi^2_{q\rho}+
 {{{\vec S}^2_q}\over {\rho^2_q}},\nonumber \\
 &&{}\nonumber \\
 S^1_q &=& - {{cos\, \theta cos\, \varphi}\over {sin\, \theta}} \pi_{q\varphi} -
 sin\, \varphi \pi_{q\theta},\nonumber \\
 S^2_q &=& - {{cos\, \theta sin\, \varphi}\over {sin\, \theta}} \pi_{q\varphi} +
 cos\, \varphi \pi_{q\theta},\nonumber \\
 S^3_q &=& \pi_{q\varphi},\qquad {\vec S}_q^2 = \pi^2_{q\theta}+{{\pi^2_{q\varphi}}
 \over {sin^2\, \theta}},
\label{III3}
\end{eqnarray}

\begin{eqnarray}
\rho_q&=&\sqrt{{\vec \rho}_q^2},\nonumber \\
cos\, \theta &=& {{\rho^3_q}\over {\rho_q}},\quad\quad sin\, \theta =
{{\sqrt{ (\rho_q)^2-(\rho^3_q)^2}}\over {\rho_q}},\quad\quad tg\, \theta =
{{\sqrt{ (\rho_q)^2-(\rho^3_q)^2}}\over {\rho_q^3}},\nonumber \\
tg\, \varphi &=& {{\rho^2_q}\over {\rho^1_q}},\quad\quad sin\, \varphi =
{{\rho_q^2}\over {\sqrt{ (\rho_q)^2-(\rho^3_q)^2}}},\quad\quad cos\, \varphi =
{{\rho^1_q}\over {\sqrt{ (\rho_q)^2-(\rho^3_q)^2}}},\nonumber \\
&&{}\nonumber \\
\pi_{q\rho}&=&{\tilde \pi}_q,\nonumber \\
\pi_{q\theta}&=&\rho_q [cos\, \theta (cos\, \varphi \pi^1_q+sin\, \varphi \pi^2
_q) -sin\, \theta \pi^3_q]={{\rho^3_q {\vec \rho}_q\cdot {\vec \pi}_q-\rho^2
_q\pi^3_q}\over {\sqrt{ (\rho_q)^2-(\rho^3_q)^2}}},\nonumber \\
\pi_{q\varphi}&=& \rho_q sin\, \theta (cos\, \varphi \pi^2_q-sin\, \varphi
\pi^1_q)=\rho^1_q\pi^2_q-\rho^2_q\pi^1_q = S^3_q.
\label{III4}
\end{eqnarray}

After this point canonical transformation, the Hamiltonian becomes
$H_{rel}={1\over {2\mu}} \Big[ \pi^2_{q\rho}+{{{\vec S}_q^2}\over
{\rho_q^2}}\Big]$, while the {\it static} shape canonical variables
are the pair $\rho_q$, $\pi_{q\rho}$.

As  shown in Ref.\cite{lucenti}, instead of this point canonical
transformation,  it is instrumental to consider the following
non-point canonical transformation adapted to the SO(3) group
\footnote{Note that in the new canonical basis the invariant $S_q=|{\vec S}_q|$
becomes one of the new canonical variables.} valid when ${\vec
S}_q\not= 0$

\begin{eqnarray}
&&\begin{minipage}[t]{1cm}
\begin{tabular}{|l|} \hline
${\vec \rho}_q$ \\ \hline ${\vec \pi}_q$ \\ \hline
\end{tabular}
\end{minipage} \ {\longrightarrow \hspace{.2cm}} \
\begin{minipage}[t]{2 cm}
\begin{tabular}{|ll|l|} \hline
$\alpha$ & $\beta$ & $\rho_q$\\ \hline
 $S_q$   & $S^3_q$ & ${\tilde \pi}_q$  \\ \hline
\end{tabular}
\end{minipage}  ,
\label{III5}
\end{eqnarray}

\noindent where

\begin{eqnarray}
\alpha &=& tg^{-1} {1\over {S_q}} \Big( {\vec \rho}_q\cdot {\vec \pi}_q-
{{(\rho_q)^2}\over {\rho^3_q}} \pi^3_q\Big) ,\nonumber \\
&&{}\nonumber \\
\beta &=& tg^{-1} {{S^2_q}\over {S^1_q}},\quad\quad sin\, \beta ={{S^2_q}\over
{\sqrt{ (S_q)^2-(S^3_q)^2}}},\quad\quad cos\, \beta
={{S^1_q}\over {\sqrt{ (S_q)^2-(S^3_q)^2}}}.
\label{III6}
\end{eqnarray}

The two pairs of canonical variables $\alpha$, $S_q$, $\beta$, $S_q^3$
form the irreducible kernel of the {\it scheme A} of a
(non-irreducible, type 3, see Ref.\cite{lucenti}) canonical
realization of the group E(3) generated by ${\vec S}_q$, $\hat R$,
with fixed values of the invariants ${\hat R}^2=1$, $\hat R
\cdot {\vec S}_q=0$, just as the variables $S^3_q$, $\beta$ and $S_q$ form the {\it scheme
A} of the SO(3) group with invariant $S_q$.

Geometrically we have:

i) the angle $\alpha$ is the angle between the plane determined by
${\vec S}_q$ and ${\hat f}_3$ and the plane determined by ${\vec S}_q$
and $\hat R$;

ii) the angle $\beta$ is the angle between the plane ${\vec S}_q$ -
${\hat f}_3$ and the plane ${\hat f}_3$ - ${\hat f}_1$.

We have

\begin{eqnarray}
S^1_q&=& \rho^2_q\pi^3_q-\rho^3_q\pi^2_q =- sin\, \varphi \pi_{q\theta} -cos\,
\varphi cotg\, \theta \pi_{q\varphi}=\sqrt{ (S_q)^2-(S^3_q)^2} cos\, \beta ,
\nonumber \\
S^2_q&=&\rho^3_q\pi^1_q-\rho^1_q\pi^3_q= cos\, \varphi \pi_{q\theta}-sin\,
\varphi cotg\, \theta \pi_{q\varphi}=\sqrt{ (S_q)^2-(S^3_q)^2} sin\, \beta ,
\nonumber \\
S^3_q&=&\rho^1_q\pi^2_q-\rho^2_q\pi^1_q= \pi_{q\varphi},\nonumber \\
&&{}\nonumber \\
S_q&=& \sqrt{ \pi^2_{q\theta} + {{\pi^2_{q\varphi}}\over {sin^2\, \theta}} },
\label{III7}
\end{eqnarray}

\begin{eqnarray}
{\hat R}^1&=&{\hat \rho}^1_q=sin\, \theta cos\, \varphi = sin\, \beta
sin\, \alpha -{{S^3_q}\over {S_q}} cos\, \beta cos\, \alpha ,\nonumber
\\ {\hat R}^2&=& {\hat \rho}^2_q=sin\, \theta sin\,
\varphi =-cos\, \beta sin\, \alpha - {{S^3_q}
\over {S_q}} sin\, \beta cos\, \alpha ,\nonumber \\
{\hat R}^3&=& {\hat \rho}^3_q=cos\, \theta ={1\over {S_q}} \sqrt{
(S_q)^2-(S^3_q)^2} cos\, \alpha ,\nonumber \\
 &&{}\nonumber \\
({\hat S}_q\times \hat R)^1&=&{\hat S}^2_q{\hat R}^3-{\hat S}^3_q{\hat
R}^2=
 sin\, \beta cos\, \alpha +{{S^3_q}\over {S_q}} cos\, \beta sin\, \alpha ,
 \nonumber \\
({\hat S}_q\times \hat R)^2&=&{\hat S}^3_q{\hat R}^1-{\hat S}^1_q{\hat
R}^3=-cos\, \beta cos\, \alpha +{{S^3_q}\over {S_q}} sin\, \beta sin\,
\alpha ,\nonumber \\
 ({\hat S}_q\times \hat R)^3&=&-{\hat S}^1_q{\hat R}^2-{\hat S}^2_q{\hat R}^1=
 {1\over {S_q}} \sqrt{(S_q)^2-(S^3_q)^2} sin\, \alpha ,\nonumber \\
 &&{}\nonumber \\
 &&{\hat S}_q \times \hat R(\alpha ) = {{\partial \hat R(\alpha )}\over
 {\partial \alpha}}= \hat R(\alpha + {{\pi}\over 2}),\nonumber \\
 &&{}\nonumber \\
 \Rightarrow && \alpha =- tg^{-1}\, {{ ({\hat S}_q \times \hat R)^3}\over
 {[{\hat S}_q \times ( {\hat S}_q \times \hat R)]^3}}.
\label{III8}
\end{eqnarray}

From the last line of this equation we see that the angle $\alpha$ can
be expressed in terms of ${\hat S}_q$ and $\hat R$. Given the
Hamiltonian description of any isolated system (a deformable body) in
its rest frame with conserved spin ${\vec S}(q,p)$ [$q,p$ denote a
canonical basis for the system], a solution $\alpha (q,p)$ of the
equations $\{ \alpha (q,p), S(q,p) \}
=1$,  $\{ \alpha (q,p),\beta (q,p) \} = \{ \alpha (q,p), S^3(q,p) \}
=0$, allows to construct the unit vector $\hat R$ associated with the
isolated system and then to build the spin frame and the E(3) group.

The following inverse canonical transformation holds true

\begin{eqnarray}
{\vec \rho}_q&=&\rho_q \hat R(\alpha ,\beta ,S_q,S^3_q),\nonumber \\
{\vec \pi}_q&=&{\tilde \pi}_q \hat R(\alpha ,\beta ,S_q,S^3_q)+
{{S_q}\over {\rho_q}}{\hat S}_q(\beta ,S_q,S^3_q)\times  \hat R(\alpha
,\beta ,S_q,S^3_q),\nonumber \\
 &&{}\nonumber \\
 \Rightarrow&& {\vec \pi}_q^2={\tilde \pi}_q^2+{{S_q^2}\over {\rho_q^2}}.
\label{III9}
\end{eqnarray}

In this degenerate case, the {\it dynamical} shape variables $\rho_q$,
${\tilde \pi}_q$ coincide with the {\it static} ones and describe the
vibration of the dipole.

The rest-frame Hamiltonian for the relative motion becomes [$\check I$
is the barycentric tensor of inertia of the dipole]

\beq
 H_{rel}= {1\over 2} \Big[ {\check I}^{-1} S^2_q +{{ {\tilde \pi}^2_q}\over {\mu}}\Big],
 \qquad \check I = \mu \rho^2,\quad\quad \mu ={{m_1m_2}\over {m_1+m_2}},
 \label{III10}
 \eeq

\noindent while the body frame angular velocity is

\beq
 \check \omega = {{\partial H_{rel}}\over {\partial {\check S}_q}} ={{ {\check S}_q}\over
 {\check I}}.
\label{III11}
\eeq

We conclude this Subsection with some more details on what has already
been anticipated in the Introduction concerning the canonical
reduction. When ${\vec S}_q \not= 0$, Eq.(\ref{III5}) explicitly shows
that a non-Abelian symmetry group like SO(3) does not allow a
canonical reduction like in the Abelian case of translations. In this
latter case we can eliminate the three Abelian constants of motion
${\vec \kappa}_{+}\approx 0$ and gauge fix the three conjugate
variables ${\vec q}_{nr}\approx 0$.  In the non-Abelian case we could
surely fix $S^3_q$, $\beta$ by imposing second class constraints
$S^3_q-a\approx 0$, $\beta -b\approx 0$, and eliminate this pair of
canonical variables by going to Dirac brackets.  However, since
$\alpha$ is not a constant of motion [see later on Eq.(\ref{III10});
we get instead ${d\over {dt}} (S_q-S)\, {\buildrel
\circ \over =}\, 0$], we can only add by hand the first class constraint $S_q-S\approx 0$
($S\not= 0$). It is only after the solution of the equations of
motion, that we could also complete the reduction by adding $\alpha -
\alpha_{solution}\approx 0$ as a gauge-fixing.

In absence of interactions the solution for $\alpha$ can be easily
worked out. The Hamilton equations, equivalent to ${\ddot {\vec
\rho}}_q\, {\buildrel \circ \over =}\, 0$, are ${\dot
\rho}_q\, {\buildrel \circ \over =}\, {\tilde \pi}_q/\mu$,
${\dot {\tilde \pi}}_q\, {\buildrel \circ \over =}\, (S_q)^2/\mu
\rho_q^3$, ${\dot S}_q\, {\buildrel \circ \over =}\, 0$, $\dot \alpha
\, {\buildrel \circ \over =}\, S_q/\mu \rho_q^2$. The solution
${\vec \rho}_q\, {\buildrel \circ \over =}\, \vec b t +\vec a$ [$\vec
a$, $\vec b$ constant vectors] implies: $\rho_q\, {\buildrel \circ
\over =}\, |\vec b t+\vec a|$, $\hat R\, {\buildrel \circ \over =}\,
(\vec bt+\vec a)/|\vec bt+\vec a|$, ${\tilde \pi}_q\, {\buildrel \circ
\over =}\, 2\mu \vec b \cdot (\vec bt+\vec a)/|\vec bt+\vec a|$,
$S_q\, {\buildrel \circ \over =}\, \mu \sqrt{2} |\vec a\times \vec
b|$, $\alpha \, {\buildrel \circ \over =}\, arccos\, \Big( {\hat R}^3
S_q /\sqrt{(S_q)^2-(S^3_q)^2}\Big)$.

In the 2-body case the condition ${\vec S}_q \approx 0$, imposed as
three first class constraints, is equivalent to ${\vec \rho}_q - k
{\vec \pi}_q \approx 0$ and selects only the solution ${\vec
\rho}_q(t)\, {\buildrel \circ \over =}\, \vec A e^{Bt}$ [$k$, $\vec A$ and $B$ are
constants].

\subsection{3-Body Systems.}

In the case N=3 the range of the indices is $i=1,2,3$, $a=1,2$. The
spin is ${\vec S}_q=\sum_{a=1}^2{\vec \rho}_{qa}\times {\vec
\pi}_{qa}=\sum_{a=1}^2 {\vec S}_{qa}$ after the canonical
transformation which separates the internal center of mass

\begin{equation}
\begin{minipage}[t]{1cm}
\begin{tabular}{|l|} \hline
${\vec \eta}_i$ \\    \hline
 ${\vec \kappa}_i$ \\ \hline
\end{tabular}
\end{minipage} \ {\longrightarrow \hspace{.2cm}} \
\begin{minipage}[t]{2 cm}
\begin{tabular}{|l|l|} \hline
${\vec q}_{+}$   & ${\vec \rho}_{qa}$  \\ \hline
 ${\vec \kappa}_{+}$ & ${\vec \pi}_{qa}$ \\ \hline
\end{tabular}
\end{minipage}
\label{III12}
\end{equation}

The relative motions are governed by the Hamiltonian

\begin{equation}
H_{rel}= {1\over 2} \sum_{a,b}^{1,2} k^{-1}_{ab}[m_i,\Gamma_{ci}]
{\vec \pi}_{qa} \cdot {\vec \pi}_{qb}.
\label{III13}
\end{equation}

Again, we shall assume ${\vec S}_q \not= 0$, because the exceptional
SO(3) orbit $S_q=0$ has to be studied separately. This is done by
adding $S_q \approx 0$ as a first class constraint and studying the
following two disjoint strata with a different number of first class
constraints separately: a) ${\vec S}_q\approx 0$, but ${\vec
S}_{q1}=-{\vec S}_{q2}\not= 0$; b) ${\vec S}_{qa}\approx 0$, $a=1,2$
[in this case we have ${\vec \rho}_{qa}- k_a {\vec \pi}_{qa} \approx
0$]\footnote{Let us remark that with the canonical basis (\ref{III14})
the degenerate case defined by imposing  the constraints ${\vec S}_q
\approx 0$ with ${\vec S}_{q1}\approx -{\vec S}_{q2}\not= 0$ implies
the three extra first class constraints $S_{q1}-S_{q2} \approx 0$,
$S^3_{q1}+S^3_{q2} \approx 0$, $\beta_1-\beta_2 \approx 0$: therefore
we get three arbitrary conjugate gauge variables ${1\over
2}(\alpha_1-\alpha_2)$, ${1\over 2}(\beta_1+\beta_2)$, ${1\over
2}(S^3_{q1}+S^3_{q2})\not= 0$. Besides these three pairs of conjugate
variables, a canonical basis adapted to ${\vec S}_q\approx 0$ (with
${\vec S}_{q1}\approx -{\vec S}_{q2}\not=0$) also contains the
physical variables $\bar \alpha =\alpha_1+\alpha_2$, $\bar S={1\over
2}(S_{q1}+S_{q2})$, $\rho_{qa}$, ${\tilde \pi}_{qa}$, and the
Hamiltonian (\ref{III13}) becomes $H_D={1\over 2}\sum_{a,b}^{1,2}
k_{ab}^{-1} {\vec \pi}_{qa}\cdot {\vec \pi}_{qb} {|}_{{\vec S}_q=0} +
\lambda_1(t) (S_{q1}-S_{q2}) +\lambda_2(t) (S^3_{q1}+S^3_{q2}) +
\lambda_3(t) (\beta_1-\beta_2)$ [the $\lambda$'s are Dirac multipliers].}.

For each value of $a=1,2$, we consider the non-point canonical
transformation (\ref{III5})

\begin{equation}
\begin{minipage}[t]{1cm}
\begin{tabular}{|l|} \hline
${\vec \rho}_{qa}$ \\ \hline ${\vec \pi}_{qa}$ \\ \hline
\end{tabular}
\end{minipage} \ {\longrightarrow \hspace{.2cm}} \
\begin{minipage}[t]{2 cm}
\begin{tabular}{|ll|l|} \hline
$\alpha_a$ & $\beta_a$ & $\rho_{qa}$\\ \hline
 $S_{qa}$   & $S^3_{qa}$ & ${\tilde \pi}_{qa}$  \\ \hline
\end{tabular}
\end{minipage}
\label{III14}
\end{equation}

\noindent where

\begin{eqnarray}
\alpha_a &=& tg^{-1} {1\over {S_{qa}}} \Big( {\vec \rho}_{qa}\cdot {\vec \pi}
_{qa}-{{(\rho_{qa})^2}\over {\rho^3_{qa}}} \pi^3_{qa}\Big) ,\nonumber \\
&&{}\nonumber \\
\beta_a &=& tg^{-1} {{S^2_{qa}}\over {S^1_{qa}}},\quad\quad sin\, \beta_a =
{{S^2_{qa}}\over {\sqrt{ (S_{qa})^2-(S^3_{qa})^2}}},\quad\quad cos\,
\beta_a ={{S^1_{qa}}\over {\sqrt{ (S_{qa})^2-(S^3_{qa})^2}}}.
\label{III15}
\end{eqnarray}

\begin{eqnarray}
{\vec \rho}_{qa}&=& \rho_{qa} {\hat R}_a,\quad\quad \rho_{qa}=\sqrt{{\vec \rho}
_{qa}^2},\quad\quad {\hat R}_a={{{\vec \rho}_{qa}}\over {\rho_{qa}}}={\hat \rho}_{qa},
\quad\quad {\hat R}_a^2=1,\nonumber \\
 &&{}\nonumber \\
 {\vec \pi}_{qa}&=&
{\tilde \pi}_{qa} {\hat R}_a +{{S_{qa}}\over {\rho_{qa}}} {\hat
S}_{qa}\times {\hat R}_a,\quad\quad {\tilde
\pi}_{qa}= {\vec \pi}_{qa}\cdot {\hat R}_a.
\label{III16}
\end{eqnarray}

\begin{eqnarray}
{\vec \rho}_{qa}&=&\rho_{qa} {\hat \rho}_{qa}(\alpha_a ,\beta_a
,S_{qa},S^3_{qa})= \rho_{qa} {\hat R}_a(\alpha_a ,\beta_a
,S_{qa},S^3_{qa}),\nonumber \\
 {\vec \pi}_{qa}&=&{\tilde \pi}_{qa} {\hat \rho}_{qa}(\alpha_a ,\beta_a
,S_{qa},S^3_{qa})+{{S_{qa}}\over {\rho_{qa}}}  {\hat S}_{qa}(\beta_a
,S_{qa},S^3_{qa}\times {\hat
\rho}_{qa}(\alpha_a ,\beta_a ,S_{qa},S^3_{qa}))=\nonumber \\
 &=& {\tilde \pi}_{qa} {\hat R}_a(\alpha_a ,\beta_a
,S_{qa},S^3_{qa})+{{S_{qa}}\over {\rho_{qa}}}  {\hat S}_{qa}(\beta_a
,S_{qa},S^3_{qa})\times {\hat R}_a(\alpha_a ,\beta_a
,S_{qa},S^3_{qa}).
\label{III17}
\end{eqnarray}

We have now  {\it two} unit vectors ${\hat R}_a$ and {\it two}  E(3)
realizations generated by ${\vec S}_{qa}$, ${\hat R}_a$ respectively
and fixed invariants ${\hat R}^2_a=1$, ${\vec S}_{qa}\cdot {\hat
R}_a=0$ (non-irreducible, type 2, see Ref.\cite{lucenti}).

Then, the {\it simplest choice}, within the existing arbitrariness
(footnote 10), for the orthonormal vectors $\vec N$ and $\vec
\chi$ functions only of the relative coordinates is
\footnote{See Ref.\cite{lucenti} with the interchange ${\vec
\rho}_{qa} \leftrightarrow {\vec \pi}_{qa}$ in the canonical
transformation introduced there.}

\begin{eqnarray}
\vec N&=& {1\over 2} ({\hat R}_1+{\hat R}_2)=
{1\over 2} ({\hat \rho}_{q1}+{\hat \rho}_{q2}),\quad\quad \hat
N={{\vec N}\over {|\vec N|}},\qquad |\vec N|=\sqrt{ {{1+{\hat \rho}_{q1}\cdot {\hat
\rho}_{q2}}\over 2} },\nonumber \\
\vec \chi &=&{1\over 2}({\hat R}_1-{\hat R}_2)=
  {1\over 2}({\hat \rho}_{q1}-{\hat \rho}_{q2})
,\quad\quad \hat \chi ={{\vec \chi}\over {|\vec \chi |}},\qquad
  |\vec \chi |=\sqrt{ { {1-{\hat \rho}_{q1}\cdot {\hat
\rho}_{q2} }\over 2} }=\sqrt{1-{\vec N}^2},\nonumber \\
 &&{}\nonumber \\
  \vec N\times \vec \chi &=&-{1\over 2}{\hat
\rho}_{q1}\times {\hat \rho}_{q2},\quad \quad |\vec N\times \vec \chi |=
|\vec N| |\vec \chi |={1\over 2}\sqrt{1-({\hat \rho}_{q1}\cdot {\hat
\rho}_{q2})^2},\nonumber \\
 &&{}\nonumber \\
 &&\vec N\cdot \vec \chi = 0,\qquad \{ N^r,N^s \}=
 \{ \chi^r,\chi^s \} = \{ N^r, \chi^s \} =0,\nonumber \\
 &&{}\nonumber \\
 {\hat R}_1&=&{\hat \rho}_{q1}= \vec N+\vec \chi ,\quad\quad {\hat R}_2=
 {\hat \rho}_{q2}=\vec N-\vec \chi ,\quad\quad {\hat R}_1\cdot {\vec R}_2=
 {\hat \rho}_{q1}\cdot {\hat \rho}_{q2}={\vec N}^2-{\vec \chi}^2.
\label{III18}
\end{eqnarray}

Likewise, we have for the spins

\begin{eqnarray}
 {\vec S}_q&=& {\vec S}_{q1}+{\vec S}_{q2},\nonumber \\
 {\vec W}_q&=&{\vec S}_{q1}-{\vec S}_{q2},\nonumber \\
 &&{}\nonumber \\
 {\vec S}_{q1}&=&{1\over 2} ({\vec S}_q+{\vec W}_q),\quad\quad {\vec
S}_{q2}= {1\over 2} ({\vec S}_q-{\vec W}_q),\nonumber \\
 &&{}\nonumber \\
 &&\{ W^r_q,W^s_q \} = \epsilon^{rsu} S^u_q.
\label{III19}
\end{eqnarray}

We  therefore succeeded in constructing an orthonormal triad (the {\it
dynamical body frame}) and two E(3) realizations (non-irreducible,
type 3, see Ref.\cite{lucenti}): one with generators ${\vec S}_q$,
$\vec N$ and non-fixed invariants $|{\vec N}|$ and $\vec S\cdot
\hat N$, the other with generators ${\vec S}_q$ and $\vec \chi$ and
non-fixed invariants $|{\vec \chi}|$ and ${\vec S}_q\cdot {\hat
\chi}$. As said in the Introduction, Eq.(\ref{I0}), this is
equivalent to the determination of the
non-conserved generators ${\check S}_q^r$ of a  Hamiltonian {\it right
action} of SO(3): ${\check S}^1_q={\vec S}_q\cdot \hat \chi = {\vec
S}_q\cdot {\hat e}_1$, ${\check S}^2_q={\vec S}_q\cdot \hat N\times
\hat \chi ={\vec S}_q\cdot {\hat e}_2$, ${\check S}^3_q={\vec S}_q\cdot \hat N
={\vec S}_q\cdot {\hat e}_3$.

The realization of the E(3) group with generators ${\vec S}_q$, $\vec
N$ and non-fixed invariants ${\vec N}^2$, ${\vec S}_q\cdot
\vec N$ leads to the final canonical transformation introduced in
Ref.\cite{lucenti}

\begin{equation}
\begin{minipage}[t]{1cm}
\begin{tabular}{|l|} \hline
${\vec \rho}_{qa}$ \\ \hline
 ${\vec \pi}_{qa}$ \\ \hline
\end{tabular}
\end{minipage}\ {\longrightarrow \hspace{.2cm}} \
\begin{minipage}[t]{4cm}
\begin{tabular}{|ll|ll|l|} \hline
$\alpha_1$ & $\beta_1$ & $\alpha_2$ & $\beta_2$ & $\rho_{qa}$\\ \hline
 $S_{q1}$   & $S^3_{q1}$ & $S_{q2}$ & $S^3_{q2}$ & ${\tilde \pi}_{qa}$\\ \hline
\end{tabular}
\end{minipage}
\ {\longrightarrow \hspace{.2cm}}\
\begin{minipage}[b]{4cm}
\begin{tabular}{|lll|l|l|} \hline
$\alpha$ & $\beta$ & $\gamma$ & $|\vec N|$ & $\rho_{qa}$ \\ \hline
 $S_q={\check S}_q$   & $S^3_q$& ${\check S}_q^3={\vec S}_q\cdot \hat
N$ & $\xi$ & ${\tilde \pi}_{qa}$ \\ \hline
\end{tabular}
\end{minipage}
\label{III20}
\end{equation}

\noindent where

\begin{eqnarray}
|\vec N|&=& \sqrt{ {{1+{\hat \rho}_{q1}\cdot {\hat \rho}_{q2}}\over 2}
},\nonumber \\
 {\check S}^3_q&=&{\vec S}_q\cdot \hat N={1\over {\sqrt{2}}} \sum_{a=1}^2 {\vec \rho}_{qa}
 \times {\vec \pi}_{qa} \cdot {{ {\hat \rho}_{q1}+{\hat \rho}_{q2}}\over
 { \sqrt{1+{\hat\rho}_{q1}\cdot {\hat \rho}_{q2}} }}\equiv S_q cos\, \psi ,
 \nonumber \\
  && cos\, \psi = {\hat S}_q\cdot \hat N ={{{\check S}^3_q}\over {S_q}},
  \quad\quad sin\, \psi = {1\over {S_q}}
  \sqrt{(S_q)^2-({\check S}^3_q)^2},\nonumber \\
  S_q&=&{\check S}_q= |\sum_{a=1}^2 {\vec \rho}_{qa}\times {\vec \pi}_{qa}|,\nonumber \\
  S^3_q&=&\sum_{a=1}^2 ({\vec \rho}_{qa}\times {\vec \pi}_{qa})^3,\nonumber \\
  &&{}\nonumber \\
\alpha &=& -tg^{-1}\, {{ ({\hat S}_q\times \hat N)^3}\over
{[{\hat S}_q\times ({\hat S}_q\times \hat N)]^3}}=\nonumber \\
  &=&-tg^{-1}\, {{ [{\hat S}_q\times ({\hat \rho}_{q1}+{\hat \rho}_{q2})]^3}\over
{[{\hat S}_q\times ({\hat S}_q\times [{\hat \rho}_{q1}+{\hat
\rho}_{q2}])]^3}},\nonumber \\
\beta &=& tg^{-1}\, {{S^2_q}\over {S^1_q}},\nonumber \\
\gamma &=& tg^{-1}\, {{{\vec S}_q\cdot (\hat N\times \hat \chi )}\over
{{\vec S}_q\cdot \hat \chi}}= tg^{-1}\, {{{\check S}_q^2}\over
{{\check S}_q^1}},\nonumber \\
  \Rightarrow && sin\, \gamma ={{ {\check S}_q^2}\over
  { \sqrt{({\check S}_q)^2-({\check S}_q^3)^2}}},\quad\quad
  cos\, \gamma ={{{\check S}_q^1}\over { \sqrt{({\check S}_q)^2-({\check S}_q^3)^2}}},
  \nonumber \\
 &=& tg^{-1}\, {{\sqrt{2} {\vec S}_q\cdot {\hat \rho}_{q2}\times {\hat \rho}_{q1}}\over
 { \sqrt{1+{\hat \rho}_{q1}\cdot {\hat \rho}_{q2}}\, {\vec S}_q\cdot
 ({\hat \rho}_{q1}-{\hat \rho}_{q2})}},\nonumber \\
\xi &=& {{{\vec W}_q\cdot (\hat N\times \hat \chi )}\over {|\vec \chi |}}=
{{{\vec W}_q\cdot (\hat N\times \hat \chi )}\over {\sqrt{1-{\vec
N}^2}}}\nonumber \\
 &=&{{ \sqrt{2} \sum_{a=1}^2(-)^{a+1}{\vec
\rho}_{qa}\times {\vec
\pi}_{qa}\cdot ({\hat \rho}_{q2}\times {\hat \rho}_{q1})}\over
{[1-{\hat \rho}_{q1}\cdot {\hat \rho}_{q2}]\sqrt{1+{\hat \rho}_{q1}
\cdot {\hat \rho}_{q2}} }}.
\label{III21}
\end{eqnarray}

For N=3 the {\it dynamical shape variables}, functions of the relative
coordinates ${\vec \rho}_{qa}$ only, are  $|\vec N|$ and $\rho_{qa}$,
while the conjugate shape momenta  are $\xi$, ${\tilde \pi}_{qa}$.

The final array (\ref{III20}) is nothing else than a {\it scheme B}
\cite{pauri2} of a realization of the E(3) group with generators
${\vec S}_q$, $\vec N$ (non-irreducible type 3). In particular, the
two canonical pairs $S^3_q$, $\beta$, $S_q$, $\alpha$, constitute the
irreducible kernel of the E(3) {\it scheme A}, whose invariants are
${\check S}^3_q$, $|\vec N|$; $\gamma$ and $\xi$ are the so-called
{\it supplementary variables} conjugated to the invariants; finally,
the two pairs $\rho_{qa}$, ${\tilde \pi}_{qa}$ are  so-called {\it
inessential variables}. Let us remark that $S^3_q$, $\beta$, $S_q$,
$\alpha$, $\gamma$, $\xi$, are a local coordinatization of every E(3)
coadjoint orbit with ${\check S}^3_q=const.$, $|\vec N| =const.$  and
fixed values of the inessential variables, present in the 3-body phase
space.

We can now reconstruct ${\vec S}_q$ and define a {\it new} unit vector
$\hat R$ orthogonal to ${\vec S}_q$ by adopting the prescription of
Eq.(\ref{III8}) as follows

\begin{eqnarray}
{\hat S}^1_q&=&{1\over {S_q}} \sqrt{(S_q)^2-(S^3_q)^2} cos\, \beta
,\nonumber \\
 {\hat S}^2_q&=&{1\over {S_q}} \sqrt{(S_q)^2-(S^3_q)^2}
sin\, \beta ,\nonumber \\
 {\hat S}^3_q&=& {{S^3_q}\over {S_q}},\nonumber \\
 &&{}\nonumber \\
 {\hat R}^1&=&sin\, \beta sin\, \alpha - {{S^3_q}\over {S_q}} cos\,
 \beta cos\, \alpha ,\nonumber \\
 {\hat R}^2&=&-cos\, \beta sin\, \alpha - {{S^3_q}\over {S_q}} sin\,
 \beta cos\, \alpha ,\nonumber \\
 {\hat R}^3&=&{1\over {S_q}} \sqrt{(S_q)^2-(S^3_q)^2} cos\, \alpha ,\nonumber \\
 &&{}\nonumber \\
 && {\hat R}^2=1,\quad\quad {\hat R}\cdot {\vec S}_q=0,\quad\quad \{
{\hat R}^r,{\hat R}^s \} =0,\nonumber \\
 &&{}\nonumber \\
 ({\hat S}_q\times \hat R)^1&=&{\hat S}^2_q{\hat R}^3-{\hat S}^3_q{\hat R}^2=
 sin\, \beta cos\, \alpha +{{S^3_q}\over {S_q}} cos\, \beta sin\, \alpha ,
 \nonumber \\
({\hat S}_q\times \hat R)^2&=&{\hat S}^3_q{\hat R}^1-{\hat S}^1_q{\hat
R}^3=-cos\, \beta cos\, \alpha +{{S^3_q}\over {S_q}} sin\, \beta sin\,
\alpha ,\nonumber \\
 ({\hat S}_q\times \hat R)^3&=&{\hat S}^1_q{\hat R}^2-{\hat S}^2_q{\hat R}^1=
 {1\over {S_q}} \sqrt{(S_q)^2-(S^3_q)^2} sin\, \alpha ,
\label{III22}
\end{eqnarray}

The vectors  ${\hat S}_q$, $\hat R$, ${\hat S}_q\times \hat R$ build
up the {\it spin frame} for N=3. The angle $\alpha$ conjugate to $S_q$
is explicitly given by\footnote{ The two expressions of $\alpha$ given
here are consistent with the fact that ${\hat S}_q$, $\hat R$ and
$\hat N$ are coplanar, so that $\hat R$ and $\hat N$ differ only by a
term in ${\hat S}_q$.}

\beq
  \alpha = -tg^{-1}\, {{ ({\hat S}_q\times \hat N)^3}\over
{[{\hat S}_q\times ({\hat S}_q\times \hat N)]^3}}=- tg^{-1}\, {{({\hat
S}_q \times \hat R)^3}\over {[{\hat S}_q \times ({\hat S}_q
\times \hat R)]^3}}.
\label{III23}
\eeq

{\it As a consequence of this definition of $\hat R$}, we get the
following expressions for the {\it dynamical body frame} $\hat N$,
$\hat \chi$, $\hat N\times \hat \chi$  in terms of the final canonical
variables

\begin{eqnarray}
 \hat N&=& cos\, \psi {\hat S}_q+sin\, \psi \hat R={{{\check S}^3_q}\over {S_q}}
 {\hat S}_q+{1\over {S_q}}\sqrt{(S_q)^2-({\check S}^3_q)^2}\hat R=\nonumber \\
 &=&\hat N [S_q,\alpha ;S^3_q,\beta ;{\check S}^3_q,\gamma ],\nonumber \\
 &&{}\nonumber \\
 \hat \chi &=&sin\, \psi cos\, \gamma {\hat S}_q-cos\, \psi cos\, \gamma \hat R +sin\,
 \gamma {\hat S}_q\times \hat R=\nonumber \\
  &=& {1\over {S_q}}\sqrt{(S_q)^2-({\check S}^3_q)^2} cos\, \gamma {\hat S}_q-
  {{{\check S}^3_q}\over {S_q}}cos\, \gamma \hat R +sin\, \gamma
  {\hat S}_q\times \hat R=\nonumber \\
  &=& {{{\check S}^1_q}\over {S_q}}\, {\hat S}_q- {{{\check S}^3_q}\over {S_q}}
  {{ {\check S}_q^1\, \hat R + {\check S}_q^2\,  {\hat S}_q\times \hat R}\over
{ \sqrt{({ S}_q)^2-({\check S}_q^3)^2}}}=\nonumber \\
  &=&\hat \chi [S_q,\alpha ;S^3_q,\beta ;{\check S}^3_q,\gamma ],\nonumber \\
 &&{}\nonumber \\
 \hat N\times \hat \chi &=&
sin\, \psi sin\, \gamma {\hat S}_q-cos\, \psi sin\, \gamma \hat R
 -cos\, \gamma  {\hat S}_q\times \hat R=\nonumber \\
  &=& {1\over {S_q}}\sqrt{(S_q)^2-({\check S}^3_q)^2} sin\, \gamma {\hat S}_q-
  {{{\check S}^3_q}\over {S_q}}sin\, \gamma \hat R -cos\, \gamma
  {\hat S}_q\times \hat R=\nonumber \\
 &=& {{{\check S}^2_q}\over {S_q}}\, {\hat S}_q- {{{\check S}^3_q}\over {S_q}}
  {{ {\check S}_q^1\, \hat R - {\check S}_q^2\,  {\hat S}_q\times \hat R}\over
{ \sqrt{({S}_q)^2-({\check S}_q^3)^2}}}=\nonumber \\
  &=&(\hat N\times \hat \chi ) [S_q,\alpha ;S^3_q,\beta ;{\check S}^3_q,\gamma ],
  \nonumber \\
  &&\Downarrow\nonumber \\
  {\hat S}_q &=& sin\, \psi cos\, \gamma \hat \chi +sin\, \psi sin\, \gamma
  \hat N \times \hat \chi + cos\, \psi \hat N\nonumber \\
  &{\buildrel {def} \over =}& {1\over {S_q}} \Big[ {\check S}^1_q \hat \chi +{\check S}^2_q
  \hat N \times \hat \chi +{\check S}^3_q \hat N\Big],\nonumber \\
  &&{}\nonumber \\
  \hat R &=& -cos\, \psi cos\, \gamma \hat \chi -cos\, \psi sin\, \gamma
  \hat N \times \hat \chi +sin\, \psi \hat N,\nonumber \\
  &&{}\nonumber \\
  \hat R \times {\hat S}_q &=& -sin\, \gamma \hat \chi +
  cos\, \gamma \hat N \times \hat \chi .
\label{III24}
\end{eqnarray}

While $\psi$ is the angle between ${\hat S}_q$ and $\hat N$, $\gamma$
is the angle between the plane $\hat N - \hat \chi$ and the plane
${\hat S}_q - \hat N$. As in the case N=2, $\alpha$ is the angle
between the plane ${\hat S}_q - {\hat f}_3$ and the plane ${\hat S}_q
- \hat R$, while $\beta$ is the angle between the plane ${\hat S}_q
- {\hat f}_3$ and the plane ${\hat f}_3 - {\hat f}_1$. See the Figure.

\begin{figure}
\begin{center}
\epsfig{file=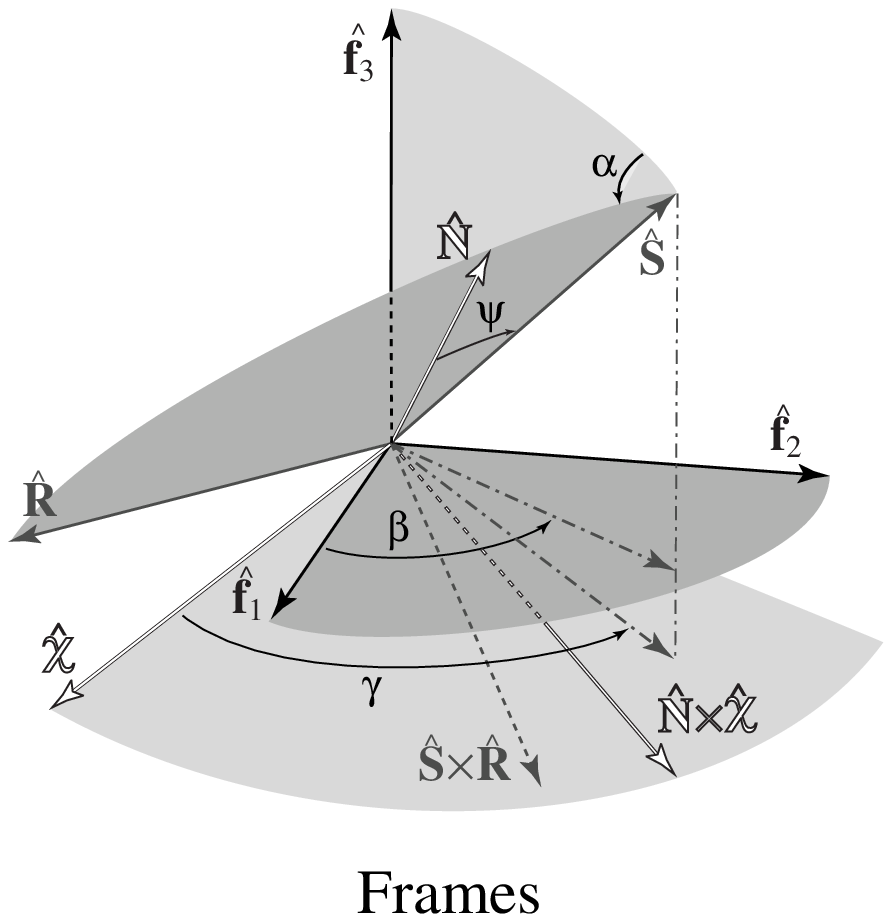, width=8.6cm}
\caption{\footnotesize
 {\it Space} frame (${\hat f}_r$), {\it spin} frame
 ($\hat S$, $\hat R$, $\hat S\times \hat R$) and
{\it dynamical body} frame ($\hat N$, $\hat \chi$, $\hat N\times \hat
\chi$) for $N \geq 3$ particles.
 $\psi$ - angle between $\hat N$ and $\hat S$.
 $\alpha$ - angle between the planes $\hat S$-${\hat f}_3$ and $\hat S$-$\hat R$.
 $\beta$ - angle between ${\hat f}_1$ and the projection of $\hat S$ onto the plane
 ${\hat f}_1$-${\hat f}_2$.
 $\gamma$ - angle between $\hat \chi$ and the projection of $\hat S$ onto the plane
 $\hat \chi$-$\hat N \times \hat \chi$.
\label{assi} }
\end{center}
\end{figure}

Owing to the results of Appendix C, which allow to re-express
$S_{qa}=|{\vec S}_{qa}|$, $S^3_{qa}$, $\beta_a=tg^{-1}
{{S^2_{qa}}\over {S^1_{qa}}}$ in terms of the final variables and
owing to Eqs.(\ref{III18}), (\ref{III23}) which allow to get $\alpha_a
=-tg^{-1} {{({\hat S}_{qa}\times {\hat R}_a )^3}\over {({\hat
S}_{qa}\times ({\hat S}_{qa}\times {\hat R}_a))^3}}$, we can
reconstruct the inverse canonical transformation.

The existence  of the {\it spin frame} and of the {\it dynamical body
frame} allows to define two decompositions of the relative variables,
which make explicit the inverse canonical transformation. For the
relative coordinates we get from Eqs. (\ref{III18}) and (\ref{cc3})

\begin{eqnarray}
{\vec \rho}_{qa}&=&\rho_{qa}{\hat R}_a=\rho_{qa}[\vec N+(-)^{a+1} \vec
\chi ]= \rho_{qa}[|\vec N|\hat N+(-)^{a+1}\sqrt{1-{\vec N}^2}\hat \chi ]=
\nonumber \\
 &=& [{\vec \rho}_{qa}\cdot {\hat  S}_q] {\hat S}_q+[{\vec
\rho}_{qa}\cdot \hat R] \hat R +[{\vec \rho}_{qa}\cdot  {\hat S}_q\times \hat R]
 {\hat S}_q\times \hat R =\nonumber \\
 &=&{{\rho_{qa}}\over {S_q}} \Big[ \Big( |\vec N| \, {\check S}^3_q+(-)^{a+1}
 \sqrt{1-{\vec N}^2}\, {\check S}^1_q\Big) {\hat S}_q+\nonumber \\
 &+&\Big( |\vec N| \sqrt{(S_q)^2-({\check S}^3_q)^2}-(-)^{a+1} \sqrt{1-{\vec N}^2}
 {{ {\check S}^1_q{\check S}^3_q}\over {\sqrt{(S_q)^2-({\check S}^3_q)^2}}}\Big) \hat R -
 \nonumber \\
 &-& (-)^{a+1} \sqrt{1-{\vec N}^2} {{{\check S}^2_q}\over {\sqrt{(S_q)^2-({\check S}^3_q)^2}}}
 {\hat S}_q\times \hat R \Big] =\nonumber \\
 &=&{\vec \rho}_{qa}[S_q,\alpha ;S^3_q,\beta ;{\check
S}^3_q,\gamma ; \rho_{qa},|\vec N|].
\label{III25}
\end{eqnarray}

The analogous formulae for the relative momenta are [see
Eq.(\ref{cc5}) for the expression of the body frame components of
${\vec \pi}_{qa}$]

\begin{eqnarray}
 {\vec \pi}_{qa}&=& {\tilde \pi}_{qa} {\hat R}_a+{{S_{qa}}\over {\rho_{qa}}}
{\hat S}_{qa}\times {\hat R}_a=
 {\tilde \pi}_{qa} {\hat \rho}_{qa}+{{S_{qa}}\over {\rho_{qa}}}
{\hat S}_{qa}\times {\hat \rho}_a=\nonumber \\
 &=&[{\vec \pi}_{qa}\cdot \hat N] \hat N +[{\vec \pi}_{qa}\cdot \hat \chi ]
  \hat \chi +[{\vec \pi}_{qa}\cdot \hat N\times \hat \chi ] \hat N\times \hat \chi
  =\nonumber \\
  &&{}\nonumber \\
&=& [{\vec \pi}_{qa}\cdot {\hat S}_q] {\hat S}_q+[{\vec \pi}_{qa}\cdot
\hat R] \hat R +[{\vec \pi}_{qa}\cdot {\hat S}_q\times \hat R]
{\hat S}_q\times \hat R=\nonumber \\
 &=&{1\over {S_q}} \Big[ \Big( ({\vec \pi}_{qa}\cdot \hat N) {\check S}^3_q+({\vec \pi}_{qa}\cdot
 \hat \chi ) {\check S}^1_q +({\vec \pi}_{qa}\cdot \hat N\times \hat \chi )
{\check S}^2_q\Big) {\hat S}_q +\nonumber \\ &+&\Big( ({\vec
\pi}_{qa}\cdot
\hat N) \sqrt{(S_q)^2-({\check S}^3_q)^2} -[({\vec \pi}_{qa}\cdot \hat \chi +
({\vec \pi}_{qa}\cdot \hat N\times \hat \chi )] {{{\check
S}^1_q{\check S}^3_q}\over {\sqrt{(S_q)^2-({\check S}^3_q)^2}}} \Big)
\hat R +\nonumber \\
 &+&[({\vec \pi}_{qa}\cdot \hat \chi ) - ({\vec \pi}_{qa}\cdot \hat N\times
 \hat \chi )] {{{\check S}^2_q{\check S}^3_q}\over {\sqrt{(S_q)^2-({\check S}^3_q)^2}}}
 \hat R \times {\hat S}_q \Big] =\nonumber \\
  &=&{\vec \pi}_{qa}[S_q,\alpha ;S^3_q,\beta ;{\check S}^3_q,\gamma ;
 |\vec N|,\xi ; \rho_{qa},{\tilde \pi}_{qa}].
\label{III26}
\end{eqnarray}

Finally, the results of Appendix D  allow to perform a sequence of a
canonical transformation to Euler angles $\tilde \alpha$, $\tilde
\beta$, $\tilde \gamma$ with their conjugate momenta, followed by a
transition to the anholonomic basis used in the orientation-shape
bundle approach\cite{little}

\begin{eqnarray}
&&\begin{minipage}[t]{3cm}
\begin{tabular}{|lll|} \hline
$\alpha$ & $\beta$ & $\gamma$ \\ \hline
 $S_q={\check S}_q$   & $S^3_q$& ${\check S}_q^3$ \\ \hline
\end{tabular}
\end{minipage}
\ {\longrightarrow \hspace{.2cm}}\
\begin{minipage}[t]{3cm}
\begin{tabular}{|lll|} \hline
$\tilde \alpha$ & $\tilde \beta$ & $\tilde \gamma$ \\ \hline
 $p_{\tilde \alpha}$   & $p_{\tilde \beta}$& $p_{\tilde \gamma}$ \\ \hline
\end{tabular}
\end{minipage}
\ {{\buildrel {non\,\, can.} \over \longrightarrow} \hspace{.2cm}}\
\begin{minipage}[b]{3cm}
\begin{tabular}{|lll|} \hline
$\tilde \alpha$ & $\tilde \beta$ & $\tilde \gamma$ \\ \hline
 ${\check S}^1_q$ & ${\check S}^2_q$ & ${\check S}^3_q$  \\
 \hline
 \end{tabular}
 \end{minipage} \nonumber \\
 &&{}\nonumber \\
 S_q&=& {\check S}_q = \sqrt{ ({\check S}^1_q)^2+({\check S}^2_q)^2
 +({\check S}^3_q)^2},\nonumber \\
 S^3_q&=&-sin\, \tilde \beta cos\, \tilde \gamma {\check S}^1_q
 +sin\, \tilde \beta sin\, \tilde \gamma {\check S}^2_q +cos\, \tilde \beta
 {\check S}^3_q,\nonumber \\
 \alpha &=& arctg\, {{ p_{\tilde \beta} tg\, \tilde \beta}\over { {\check S}_q -
 {{p_{\tilde \alpha}p_{\tilde \gamma}}\over {{\check S}_q cos\, \tilde \beta}} }},
 \nonumber \\
 \gamma &=& {{\pi}\over 2} -\tilde \gamma - arctg\, {{ctg\, \tilde \beta p_{\tilde \gamma}-
 {{p_{\tilde \alpha}}\over {sin\, \tilde \beta}} }\over {p_{\tilde \beta}}},
 \nonumber \\
 \beta &=& \tilde \alpha + arctg\, {{ ctg\, \tilde \beta p_{\tilde \alpha} -
 {{p_{\tilde \gamma}}\over {sin\, \tilde \beta}} }\over {p_{\tilde \beta}}} -
 {{\pi}\over 2},
\label{III27}
\end{eqnarray}

\noindent Here $p_{\tilde \alpha}$, $p_{\tilde \beta}$, $p_{\tilde \gamma}$
are the functions of $\tilde \alpha$, $\tilde \beta$, $\tilde \gamma$,
${\check S}^r_q$ given in Eqs.(\ref{c3}). The equations (\ref{c3}),
(\ref{III27}), (\ref{III18}) and ${\check S}^2_q={\vec S}_q\cdot \hat
N\times \hat \chi$ lead to the determination of the {\it dynamical
orientation variables} $\tilde \alpha$, $\tilde \beta$, $\tilde
\gamma$ in terms of ${\vec \rho}_{qa}$, ${\vec \pi}_{qa}$. Let us stress that,
while in the orientation-shape bundle approach the orientation
variables $\theta^{\alpha}$ are gauge variables, the Euler angles
$\tilde \alpha$, $\tilde \beta$, $\tilde \gamma$ are {\it uniquely
determined} in terms of the original configurations and momenta.

In conclusion,  the complete transition to the anholonomic basis used
in the {\it static} theory of the orientation-shape bundle is

\begin{equation}
\begin{minipage}[t]{5cm}
\begin{tabular}{|lll|l|l|} \hline
$\alpha$ & $\beta$ & $\gamma$& $|\vec N|$ & $\rho_{qa}$ \\ \hline
 $S_q={\check S}_q$   & $S^3_q$& ${\check S}_q^3$ & $\xi$ & ${\tilde \pi}_{qa}$ \\ \hline
\end{tabular}
\end{minipage}
\ {{\buildrel {non\,\, can.} \over \longrightarrow} \hspace{.2cm}}\
\begin{minipage}[b]{5cm}
\begin{tabular}{|lll|l|l|} \hline
$\tilde \alpha$ & $\tilde \beta$ & $\tilde \gamma$& $|\vec N|$ &
$\rho_{qa}$ \\ \hline
 ${\check S}^1_q$ & ${\check S}^2_q$ & ${\check S}^3_q$ & $\xi$ & ${\tilde \pi}_{qa}$ \\
 \hline
 \end{tabular}
 \end{minipage}.
\label{III28}
\end{equation}

In order to further the comparison with the orientation-shape bundle
approach, let us note the following relation between the space and
body components of the relative coordinates. Eqs.(\ref{III26}),
(\ref{III28}), (\ref{III24}) and (\ref{c2}) imply

\bea
\rho^r_{qa} &=& {\cal R}^r{}_s(\tilde \alpha ,\tilde \beta ,\tilde \gamma ) {\check
\rho}^s_{qa}(q),\quad\quad with\nonumber\\
&&{}\nonumber \\
 &&{\check \rho}^1_{qa}(q) = (-)^{a+1} \rho_{qa}
\sqrt{1-{\vec N}^2},\quad\quad {\check \rho}^2_{qa}(q)=0,\quad\quad
{\check \rho}^3_{qa}(q) = \rho_{qa} |\vec N|,\nonumber \\
 &&{}\nonumber \\
 &&and\nonumber \\
 S^r_q &=&  {\cal R}^r{}_s(\tilde \alpha ,\tilde \beta ,\tilde \gamma ) {\check
 S}^s_q,
 \label{III29}
 \eea

\noindent so that the final visualization of our sequence of transformations is

\beq
 \begin{minipage}[t]{5cm}
\begin{tabular}{|l|} \hline
${\vec \rho}_{qa}$ \\ \hline
 ${\vec \pi}_{qa}$ \\ \hline
\end{tabular}
\end{minipage}
\ {{\buildrel {non\,\, can.} \over \longrightarrow} \hspace{.2cm}}\
\begin{minipage}[b]{5cm}
\begin{tabular}{|lll|l|} \hline
$\tilde \alpha$ & $\tilde \beta$ & $\tilde \gamma$& $q^{\mu}({\vec
\rho}_{qa})$
\\ \hline
 ${\check S}^1_q$ & ${\check S}^2_q$ & ${\check S}^3_q$ &
 $p_{\mu}({\vec \rho}_{qa},{\vec \pi}_{qa})$ \\ \hline
 \end{tabular}
 \end{minipage}.
\label{III30}
\eeq

Note furthermore that  we get ${\check \rho}^2_{qa}={\vec
\rho}_{qa}\cdot \hat N\times \hat \chi =0$ by construction and this
entails that using our {\it dynamical body frame} is equivalent to a
convention ({\it xxzz gauge}) about the body frame of the type of {\it
xxz} and similar gauges quoted in Ref.\cite{little}
\footnote{These gauges utilize the `static' shape variables $\rho_{qa}$, $\phi$
with $cos\, \phi = 2{\vec N}^2-1$, which can be expressed in terms of
the dynamical shape variables $\rho_{qa}$, $|\vec N|$.}.

Finally, we can give the expression of the Hamiltonian for relative
motions\footnote{The Hamiltonian in the basis (\ref{III20}) can be
obtained with the following replacements ${\check
S}^1_q=\sqrt{(S_q)^2-({\check S}^3_q)^2} cos\,
\gamma$ and ${\check S}^2_q=\sqrt{(S_q)^2-({\check S}^3_q)^2} sin\,
\gamma$.} in terms of the anholonomic Darboux basis of Eqs.(\ref{III27}).
By using Eq.(\ref{cc7}) we get

\begin{eqnarray}
H_{rel}&=&  {1\over {2{\vec N}^2}}\Big[ {1\over 2}
({{k^{-1}{}_{11}}\over {\rho^2_{q1}}}+{{k^{-1}{}_{22}}\over
{\rho^2_{q2}}})+{{k^{-1}{}_{12}}\over {\rho_{q1}\rho_{q2}}}\Big]
({\check S}^1_q)^2+\nonumber \\
 &+& {1\over 2} \Big[ {1\over 2}({{k^{-1}{}_{11}}\over {\rho^2_{q1}}}+{{k^{-1}{}_{22}}\over
{\rho^2_{q2}}})+{{k^{-1}{}_{12}(2{\vec N}^2-1)}\over
{\rho_{q1}\rho_{q2}}}\Big] ({\check S}^2_q)^2+\nonumber \\
 &+&{1\over {2(1-{\vec N}^2)}} \Big[ {1\over 2}
({{k^{-1}{}_{11}}\over {\rho^2_{q1}}}+{{k^{-1}{}_{22}}\over
{\rho^2_{q2}}})-{{k^{-1}{}_{12}}\over {\rho_{q1}\rho_{q2}}}\Big]
({\check S}^3_q)^2+\nonumber \\
 &+&\sqrt{1-{\vec N}^2} \Big[ {{\xi}\over 2} ({{k^{-1}{}_{11}}\over {\rho^2_{q1}}}
 -{{k^{-1}{}_{22}}\over {\rho^2_{q2}}} ) +2k^{-1}{}_{12}
  |\vec N| \sqrt{1-{\vec N}^2} ({{{\tilde \pi}_{q1}}\over {\rho_{q2}}}-{{{\tilde \pi}_{q2}}
  \over {\rho_{q1}}})\Big] {\check S}_q^2 -\nonumber \\
  &-& {1\over {2|\vec N| \sqrt{1-{\vec N}^2}}}
  ({{k^{-1}{}_{11}}\over {\rho^2_{q1}}}-{{k^{-1}{}_{22}}\over
  {\rho^2_{q2}}}) {\check S}^1_q {\check S}^3_q+\nonumber \\
  &+&k^{-1}{}_{11} \Big[ {\tilde \pi}^2_{q1}+{{\xi^2 (1-{\vec N}^2)}\over {4\rho^2_{q1}}}\Big] +
  k^{-1}{}_{22} \Big[ {\tilde \pi}^2_{q2}+{{\xi^2 (1-{\vec N}^2)}\over {4\rho^2_{q2}}}\Big]
+\nonumber \\
 &+& 2k^{-1}{}_{12} \Big[ (2{\vec N}^2-1) {\tilde \pi}_{q1} {\tilde \pi}_{q2}
 -|\vec N| (1-{\vec N}^2) \xi ({{{\tilde \pi}_{q1}}\over {\rho_{q2}}}+{{{\tilde \pi}_{q2}}\over
 {\rho_{q1}}}) +\nonumber \\
 &+&{{\xi^2 (1-{\vec N}^2)(2{\vec N}^2-1)}\over {4\rho_{q1}\rho_{q2}}}\Big]
 \, {\buildrel {def} \over =}\nonumber \\
  &{\buildrel {def} \over =}\,& {1\over 2}\Big[ {\check S}_q^r
  ({\check {\cal I}}^{-1})^{rs} {\check
  S}_q^s +{\tilde g}^{\mu\nu} (p_{\mu}-{\vec S}_q\cdot {\vec {\cal A}}_{\mu})
  (p_{\nu}-{\check {\vec S}}_q\cdot {\check {\vec {\cal A}}}_{\nu})\Big]=\nonumber \\
  &=&{1\over 2} ({\check S}_q^r,\, p_{\mu})
  \left( \begin{array}{cc} ({\check {\cal I}}^{-1})^{rs}+{\tilde g}^{\alpha\beta}{\check {\cal
  A}}^r_{\mu}{\check {\cal A}}^s_{\nu} & -{\tilde g}^{\nu\alpha}{\check {\cal A}}^r_{\alpha} \\
  -{\tilde g}^{\mu\alpha}{\check {\cal A}}^s_{\alpha} & {\tilde g}^{\mu\nu} \end{array} \right)
  \left( \begin{array}{c} {\check S}_q^s \\ p_{\nu} \end{array} \right) .
\label{III31}
\end{eqnarray}

\noindent where $q^{\mu}= (\rho_{q1}, \rho_{q2}, |\vec N|)$, $p_{\mu}=({\tilde \pi}_{q1},
{\tilde \pi}_{q2}, \xi )$ are the dynamical  shape variables. In the
last two lines we have rewritten the Hamiltonian in the form of
Eq.(\ref{b12}).

In Appendix E we  evaluate the quantities ${\check  {\cal
A}}^r_{\mu}(q)$, ${\tilde g}^{\mu\nu}(q)$, ${\check {\cal I}}^{-1
rs}(q)$ appearing in the standard {\it static} theory of the
orientation-shape bundle in the {\it xxzz gauge}, adopting the
convention induced by the dynamical body frame. Recall that the
special {\it xxzz} gauge potentials ${\check {\cal A}}^r_{\mu}(q)$ are
measurable quantities in our approach. The same holds for the angular
velocity in the evolving dynamical body frame.

In the static orientation-shape trivial SO(3) principal bundle
approach the Hamiltonian version  of the conditions ${\check S}^r_q=0$
and ${\dot q}^{\mu}=0$, definitory of {\it C-horizontality} and {\it
verticality} respectively, is given by Eq.(\ref{II32}). Note that the
definition $({\check S}^r_q,\, p_{\mu})_v=({\check S}^r_q,\,
p_{\mu}{|}_{\dot q=0}= {\vec S}_q\cdot {\vec {\cal A}}_{\mu}(q))$ of
vertical rotational motion ${\dot q}^{\mu}=0$ is still valid in our
{\it dynamical body frame} approach. By using Eq.(\ref{d2}) to find
$p_{\mu}{|}_{\dot q=0}$, we recover the rotational kinetic energy (the
centrifugal potential) $H^{(rot)}_{rel} = {1\over 2}{\check
S}_q^r({\check {\cal I}}^{-1})^{rs} {\check S}_q^s$ of the {\it xxzz
gauge}.

On the other hand, the C-horizontal component should be determined by
the condition ${\check S}^r_q=0$. Since our construction  requires
$S_q\not= 0$, we cannot utilize it. In our approach the measurable
{\it vibrational} kinetic energy $H^{(vib)}_{rel}$ for ${\vec S}\not=
0$ non-singular N=3 configurations is obtained by restricting ${\check
S}^r_q$ in $H_{rel}$ to the value ${\check S}^r_q{|}_{{\check
\omega}^s=0}$ of Eq.(\ref{d5}), upon the requirement that the
dynamical angular velocity vanishes in the {\it xxzz gauge}. From
Eq.(\ref{d5}) we get

\bea
&&{\check S}^1_q {|}_{{\check \omega}^r=0} = {\check S}^3_q
{|}_{{\check \omega}^r=0}=0,\nonumber \\
 &&{\check S}^2_q {|}_{{\check \omega}^r=0} =- {{ \sqrt{1-{\vec N}^2}
 \Big[{{\xi}\over 2}\Big({{k^{-1}_{11}}\over {\rho_{q1}^2}}-{{k^{-1}_{22}}\over
 {\rho_{q2}^2}}\Big) +2k^{-1}_{12} |\vec N| \sqrt{1-{\vec N}^2}\Big( {{{\tilde \pi}_{q1}}\over
 {\rho_{q2}}}-{{{\tilde \pi}_{q2}}\over {\rho_{q1}}}\Big)\Big]}\over
 { {1\over 2}\Big({{k^{-1}_{11}}\over {\rho_{q1}^2}}+{{k^{-1}_{22}}\over
 {\rho_{q2}^2}}\Big)  + {{k^{-1}_{12}(2{\vec N}^2-1)}\over {\rho_{q1}\rho_{q2}}} }},
 \nonumber \\
 &&{}\nonumber \\
 &&\Downarrow \nonumber \\
 &&{}\nonumber \\
 H_{rel}&&{|}_{{\check \omega}^r=0}= k^{-1}_{11} \Big[ {\tilde \pi}^2_{q1}+
 {{\xi^2 (1-{\vec N}^2)}\over {4\rho_{q1}^2}}\Big] + k^{-1}_{22} \Big[ {\tilde \pi}^2_{q2}+
 {{\xi^2 (1-{\vec N}^2)}\over {4\rho_{q2}^2}}\Big]+\nonumber \\
 &&+2k^{-1}_{12}\Big[ (2{\vec N}^2-1){\tilde \pi}_{q1}{\tilde \pi}_{q2}-|\vec N| (1-{\vec N}^2)\xi
\Big( {{{\tilde \pi}_{q1}}\over {\rho_{q2}}}+{{{\tilde \pi}_{q2}}\over {\rho_{q1}}}\Big)+
{{\xi^2 (1-{\vec N}^2)(2{\vec N}^2-1)}\over
{4\rho_{q1}\rho_{q2}}}\Big]-\nonumber \\
 &&-{1\over 2} {{ (1-{\vec N}^2)\Big[ {{\xi}\over 2}
 \Big({{k^{-1}_{11}}\over {\rho_{q1}^2}}-{{k^{-1}_{22}}\over
 {\rho_{q2}^2}}\Big) +2k^{-1}_{12} |\vec N| \sqrt{1-{\vec N}^2}\Big( {{{\tilde \pi}_{q1}}\over
 {\rho_{q2}}}-{{{\tilde \pi}_{q2}}\over {\rho_{q1}}}\Big)\Big]}\over
 { {1\over 2}\Big({{k^{-1}_{11}}\over {\rho_{q1}^2}}+{{k^{-1}_{22}}\over
 {\rho_{q2}^2}}\Big)  + {{k^{-1}_{12}(2{\vec N}^2-1)}\over {\rho_{q1}\rho_{q2}}} }}.
\label{III32}
\eea

Let us remark that with this definition we get $H_{rel} \not=
H^{(rot)}_{rel}+H^{(vib)}_{rel}$, differently from the static
orientation-shape bundle result associated with the C-connection
\footnote{See the interpretation of the term ${1\over 2} {\tilde
g}^{\mu\nu}(q)
\Big[ p_{\mu}-{\vec S}_q\cdot {\vec {\cal A}}_{\mu}(q)\Big] \Big[
p_{\nu} - {\vec S}_q\cdot {\vec {\cal A}}_{\nu}(q)\Big]$ in
Eq.(\ref{II34}).}. In order to get the theory with the Jacobi normal
coordinates one has to perform our sequence of canonical
transformations after having diagonalized $k_{ab}$.

Let us end this Subsection by recalling that in Ref.\cite{little1} the
N=3 shape space\footnote{It is defined for normal Jacobi coordinates
${\vec \rho}_{qa}={\vec s}_a^{(k)}$, but it can be extended to
non-Jacobi ones.} is parametrized with the following configurational
coordinate system $(w,w_1,w_2)$: $w=\rho_{q1}^2+\rho_{q2}^2$
($\sqrt{w}$ is the hyperradius), $w_1=\rho_{q1}^2-\rho_{q2}^2$, $w_2=2
{\vec \rho}_{q1}\cdot {\vec \rho}_{q2}$ with the physical region of
non-singular configurations defined by $\sqrt{w_1^2+w_2^2}\leq w \leq
\infty$, $-\infty \leq w_1,w_2 \leq \infty$. The variable $w$ may be
replaced by $w_3=2| {\vec \rho}_{q1}\times {\vec \rho}_{q2}|
\geq 0$ ($w_3\geq 0$ is the physical region), because $w^2=w^2_1+w_2^2+w_3^2$.
Thus, another basis for the N=3 shape space is $(w_1,w_2,w_3)$. A
further basis  $(w, \chi ,\psi )$ is obtained by  putting $w_1=w cos\,
\chi cos\, \psi$, $w_1=w cos\, \chi sin\, \psi$, $w_2=w sin\, \chi$,
($\chi ,\psi$ hyperspherical angles), The quantities $w$ and
$a=w_1^2+w_2^2=(\rho_{q1}^2-\rho_{q2}^2)^2+4({\vec
\rho}_{q1}\cdot {\vec \rho}_{q2})^2$ [$w_3=\sqrt{w^2-a}$] are {\it
democratic invariants}, and still another basis is $(w,w_3, \alpha )$,
with $\alpha$ the angle parametrizing the democracy group SO(2). In
our spin basis we have the 3 purely configurational shape variables
$\rho_{q1}$, $\rho_{q2}$ and $|\vec N|=\sqrt{ {{1+{\hat
\rho}_{q1}\cdot {\hat \rho}_{q2}}\over 2} }$. We can define a point canonical transformation
to $w=\rho_{q1}^2+\rho_{q2}^2$, $w_1=\rho_{q1}^2-\rho_{q2}^2$,
$w_2=2\rho_{q1}\rho_{q2} (2{\vec N}^2-1)$, and then find the
corresponding momenta.

\subsection{N-Body Systems.}

Let us now consider the general case with $N \geq 4$ without
introducing Jacobi normal coordinates. Instead of coupling the centers
of mass of particle clusters as it is done with  Jacobi coordinates
({\it center-of-mass clusters}), the {\it canonical spin bases} will
be obtained by coupling the spins of the 2-body subsystems ({\it
relative particles}) ${\vec \rho}_{qa}$, ${\vec \pi}_{qa}$,
$a=1,..,N-1$, defined in Eqs.(\ref{II7}), in all possible ways ({\it
spin clusters} from the addition of angular momenta). Let us stress
that we can build a {\it spin basis} with a pattern of {\it spin
clusters} completely unrelated to a possible pre-existing {\it
center-of-mass clustering}.

Let us consider the case $N=4$ as a prototype of the general
construction. We have now three relative variables ${\vec \rho}_{q1}$,
${\vec \rho}_{q2}$, ${\vec \rho}_{q3}$ and related momenta ${\vec
\pi}_{q1}$, ${\vec \pi}_{q2}$, ${\vec \pi}_{q3}$. In the following
formulas we use the convention that the subscripts $a,b,c$ mean any
permutation of $1,2,3$.

By using the explicit construction given in Appendix  F, we define the
following sequence of canonical transformations (we assume $S_q\not=
0$; $S_{qA}\not= 0$, $A=a,b,c$) corresponding to the {\it spin
clustering} pattern $abc \mapsto (ab) c \mapsto ((ab)c)$ [build first
the spin cluster $(ab)$, then the spin cluster $((ab)c)$]:

\begin{eqnarray}
&&\begin{minipage}[t]{3cm}
\begin{tabular}{|lll|} \hline
${\vec \rho}_{qa}$ & ${\vec \rho}_{qb}$ & ${\vec \rho}_{qc}$\\ ${\vec
\pi}_{qa}$ & ${\vec \pi}_{qb}$ & ${\vec \pi}_{qc}$ \\ \hline
\end{tabular}
\end{minipage}
\ {\longrightarrow \hspace{.2cm}}\      \nonumber \\
\ {\longrightarrow \hspace{.2cm}}\
&&\begin{minipage}[t]{7cm}
\begin{tabular}{|ll|ll|ll|lll|} \hline
$\alpha_a$ & $\beta_a$ & $\alpha_b$& $\beta_b$ & $\alpha_c$ &
$\beta_c$ & $\rho_{qa}$ & $\rho_{qb}$ & $\rho_{qc}$ \\
\hline
 $S_{qa}$   & $S^3_{qa}$& $S_{qb}$ & $S^3_{qb}$ & $S_{qc}$ & $S^3_{qc}$ &
 ${\tilde \pi}_{qa}$ & ${\tilde \pi}_{qb}$ & ${\tilde \pi}_{qc}$ \\ \hline
\end{tabular}
\end{minipage}
\ {\longrightarrow \hspace{.2cm}}\     \nonumber \\
\ { {\buildrel {(ab)c} \over \longrightarrow} \hspace{.2cm}} \
&&\begin{minipage}[t]{12cm}
\begin{tabular}{|ccc|cc|cccc|} \hline
$\alpha_{(ab)}$ & $\beta_{(ab)}$ & $\gamma_{(ab)}$& $\alpha_c$ &
$\beta_c$ & $|{\vec N}_{(ab)}|$ & $\rho_{qa}$ & $\rho_{qb}$ &
$\rho_{qc}$\\
 \hline
 $S_{q(ab)}$ & $S^3_{q(ab)}$   &  ${\check S}_{q(ab)}^3=
 {\vec S}_{q(ab)}\cdot {\hat N}_{(ab)}$ & $S_{qc}$ & $S^3_{qc}$
  & $\xi_{(ab)}$ & ${\tilde \pi}_{qa}$ & ${\tilde \pi}_{qb}$ & ${\tilde \pi}_{qc}$\\ \hline
\end{tabular}
\end{minipage}
\ {\longrightarrow \hspace{.2cm}}\     \nonumber \\
\ {\longrightarrow \hspace{.2cm}}\
&&\begin{minipage}[t]{13cm}
\begin{tabular}{|ccc|cccccc|} \hline
$\alpha_{((ab)c)}$ & $\beta_{((ab)c)}$ & $\gamma_{((ab)c)}$& $|{\vec
N}_{((ab)c)}|$ & $\gamma_{(ab)}$& $|{\vec N}_{(ab)}|$ & $\rho_{qa}$ &
$\rho_{qb}$ & $\rho_{qc}$\\
\hline
 $S_q={\check S}_q$   & $S^3_q$& ${\check S}_q^3={\vec S}_q\cdot {\hat N}_{((ab)c)}$
 & $\xi_{((ab)c)}$ & ${\vec S}_{q(ab)}\cdot {\hat N}_{(ab)}$ &
 $\xi_{(ab)}$ & ${\tilde \pi}_{qa}$ & ${\tilde \pi}_{qb}$ & $ {\tilde \pi}_{qc}$\\ \hline
\end{tabular}
\end{minipage}
\ {\rightarrow \hspace{.2cm}}\     \nonumber \\
\ {{\buildrel {non\, can.} \over \longrightarrow} \hspace{.2cm}}\
&&\begin{minipage}[t]{12cm}
\begin{tabular}{|ccc|cccccc|} \hline
$\tilde \alpha$ & $\tilde \beta$ & $\tilde \gamma$& $|{\vec
N}_{((ab)c)}|$ & $\gamma_{(ab)}$& $|{\vec N}_{(ab)}|$ & $\rho_{qa}$ &
$\rho_{qb}$ & $\rho_{qc}$\\
\hline
 ${\check S}_q^1$   & ${\check S}^2_q$& ${\check S}_q^3$
 & $\xi_{((ab)c)}$ & $\Omega_{(ab)}={\vec S}_{q(ab)}\cdot {\hat N}_{(ab)}$ &
 $\xi_{(ab)}$ & ${\tilde \pi}_{qa}$ & ${\tilde \pi}_{qb}$ & $ {\tilde \pi}_{qc}$\\ \hline
\end{tabular}
\end{minipage}.  \nonumber \\
&&{}
\label{III33}
\end{eqnarray}

The first non-point canonical transformation is based on the existence
of the three unit vectors ${\hat R}_A$, $A=a,b,c$, and of three E(3)
realizations with generators ${\vec S}_{qA}$, ${\hat R}_A$ and fixed
values (${\hat R}_A^2=1$, ${\vec S}_A\cdot {\hat R}_A=0$) of the
invariants. Use Eqs.(\ref{III15}), (\ref{III16}) and (\ref{III17}).

In the next canonical transformation the  spins of the {\it relative
particles} $a$ and $b$ are coupled to form the spin cluster $(ab)$,
leaving the {\it relative particle} $c$ as a spectator. We use
Eq.(\ref{III18}) to define ${\vec N}_{(ab)}={1\over 2}({\hat
R}_a+{\hat R}_b)$, ${\vec \chi}_{(ab)}={1\over 2}({\hat R}_a-{\hat
R}_b)$, ${\vec S}_{(ab)}={\vec S}_{qa}+{\vec S}_{qb}$, ${\vec
W}_{q(ab)}={\vec S}_{qa}-{\vec S}_{qb}$. We  get ${\vec N}_{(ab)}\cdot
{\vec \chi}_{(ab)}=0$, $\{ N^r_{(ab)},N^s_{(ab)} \} = \{ N^r_{(ab)},
\chi^s_{(ab)} \} = \{ \chi^r_{(ab)}, \chi^s_{(ab)} \} =0$ and a new
E(3) realization generated by ${\vec S}_{(ab)}$ and ${\vec N}_{(ab)}$,
with non-fixed invariants $|{\vec N}_{(ab)}|$, ${\vec S}_{(ab)}\cdot
{\hat N}_{(ab)}\, {\buildrel {def} \over =}\, \Omega_{(ab)}$. From
Eqs.(\ref{III26}) it follows

\bea
&&{\vec \rho}_{qa} = \rho_{qa} \Big[ |{\vec N}_{(ab)}| {\hat N}_{(ab)}
+ \sqrt{1-{\vec N}^2_{(ab)}} {\hat \chi}_{(ab)}\Big] ,\nonumber \\
&&{\vec \rho}_{qb} = \rho_{qb} \Big[ |{\vec N}_{(ab)}| {\hat N}_{(ab)}
- \sqrt{1-{\vec N}^2_{(ab)}} {\hat \chi}_{(ab)}\Big] ,\nonumber \\
 &&{\vec \rho}_{qc} = \rho_{qc} {\hat R}_c.
 \label{III34}
 \eea

 \noindent Eq.(\ref{III21}) defines $\alpha_{(ab)}$ and $\beta_{(ab)}$, so that
 Eq.(\ref{III22}) defines a unit vector ${\hat R}_{(ab)}$ with ${\vec
 S}_{(ab)}\cdot {\hat R}_{(ab)}=0$, $\{ {\hat R}^r_{(ab)},{\hat R}^s_{(ab)} \} =0$.
 This unit vector identifies the {\it spin cluster} $(ab)$ in
 the same way as the unit vectors ${\hat R}_A={\hat {\vec \rho}}_{qA}$ identify
 the {\it relative particles} $A$.

 The next step is the coupling of the {\it spin cluster} $(ab)$
 with unit vector ${\hat R}_{(ab)}$ [described by the canonical variables $\alpha_{(ab)}$,
 $S_{(ab)}$, $\beta_{(ab)}$ $S^3_{(ab)}$] with the {\it relative particle} $c$
 with unit vector ${\hat R}_c$ and  described by $\alpha_c$, $S_{qc}$,
 $\beta_c$, $S^3_{qc}$: this builds the {\it spin cluster} $((ab)c)$.

 Again, Eq.(\ref{III18}) allows to define ${\vec N}_{((ab)c)}=
 {1\over 2}({\hat R}_{(ab)}+{\hat R}_c)$,
 ${\vec \chi}_{((ab)c)}={1\over 2}({\hat R}_{(ab)}-{\hat R}_c)$, ${\vec S}_q
 ={\vec S}_{q((ab)c)}= {\vec S}_{q(ab)}+{\vec S}_{qc}$, ${\vec W}_{q((ab)c)}=
 {\vec S}_{q(ab)}-{\vec S}_{qc}$. Since we have ${\vec N}_{((ab)c)}\cdot {\vec \chi}_{((ab)c)}=0$
 and  $\{ N^r_{((ab)c)},N^s_{((ab)c)} \} = \{ N^r_{((ab)c)}, \chi^s_{((ab)c)} \} =
 \{ \chi^r_{((ab)c)}, \chi^s_{((ab)c)} \} =0$ due to $\{ {\hat R}^r_{(ab)}, {\hat R}_c^s \} =0$,
  a new E(3) realization generated by ${\vec S}_q$
 and ${\vec N}_{((ab)c)}$ with non-fixed invariants $|{\vec N}_{((ab)c)}|$,
 ${\vec S}_q\cdot {\hat N}_{((ab)c)}= {\check S}^3_q$ emerges.
 Eq.(\ref{III21}) defines $\alpha_{((ab)c)}$ and
 $\beta_{((ab)c)}$, so that Eq.(\ref{III22}) allows to identify a final unit
 vector ${\hat R}_{((ab)c)}$ with ${\vec S}_q\cdot {\hat R}_{((ab)c)}=0$ and
 $\{ {\hat R}_{((ab)c)}^r,{\hat R}^s_{((ab)c)} \} =0$.

In conclusion, when $S_q\not= 0$, we find both a {\it spin frame}
${\hat S}_q$, ${\hat R}_{((ab)c)}$, ${\hat R}_{((ab)c)}\times {\hat
S}_q$ and a {\it dynamical body frame} ${\hat \chi}_{((ab)c)}$, ${\hat
N}_{((ab)c)}\times {\hat \chi}_{((ab)c)}$, ${\hat N}_{((ab)c)}$, like
in the 3-body case. There is an {\it important difference}, however:
the orthonormal vectors ${\vec N}_{((ab)c)}$ and ${\vec
\chi}_{((ab)c)}$ depend on the momenta of the relative particles $a$
and $b$ through ${\hat R}_{(ab)}$, so that our results do not share
any relation with the N=4 non-trivial SO(3) principal bundle of the
orientation-shape bundle approach.

 The final 6 {\it dynamical shape variables} are $q^{\mu} = \{ |{\vec N}_{((ab)c)}|, \gamma_{(ab)},
 |{\vec N}_{(ab)}|, \rho_{qa}, \rho_{qb}, \rho_{qc} \}$. While the last four
 depend only on the original relative coordinates ${\vec \rho}_{qA}$, $A=a,b,c$,
 the first two depend also on the original momenta ${\vec \pi}_{qA}$:
 therefore they are {\it generalized shape variables}.
 By using Appendix D, we obtain

\beq
  \rho^r_{qA}={\cal R}^{rs}(\tilde \alpha ,
 \tilde \beta ,\tilde \gamma )\, {\check \rho}^s_{qA}(q^{\mu},
 p_{\mu}, {\check S}^r_q),\qquad A=a,b,c.
 \label{III35}
 \eeq

\noindent This means that for N=4 the dynamical body frame components
 ${\check \rho}^r_{qA}$ depend also on the dynamical shape momenta
 and on the dynamical body frame components of the spin. It is clear that
 this result stands completely outside the orientation-shape bundle approach.

As shown in Appendix F, starting from the Hamiltonian $H_{rel
((ab)c)}$ expressed in the final variables, we can define a rotational
Hamiltonian $H^{(rot)}_{rel ((ab)c)}$ (for ${\dot q}^{\mu}=0$, see
Eqs.(\ref{e18})) and a vibrational Hamiltonian $H^{(vib)}_{rel
((ab)c)}$ (vanishing of the physical dynamical angular velocity
${\check \omega}^r_{((ab)c)}=0$, see Eqs.(\ref{e21})), but $H_{rel
((ab)c)}$ fails to be the sum of these two Hamiltonians showing once
again the non separability of rotations and vibrations. Let us stress
that in the rotational Hamiltonian (\ref{e18}) we find tan {\it
inertia-like tensor} depending only on the dynamical shape variables.
A similar result, however, does not hold for the spin-angular velocity
relation (\ref{e19}) due to the presence of the shape momenta
$\xi_{(ab)}$ and $\xi_{((ab)c)}$.

The price to be paid for the existence of 3 global {\it dynamical body
frames} for N=4 is a more complicated form of the Hamiltonian kinetic
energy. On the other hand, {\it dynamical vibrations} and {\it
dynamical angular velocity} are  measurable quantities in each
dynamical body frame.

For N=5 we can repeat the previous construction either with the
sequence of spin clusterings $abcd \mapsto (ab)cd \mapsto ((ab)c)d)
\mapsto (((ab)c)d)$ or with the sequence $abcd \mapsto (ab)(cd) \mapsto
((ab)(cd))$ [$a,b,c,d$ any permutation of 1,2,3,4] as said in the
Introduction. Each {\it spin cluster} $(...)$ will be identified by
the unit vector ${\hat R}_{(...)}$, axis of the {\it spin frame} of
the cluster. All the final {\it dynamical body frames} built  with
this construction will have their axes depending on both the original
configurations and momenta.

This construction is trivially generalized to any N: we have only to
classify all the possible {\it spin clustering patterns}.

Therefore, for $N\geq 4$ our sequence of canonical and non-canonical
transformations leads to the following result, to be compared with
Eq.(\ref{III28}) of the 3-body case

\beq
\begin{minipage}[t]{5cm}
\begin{tabular}{|l|} \hline
${\vec \rho}_{qA}$ \\ \hline
 ${\vec \pi}_{qA}$ \\ \hline
\end{tabular}
\end{minipage}
\ {{\buildrel {non\,\, can.} \over \longrightarrow} \hspace{.2cm}}\
\begin{minipage}[b]{5cm}
\begin{tabular}{|lll|l|} \hline
$\tilde \alpha$ & $\tilde \beta$ & $\tilde \gamma$& $q^{\mu}({\vec
\rho}_{qA},{\vec \pi}_{qA})$
\\ \hline
 ${\check S}^1_q$ & ${\check S}^2_q$ & ${\check S}^3_q$ &
 $p_{\mu}({\vec \rho}_{qA},{\vec \pi}_{qA}$) \\ \hline
 \end{tabular}
 \end{minipage}.
\label{III36}
\eeq

This state of affairs suggests that for $N\geq 4$ and with $S_q\not=
0$, $S_{qA}\not= 0$, $A=a,b,c$, namely when the standard
(3N-3)-dimensional orientation-shape bundle is not trivial,  the
original (6N-6)-dimensional relative phase space admits the definition
of as many {\it dynamical body frames} as spin canonical bases
\footnote{To each one of them corresponds a different Hamiltonian {\it right} SO(3)
action on the relative phase space.}, which are globally defined
(apart isolated coordinate singularities) for the non-singular N-body
configurations with ${\vec S}_q\not= 0$ (and with nonzero spin for
each spin subcluster).

These {\it dynamical body frames} do not correspond to local cross
sections of the static non-trivial orientation-shape SO(3) principal
bundle and the spin canonical bases do not coincide with the canonical
bases associated with the static theory. Therefore, we do not find
gauge potentials as well as all the other quantities evaluated in
Appendix E for N=3.

\vfill\eject

\section{The Case of Interacting Particles.}

Although we have defined the kinematics for free particles, the
introduction of potentials is straightforward.

In the case N=2, the  general form of a 2-body potential  is $V({\vec
\rho}_q^2, {\vec \rho}_q\cdot {\vec \pi}_q, {\vec \pi}^2_q)= V(\rho^2_q,
\rho_q{\tilde \pi}_q, {\tilde \pi}^2_q+{{S^2_q}\over {\rho^2_q}})$.

In the case N=3, suppose there is a potential $V({\vec \eta}_{ij}\cdot
{\vec \eta}_{hk})$ with ${\vec \eta}_{ij}={\vec \eta}_i-{\vec \eta}_j$
. Then, since Eqs. (\ref{II8}) and (\ref{III26}) imply

\bea
{\vec \eta}_{ij}&=&{1\over {\sqrt{3}}} \sum_{a=1}^2
(\Gamma_{ai}-\Gamma_{aj}) {\vec \rho}_{qa}= {1\over {\sqrt{3}}}
\sum_{a=1}^2 (\Gamma_{ai}-\Gamma_{aj}) \rho_{qa} [\vec N +(-)^{a+1}
\vec \chi ],\nonumber \\
 &&{}\nonumber \\
 {\vec \eta}_{ij}\cdot {\vec \eta}_{hk} &=&
  {1\over 3} \sum_{a,b}^{1,2} (\Gamma_{ai}-\Gamma_{aj})(\Gamma_{bh}-\Gamma_{bk})
\rho_{qa} \rho_{qb} [{\vec N}^2 +(-)^{a+b} (1-{\vec N}^2)]=\nonumber \\
 &=&{1\over 3}\sum_{a=1}^2(\Gamma_{ai}-\Gamma_{aj})(\Gamma_{ah}-\Gamma_{ak})
 (\rho_{qa})^2 + \nonumber \\
 &+&{1\over 3}[(\Gamma_{1i}-\Gamma_{1j})(\Gamma_{2h}-\Gamma_{2k})+
(\Gamma_{2i}-\Gamma_{2j})(\Gamma_{1h}-\Gamma_{1k})
\rho_{q1}\rho_{q2}(2{\vec N}^2-1),
\label{IV1}
\eea

\noindent it turns out that the potential is a function of
$\rho_{q1}^2$, $\rho_{q2}^2$, $(2{\vec N}^2-1)\rho_{q1}\rho_{q2}$. On
the other hand, if the potential is $V({\vec
\eta}_{ij}\cdot {\vec \eta}_{hk}, {\vec \kappa}_i\cdot {\vec
\eta}_{hk})$, Eqs.(\ref{III26}) and (\ref{cc5}) give the momentum
dependence

\bea
{\vec \kappa}_i\cdot {\vec \eta}_{hk} &\approx& \sum_{a,b}^{1,2}
\gamma_{ai}(\Gamma_{bh}-\Gamma_{bk}) \rho_{qb} \Big( {\tilde \pi}_{qa}
[{\vec N}^2 + (-)^{a+b+1} (1-{\vec N}^2)] +\nonumber \\
 &+&{{|\vec N| \sqrt{1-{\vec N}^2}}\over {2\rho_{qa}}} [(-)^a+(-)^{b+1}] ({\check
S}^2_q+(-)^{a+1} \xi \sqrt{1-{\vec N}^2})\Big).
\label{IV2}
\eea

\noindent By means of the same equations it is possible to study a
dependence upon ${\vec \kappa}_i\cdot {\vec \kappa}_j$.

For $N \geq 4$, even a potential of the  form $V({\vec \eta}_{ij}\cdot
{\vec \eta}_{hk}))$ with ${\vec \eta}_{ij}={\vec \eta}_i-{\vec
\eta}_j$ turns out to be  {\it shape momentum} dependent, since we have

\beq
V({\vec
\eta}_{ij}\cdot {\vec \eta}_{hk})= V[{1\over N} \sum_{a,b}^{1,..,N-1}
(\Gamma_{ai}-\Gamma_{aj})(\Gamma_{bh}-\Gamma_{bk}) {\vec
\rho}_{qa}\cdot {\vec \rho}_{qb}].
\label{IV3}
\eeq

\noindent In particular, for N=4, due to Eq.(\ref{e7}), in the pattern $((ab)c)$
we get $V={\tilde V}_{((ab)c)}[\rho_{qa}, \rho_{qb}, \rho_{qc}, |{\vec
N}_{((ab)c)}|, \gamma_{(ab)}, |{\vec N}_{(ab)}|;\, \xi_{((ab)c)},
\Omega_{(ab)};\, {\check S}^r_q]$.

For more general potentials $V({\vec \eta}_{ij}\cdot {\vec \eta}_{hk},
{\vec \kappa}_i\cdot {\vec
\eta}_{hk}, {\vec \kappa}_i\cdot {\vec \kappa}_j)$, like the non-relativistic
limit of the relativistic Darwin potential of Ref.\cite{crater}, more
complicated expressions are obtained.

Finally, let us remark that both in the free and the interacting cases
it is easier to solve the original Hamilton equations, rather than the
equations for the orientation-shape coordinates, and then  use the
canonical transformation to find the time evolution of the dynamical
body frame and related  variables.

\vfill\eject

\section{Conclusions.}

While it is possible to separate out absolute translations and
relative motions in a global way, a global separation of absolute
rotations from the associated notion of vibrations is not feasible
because  of the non-Abelian nature of rotations.

This fact gave origin to the orientation-shape SO(3) principal bundle
approach described in Ref.\cite{little}. While the treatment of the
non-relativistic ideal {\it rigid body} is based on the notion of {\it
body frame} \footnote{This schematization does not exist in special
relativity: the only relativistic concept of rigidity are Born's rigid
motions \cite{born}.}, this and related concepts like angular
velocities turn out  to be {\it non-measurable gauge} quantities in
this approach.

In this paper we have shown that, by using geometric and
group-theoretical methods and by employing non-point canonical
transformations, it is possible to introduce both a {\it spin frame}
and a N-dependent discrete number of {\it evolving dynamical body
frames} for the N-body problem, with associated {\it dynamical
orientation and shape variables}. Since all these concepts are
introduced by means of a special class of canonical transformations
defining the {\it canonical spin bases}, they, and the related angular
velocities, become {\it measurable} quantities defined in terms of the
particle relative coordinates and momenta. A notion of {\it dynamical
vibrations} is associated  to each spin basis. {\it The spin bases are
defined by a method based on the coupling of spins}, which does not
require the use of Jacobi normal relative coordinates. However, the
Hamiltonian is not the sum of the dynamical {\it rotational} and {\it
vibrational} energies: this shows once again that rotations and
vibrations cannot be separated in a unique global way.

Even if, in differential geometry, Darboux canonical bases have no
intrinsic meaning, the physical requirement of the existence of the
extended Galilei group with the spin invariant $S$ selects a preferred
class of canonical coordinates adapted to the rotation subgroup SO(3).
Moreover, various E(3) realizations emerge whose invariants have a
physical meaning. The adaptation to these groups selects the final
class of physically preferred Darboux canonical bases.

Our definition of {\it dynamical body frames} is based on the
existence of {\it Hamiltonian non-symmetry right actions} of SO(3). We
begin assuming that, like for rigid bodies, the body frame axes in the
case N=3 be functions only of the particle relative coordinates.

For N=3, the orientation-shape SO(3) principal bundle is trivial and
we recover its results in a {\it xxzz gauge}. Interestingly we have
found, however, that already for N=4 the {it dynamical body frame}
axes depend also on the momenta so that any connection with the
orientation-shape bundle approach is lost.

It is an open problem whether the use of more general body frames for
N=3 and $N \geq 4$ obtained by using the freedom of making arbitrary
configurations dependent rotations (see footnote 9) may be used to
simplify the free Hamiltonian and/or some type of interaction.

It is hoped that our results may be instrumental for nuclear, atomic
and molecular physics, since a description based on spin clustering
rather than on the standard Jacobi center-of-mass clusterings was
lacking till now.

Let us observe that the extension of the {\it dynamical body frames}
to continuous deformable bodies (see Ref.\cite{mate} for an initial
study of the relativistic configurations of a Klein-Gordon field from
this point of view) is a lacking piece of kinematical information and
will be studied elsewhere.

The fact that we use non-point canonical transformations will make the
quantization more difficult than in the orientation-shape bundle
approach, where a separation of rotations from vibrations in the
Schroedinger equation is reviewed in Ref.\cite{little}. The
quantizations of the original canonical relative variables and of the
canonical spin bases will give equivalent quantum theories only if the
non-point canonical transformations are unitarily implementable. These
problems are completely unexplored.

In a future paper\cite{iten1} we shall study the relativistic N-body
problem, where the definition and separation of the center-of-mass
motion are known to be a complicated issue.  This problem has found a
solution within the Wigner-covariant rest-frame instant form of
dynamics \cite{lus}. It will be shown that concepts like reduced
masses, Jacobi normal relative coordinates and tensor of inertia do
not exist at the relativistic level. Yet, in the framework of the
rest-frame instant form, both the oriantation-shape SO(3) principal
bundle approach and the canonical spin bases can be defined just as in
the non-relativistic case.

ACKNOWLEDGMENTS: We are warmly indebted to our friend Michele
Vallisneri for his irreplaceable help in drawing our {\it frames}.

\vfill\eject

\appendix

\section{The Equations of Motion in the Orientation-Shape Bundle Approach.}

As shown in Ref.\cite{little} we get the following Euler-Lagrange
equations:\hfill\break
 i) for the orientation degrees of freedom

\begin{eqnarray}
{d\over {dt}} {\check S}_{qr}&\, {\buildrel \circ \over =}\,& {\check
X}^{(R)}{}_{(\alpha =r)}^{\beta} {{\partial L_{rel}}\over {\partial
\theta^{\beta}}} -\epsilon_{rsu}{\check \omega}_s{\check S}_{qu},\qquad
 (i.e.\,  {{d{\vec S}_q}\over {dt}}\, {\buildrel \circ \over =}\,0),
\nonumber \\
 &&\Downarrow \nonumber \\
 {{d {\vec S}_q^2}\over {dt}}&{\buildrel \circ \over =}&\, 0,
 \nonumber \\
 &&or\nonumber \\
  {{D{\check S}^r_q }\over {Dt}} &=&{d\over {dt}} {\check
S}^r_q- \Big( {\vec A}^{(k)}_{\mu}\times {\vec S}_q\Big)^r\, {\dot
q}^{\mu}\,
 {\buildrel \circ \over =}\nonumber \\
 &&{\buildrel \circ \over =}\, -\epsilon^{ruv}[{\check \omega}^u +
 {\check A}^{(k)u}_{\mu} {\dot q}^{\mu}]  {\check  S}^v_q, \nonumber \\
 \Rightarrow&& {{D {\check S}^r_q}\over {Dt}}\, {\buildrel \circ \over =}\, -\epsilon^{rsu}
 [{\check I}^{-1,sv} {\check S}^v_q] {\check S}^u_q,
 \label{a1}
 \end{eqnarray}

\noindent where in the last lines $D/Dt$ stands for the SO(3)-covariant derivative.

For a rigid body one could use ${\check S}_q^r={\check
I}^{(k)rs}{\check \omega}^s$ to eliminate either ${\vec S}_q$ or
${\vec \omega}$ from these equations, to obtain an autonomous system
of equations (Euler's equations for a torque-free rigid body). This
cannot be done for deformable bodies, because the relation between
body frame spin and angular velocity involves the shape variables.
Therefore, these equations are coupled to those for the shape
variables. The motion of ${\check S}^r_q$ takes place on the surface
of the sphere ${\vec S}_q^2=const.$ in body frame angular momentum
space. In the case of rigid bodies, this motion usually follows closed
orbits, but for deformable bodies the orbits will not close due to the
coupling to the shape variables. The body frame angular momentum
sphere is a two-dimensional surface, but counts for only a single
degree of freedom, since it is a phase space, not a configuration
space.

ii) for the shape degrees of freedom \footnote{The Lagrangian is that
of free particle in a non-Euclidean manifold with metric
$g^{(k)}_{\mu\nu}(q)$.}

\begin{eqnarray}
g_{\mu\nu}^{(k)} {{D {\dot q}^{\nu}}\over {Dt}}&=&
g_{\mu\nu}^{(k)}[{\ddot q}^{\nu} +\Gamma^{(k)\nu}_{\rho\sigma} {\dot
q}^{\rho}{\dot q}^{\sigma}]\, {\buildrel \circ \over =}\nonumber \\
 &{\buildrel \circ \over =}\,& {\vec S}_q\cdot {\vec B}^{(k)}_{\mu\nu}
 {\dot q}_{\nu}-{1\over 2} {\check S}^r_q [({\check I}^{(k)-1})^{rs}]_{;\mu}
 {\check S}^s_q,\nonumber \\
 &&{}\nonumber \\
 &&\Gamma^{(k)\mu}_{\rho\sigma}={1\over 2}g^{(k)\mu\nu}[g^{(k)}_{\nu\rho ,\sigma}
 +g^{(k)}_{\nu\sigma ,\rho}-g^{(k)}_{\rho\sigma ,\nu}],
\label{a2}
\end{eqnarray}

\noindent where $({\check I}^{(k)-1})_{;\mu}={{\partial {\check I}^{(k)-1}}\over {\partial
q^{\mu}}} -[{\vec A}^{(k)}_{\mu},{\check I}^{(k)-1}]$ is the covariant
derivative of the inverse of the inertia tensor.

While the geodesic equations (\ref{a2}) are $6N-6$ equations for the
$6N-6$ shape variables $q^{\mu}$, Eqs.(\ref{a1}) are only one equation
for one rotational degree of freedom. Therefore, these equations are a
system of equations for $6N-5$ degrees of freedom. The equations for
the other two orientational degrees of freedom  and their velocities
are represented by four first-order differential
equations:\hfill\break
\hfill\break
a) The restriction of the motion of ${\check S}^r_q$ to the surface of
a sphere $S_q=const.$, since ${{d {\vec S}^2_q}\over {dt}}\,
{\buildrel \circ \over =}\, 0$;\hfill\break
\hfill\break
b) The three equations

\begin{equation}
{d\over {dt}} R(\theta^{\alpha})\, {\buildrel \circ \over =}\,
R(\theta^{\alpha}) \check \Omega (\theta^{\alpha},{\dot
\theta}^{\alpha}) ,
\label{a3}
\end{equation}

\noindent
with $\check \Omega =\left( \begin{array}{ccc} 0& -{\check \omega}_3 &
{\check \omega}_2\\
 {\check \omega}_3 & 0& -{\check \omega}_1\\
  -{\check \omega}_2 & {\check \omega}_1& 0 \\
  \end{array} \right)$, which can be regarded as three equations for the
  three Euler angles.  Their solution requires a time ordering: $R(t)= R_o\, T
  e^{\int_{t_o}^{t_1}\check \Omega (t^{'}) dt^{'}}$ [see Refs.\cite{little,con4}].

 The Hamilton equations, whose content is equivalent to the Euler-Lagrange ones,
 are [Eqs.(\ref{II33}) are used]

 \begin{eqnarray}
 {\dot \theta}^{\alpha}\, &{\buildrel \circ \over =}\,& {\check X}^{(R)}{}^{\alpha}
 _{(\beta =r)} {{\partial H_{rel}}\over {\partial {\check S}_{qr}}}=
 {\check X}^{(R)}{}^{\alpha}_{(\beta =r)}  {\check \omega}^r=\nonumber \\
 &=&{\check X}^{(R)}{}^{\alpha}_{(\beta =r)}\, \Big( ({\check I}^{(k) -1})^{rs}{\check S}_q^s
 -g^{(k)\mu\nu} {\check A}^{(k)\alpha =r}_{\mu}[p_{\nu}-{\vec S}_q\cdot
 {\vec A}^{(k)}_{\nu}]\Big) ,\nonumber \\
 {d\over {dt}}{\check S}_{qr}\, &{\buildrel \circ \over =}\,&-{{\partial H_{rel}}\over
 {\partial \theta^{\alpha =r}}} -{\check S}_{qs}\epsilon^{sru}{{\partial H_{rel}}\over
 {\partial {\check S}_{qu}}}=\nonumber \\
 &=&\epsilon^{rsu} {\check S}_q^s[({\check I}^{(k)-1})^{uv}{\check S}_q^v-g^{(k)\mu\nu}
 {\check A}^{(k)u}_{\mu}(p_{\nu}-{\vec S}_q\cdot {\vec A}^{(k)}_{\nu})],
 \nonumber \\
 &&{}\nonumber \\
 {\dot q}^{\mu}\, &{\buildrel \circ \over =}\,&   {{\partial H_{rel}}\over {\partial p_{\mu}}},
 \nonumber \\
 {\dot p}_{\mu}\, &{\buildrel \circ \over =}\,& -{{\partial H_{rel}}\over {\partial q^{\mu}}}.
 \label{a4}
 \end{eqnarray}

\vfill\eject

\section{Rotational Kinematics in General Coordinates}

In this Appendix we shall reformulate the results of Subsection D of
Section II by using arbitrary non Jacobi relative coordinates ${\vec
\rho}_{qa}$. The body frame components of coordinates and velocities are

\begin{eqnarray}
\rho^r_{qa}&=&R^{rs}(\theta^{\alpha}) {\check \rho}^s_{qa}(q),\nonumber \\
{\dot \rho}^r_{qa}\, &{\buildrel {def} \over =}\,&
R^{rs}(\theta^{\alpha}) {\check v}^s_a,\nonumber \\
 {\vec v}_a&=&{\vec \omega}\times {\vec \rho}_{qa}+
{{\partial {\vec \rho}_{qa}}\over {\partial q^{\mu}}}{\dot
q}^{\mu},\quad\quad \omega^r{\buildrel {def} \over =}\,
R^{rs}(\theta^{\alpha}){\check \omega}^s,\nonumber \\
\Rightarrow && {\dot {\vec \rho}}_{qa}=\vec \omega \times {\vec \rho}_{qa}+{{\partial
{\vec \rho}_{qa}}\over {\partial q^{\mu}}}{\dot q}^{\mu},
\label{b1}
\end{eqnarray}

Let us remark that for the original particle positions ${\vec \eta}_i$
of Eqs.(\ref{II8}) we get the  result ${\dot {\vec
\eta}}_i={\dot {\vec q}}_{nr}+{1\over {\sqrt{N}}}\sum_{a=1}^{N-1}
\Gamma_{ai} {\dot {\vec \rho}}_{qa}={\dot {\vec q}}_{nr}+\vec \omega \times
({\vec \eta}_i-{\vec q}_{nr})+{{\partial {\vec \eta}_i}\over {\partial
q^{\mu}}} {\dot q}^{\mu}$ \footnote{For ${\dot q}^{\mu}=0$ we get the
rigid body result $\delta {\vec \eta}_i= {\dot {\vec
\eta}}_i\delta \tau =\delta {\vec q}_{nr}+\vec \omega \times ({\vec
\eta}_i-{\vec q}_{nr})\delta \tau$.}.

For the momenta we get

\begin{eqnarray}
\pi^r_{qa}&=&\sum_{b=1}^{N-1} k_{ab}{\dot \rho}^r_{qb}=R^{rs}(\theta^{\alpha}) {\check
\pi}^s_{qa},\nonumber \\
{\check \pi}^r_{qa}&=&R^{T\, rs}(\theta^{\alpha})
\pi^s_{qa}=\sum_{b=1}^{N-1}k_{ab} {\check v}^r_b=\nonumber \\
&=&\sum_{b=1}^{N-1} k_{ab} [({\vec \omega }\times {\vec
\rho}_{qb})^r+{{\partial {\check \rho}^r_{qb}}\over {\partial q^{\mu}}}{\dot q}
^{\mu}]=\nonumber \\
&=&({\vec \omega}\times \sum_{b=1}^{N-1}k_{ab}{\vec \rho}
_{qb}(q))^r+\sum_{b=1}^{N-1} k_{ab} {{\partial {\check \rho}^r_{qb}}\over {\partial
q^{\mu}}}{\dot q}^{\mu},
\label{b2}
\end{eqnarray}

\noindent while for the spin we get

\begin{eqnarray}
{\vec S}_q&=& \sum_{a=1}^{N-1} {\vec \rho}_{qa}\times {\vec
\pi}_{qa}= S^r_q {\hat f}_r ={\check S}^r_q {\hat e}_r,
\quad\quad S_q^r=R^{rs}(\theta ) {\check S}_q^s,\nonumber \\
 {\check S}^r_q  &=&
\sum_{a=1}^{N-1}\epsilon^{ruv}{\check \rho}_{qa}^u {\check
\pi}^v_{qa}=\sum_{ab}^{1..N-1}k_{ab} \epsilon^{ruv} {\check \rho}^u_{qa}
{\check  v}^v_b=\nonumber \\
 &=&\sum_{ab}^{1..N-1}k_{ab} \Big( {\vec
\rho}_{qa}\times \Big[ {\vec
\omega}\times {\vec \rho}_{qb}+{{\partial {\vec \rho}_{qb}}\over
{\partial q^{\mu}}}{\dot q}^{\mu}\Big] \Big)^r .
\label{b3}
\end{eqnarray}

By introducing the following body frame inertia tensor [$(ab)$ means
symmetrization]

\begin{eqnarray}
{\check {\cal I}}^{rs}(q,m)&=&\sum_{ab}^{1..N-1} k_{ab} {\check
I}^{rs}_{(ab)}(q),\nonumber \\
  {\check I}^{uv}_{(ab)}(q)&=&{\vec
\rho}_{qa}\cdot {\vec \rho}_{qb}\delta
^{uv}-{1\over 2}\Big( {\check \rho}^u_{qa}{\check \rho}^v_{qb}+{\check \rho}^v
_{qa}{\check \rho}^u_{qb}\Big),\nonumber \\
 &&{}\nonumber \\
 &&{\check I}^{uv}_{(ab)}({\check I}^{-1}_{(ab)})^{vt}=\delta^{ut}\qquad
for\, every\, pair\, (ab),
\label{b4}
\end{eqnarray}

\noindent and the following potentials

\begin{eqnarray}
{\check a}^u_{(ab)\mu}(q)&=&{1\over 2}\Big[ {\vec
\rho}_{qa}\times {{\partial {\vec \rho}_{qb}}\over {\partial
q^{\mu}}}+{\check {\vec \rho}}_b\times {{\partial {\check {\vec
\rho}}_a}\over {\partial q^{\mu}}}
\Big] {}^u \,\,\, {\buildrel {def} \over =}\, {\check I}^{uv}_{(ab)}(q)\,\,
{\check A}^v_{(ab)\mu}(q),\nonumber \\
 {\check a}^u_{\mu}(q,m)&=&\sum_{ab}^{1..N-1} k_{ab}\, {\check a}^u_{(ab)\mu}(q)=
 {\check {\cal I}}^{uv}(q,m)\,\, {\check {\cal A}}^v_{\mu}(q,m),
\label{b5}
\end{eqnarray}

\noindent we get the following gauge potentials

\begin{eqnarray}
{\check {\cal A}}^r_{\mu}(q,m)&=&[{\check {\cal
I}}^{-1}(q,m)]^{rs}\,\, {\check a}^s_{\mu}(q,m),\nonumber \\
 {\check A}^u_{(ab)\mu}(q)&=&({\check I}^{-1}_{(ab)}(q))^{uv}\,\,
{\check a}^v_{(ab)\mu}(q),
\label{b6}
\end{eqnarray}

For the body frame spin and angular velocity components we arrive at
the same results as in Subsection D of Section II where Jacobi
relative coordinates are used

\begin{eqnarray}
{\check S}_q^r&=&\sum_{ab}k_{ab} \Big[ {\vec \rho}_{qa}\cdot  {\vec
\rho}_{qb}{\check \omega}^r-{\vec \rho}_{qa}\cdot {\vec \omega}
{\check \rho}_{qb}^r+({\vec \rho}_{qa}\times {{\partial {\vec
\rho}_{qb}}\over {\partial q^{\mu}}})^r {\dot q}^{\mu}\Big] =\nonumber \\
 &=&\sum_{ab}^{1..N-1} k_{ab} {\check I}^{rs}_{(ab)}(q)
\Big[ {\check
\omega}^s + {\check A}^s_{(ab)\mu}(q) {\dot q}^{\mu} \Big] =\nonumber \\
&=&{\check {\cal I}}^{rs}(q,m) {\check
\omega}^s+\sum_{ab}^{1..N-1}k_{ab}{\check I}^{rs}_{(ab)}(q) {\check A}^s_{(ab)\mu}(q)
{\dot q}^{\mu}=\nonumber \\
 &=&{\check {\cal I}}^{rs}(q,m){\check
\omega}^s+{\check a}^r_{\mu}(q,m){\dot q}^{\mu}={\check {\cal I}}
^{rs}(q,m)[{\check \omega}^s+{\check {\cal A}}^s_{\mu}(q,m){\dot q}^{\mu}],
\label{b7}
\end{eqnarray}

\begin{eqnarray}
{\check \omega}^r&=&[{\check {\cal I}}^{-1}(q,m)]^{rs} {\check S}_q^s-
[{\check {\cal I}}^{-1}(q,m)]^{rs} \Big( \sum_{ab}^{1..N-1}k_{ab}
{\check a}^s_{(ab)\mu}(q)\Big) {\dot q}^{\mu} =\nonumber \\
 &=&[{\check {\cal I}}^{-1}(q,m)]^{rs} {\check S}^s_q- [{\check {\cal I}}^{-1}(q,m)]^{rs}
\,\, {\check a}^s_{\mu}(q,m) {\dot q}^{\mu}=\nonumber \\
 &=&[{\check {\cal I}}^{-1}(q,m)]^{rs} {\check S}^s_q- {\check {\cal A}}^r_{\mu}(q,m) {\dot
q}^{\mu},
\label{b8}
\end{eqnarray}

For the Lagrangian we get

\begin{eqnarray}
 L_{rel}({\vec \rho}_{qa},{\dot {\vec
\rho}}_{qa})&=&{1\over 2}\sum_{ab}^{1..N-1} k_{ab} {\dot {\vec
\rho}}_{qa}\cdot {\dot {\vec \rho}}_{qb}={1\over 2}\sum_{ab}
^{1..N-1}k_{ab} {\vec v}_a\cdot {\vec v}_b=\nonumber \\
&=&{1\over 2}\sum_{ab}^{1..N-1}k_{ab} \Big[ {\vec \omega}^2  {\vec
\rho}_{qa}\cdot {\vec \rho}_{qb} - {\vec
\omega}\cdot {\vec \rho}_{qa} {\vec \omega}\cdot {\vec \rho}_{qb}+
\nonumber \\
&+&{\vec \omega}\cdot \Big( {\vec \rho}_{qa}\times {{\partial  {\vec
\rho}_{qb}}\over {\partial q^{\mu}}}+{\vec
\rho}_{qb}\times {{\partial {\vec \rho}_{qa}}\over {\partial
q^{\mu}}}
\Big) {\dot q}^{\mu}+{{\partial {\vec \rho}_{qa}}\over {\partial q^{\mu}}}
\cdot {{\partial {\vec \rho}_{qb}}\over {\partial q^{\nu}}} {\dot q}^{\mu}
{\dot q}^{\nu} \Big] =\nonumber \\
 &=&{1\over 2}
\sum_{ab}^{1..N-1}k_{ab} \Big[ {\check I}^{uv}_{(ab)}(q){\check \omega}^u
{\check \omega}^v+2{\check a}^u_{(ab)\mu}(q){\check \omega}^u{\dot
q}^{\mu}+ {{\partial {\vec \rho}_{qa}}\over {\partial q^{\mu}}}
\cdot {{\partial {\vec \rho}_{qb}}\over {\partial q^{\nu}}} {\dot q}^{\mu}
{\dot q}^{\nu} \Big] =\nonumber \\
 &=&{1\over 2}\Big[ {\check {\cal
I}}^{uv}(q,m){\check \omega}^u{\check \omega}^v+
\sum_{ab}^{1..N-1}k_{ab} \Big( 2{\check a}^u_{(ab)\mu}(q){\check \omega}^u{\dot q}^{\mu}+
{{\partial {\vec \rho}_{qa}}\over {\partial q^{\mu}}}
\cdot {{\partial {\vec \rho}_{qb}}\over {\partial q^{\nu}}} {\dot q}^{\mu}
{\dot q}^{\nu} \Big) \Big] =\nonumber \\
 &=&{1\over 2} \Big[ {\check {\cal
I}}^{uv}(q,m){\check \omega}^u{\check \omega}^v+ 2{\check
a}^u_{\mu}(q,m) {\check \omega}^u {\dot q}^{\mu}+ h_{\mu\nu}(q,m)
{\dot q}^{\mu} {\dot q}^{\nu} \Big] =\nonumber\\
 &=&{1\over 2}\sum_{ab}^{1..N-1}k_{ab}
\Big[ {\check I}^{uv}_{(ab)}(q) ({\check \omega}^u +{\check A}^u_{(ab)\mu}(q){\dot
q}^{\mu}) ({\check \omega}^v+{\check A}^v_{(ab)\nu}(q){\dot q}
^{\nu})+\nonumber \\
&+&\Big( {{\partial {\vec \rho}_{qa}}\over {\partial q^{\mu}}}
\cdot {{\partial {\vec \rho}_{qb}}\over {\partial q^{\nu}}} -
{\check I}^{uv}_{(ab)} (q) {\check A}^u_{(ab)\mu}(q){\check
A}^v_{(ab)\nu}(q)\Big) {\dot q}^{\mu} {\dot q}^{\nu}
\Big] =\nonumber \\
 &=&{1\over 2}\sum_{ab}^{1..N-1}k_{ab}
\Big[ {\check I}^{uv}_{(ab)}(q) ({\check \omega}^u +{\check A}^u_{(ab)\mu}(q){\dot
q}^{\mu}) ({\check \omega}^v+{\check A}^v_{(ab)\nu}(q){\dot q}
^{\nu})+\nonumber \\
&+&g_{(ab)\mu\nu}(q) {\dot q}^{\mu} {\dot q}^{\nu} \Big] =\nonumber \\
&=&{1\over 2}\Big[ \sum_{ab}^{1..N-1}k_{ab} {\check I}^{uv}_{(ab)}(q)
({\check \omega}^u +{\check A}^u_{(ab)\mu}(q){\dot q}^{\mu}) ({\check
\omega}^v+{\check A}^v_{(ab)\nu}(q){\dot q}^{\nu})+\nonumber \\
&+&g_{\mu\nu}(q,m) {\dot q}^{\mu} {\dot q}^{\nu} \Big] =\nonumber \\
&=&{1\over 2} \Big[ ({\check {\cal I}}^{-1}(q,m))^{uv} {\check
S}_q^u{\check S}_q^v+{\tilde g}_{\mu\nu}(q,m) {\dot q}^{\mu} {\dot
q}^{\nu} \Big] =\nonumber \\
 &=&{1\over 2} \Big[ {\check {\cal I}}^{uv}(q,m) ({\check
\omega}^u+{\check {\cal A}}^u_{\mu}(q,m) {\dot q}^{\mu}) ({\check
\omega}^v+{\check {\cal A}}^v_{\nu}(q,m) {\dot q}^{\nu})+
\nonumber \\
&+&{\tilde g}_{\mu\nu}(q,m) {\dot q}^{\mu} {\dot q}^{\nu} \Big]
{\buildrel {def} \over =}\,
 L_{rel}(\check \omega (\theta ,\dot \theta ),q,\dot q),
\label{b9}
\end{eqnarray}

\noindent where we have introduced the following pseudo-metric and metric on shape space

\begin{eqnarray}
 h_{\mu\nu}(q,m)&=&\sum_{ab}^{1..N-1} k_{ab} {{\partial {\vec
\rho}_{qa}(q)}\over {\partial q^{\mu}}}
\cdot {{\partial {\vec \rho}_{qb}(q)}\over {\partial q^{\nu}}},
\nonumber \\
 &&{}\nonumber \\
 g_{\mu\nu}(q,m)&=&\sum_{ab}^{1..N-1}
k_{ab} g_{(ab)\mu\nu}(q)=h_{\mu\nu}(q,m)
-\sum_{ab}^{1..N-1}k_{ab}{\check I}^{uv}_{(ab)}(q)
{\check A}^u_{(ab)\mu}(q){\check A}^v_{(ab)\nu}(q),
\nonumber \\
&&g_{(ab)\mu\nu}(q)= {{\partial {\vec \rho}_{(qa}(q)}\over {\partial
q^{\mu}}}\cdot {{\partial {\vec \rho}_{qb)}(q)}\over {\partial
q^{\nu}}}-{\check I}^{uv}_{(ab)}(q){\check A}^u_{(ab)\mu}(q) {\check
A}^v_{(ab)\nu}(q),\nonumber \\ {\tilde
g}_{\mu\nu}(q,m)&=&h_{\mu\nu}(q,m)-{\check {\cal
A}}^u_{\mu}(q,m){\check {\cal I}}
^{uv}(q,m){\check {\cal A}}^v_{\nu}(q,m).
\label{b10}
\end{eqnarray}

The anholonomic momenta are

\begin{eqnarray}
{\check S}_q^r&=&{{\partial L_{rel}}\over {\partial {\check
\omega}_r }}={\check {\cal I}}^{rs}(q,m)({\check \omega}^s+{\check {\cal
A}}^s_{\mu}(q,m){\dot q}^{\mu}) ,\nonumber \\
 p_{\mu}&=&{{\partial L_{rel}}\over
{\partial {\dot q}^{\mu}}}={\tilde g}_{\mu\nu}(q,m){\dot
q}^{\nu}+{\vec S}_q\cdot {\vec {\cal A}}_{\mu}(q,m),\nonumber
\\
 &&{}\nonumber \\
{\check \omega}^u&=&[{\cal I}^{-1}(q,m)]^{uv} {\check S}_q^v-({\check
{\cal A}}^u_{\mu} {\tilde g}^{\mu\nu}[p_{\nu}-{\vec S}_q\cdot  {\vec
{\cal A}}_{\nu}])(q,m),\nonumber \\
 {\dot q}^{\mu}&=&{\tilde
g}^{\mu\nu}(q,m) [p_{\nu}-{\vec S}_q\cdot {\vec {\cal
A}}_{\nu}(q,m)],\nonumber \\
 &&\Downarrow   \nonumber \\
{\check \pi}^r_{qa}&=&\epsilon^{rus} \Big( [{\check {\cal
I}}^{-1}(q,m)]^{uv} {\check S}^v_q-({\check {\cal A}}^u_{\mu} {\tilde
g}^{\mu\nu}[p_{\nu}-{\vec S}_q\cdot {\vec {\cal A}}_{\nu}])(q,m)
\Big) \sum_{b=1}^{N-1}k_{ab}{\check \rho}^s_{qb}(q)+\nonumber \\
&+&\sum_{b=1}^{N-1}k_{ab} {{\partial {\check \rho}^r_{q}b}\over
{\partial q^{\mu}}}{\dot q}
^{\mu}{\tilde g}^{\mu\nu}(q,m) [p_{\nu}-{\vec S}_q\cdot
{\vec {\cal A}}_{\nu}(q,m)] ,\nonumber \\ &&{}\nonumber \\
 ({\check S}^r_q,\quad p_{\mu}) &=& ({\check S}^r_q,\quad p_{\mu})_v +
 ({\check S}^r_q,\quad p_{\mu})_{Ch}, \nonumber \\
 &&({\check S}^r_q,\quad p_{\mu})_v=
 ({\check S}^r_q,\quad {\vec S}_q\cdot {\vec
 A}_{\mu}(q,m) ),\quad p_{\mu}{|}_{{\dot q}^{\nu}=0}=
 {\vec S}_q\cdot {\vec A}_{\mu},\nonumber \\
 &&({\check S}^r_q,\quad p_{\mu})_{Ch}=
 ( 0,\quad p_{\mu}-{\vec S}_q\cdot {\vec
 A}_{\mu}(q,m) ),\quad p_{\mu}{|}_{{\check S}^r_q=o}=p_{\mu}-
 {\vec S}_q\cdot {\vec A}_{\mu}(q,m).
\label{b11}
\end{eqnarray}

The last lines show the phase space decomposition of generic momenta
into vertical and C-horizontal components.

Also the Hamiltonian takes the same form as with Jacobi coordinates

\begin{eqnarray}
H_{rel}&=&{1\over 2}\sum_{ab}^{1..N-1} k^{-1}_{ab} {\vec
\pi}_{qa}\cdot {\vec \pi}_{qb}= {\vec S}_q\cdot {\vec
\omega}+p_{\mu}{\dot q}^{\mu} -L_{rel}= \nonumber \\
 &=&{1\over 2}\Big[ {\check S}_q^u ({\check {\cal I}}^{-1}(q,m))^{uv} {\check S}
^v_q+\Big( {\tilde g}^{\mu\nu}(p_{\mu}-{\vec S}_q\cdot {\vec {\cal A}}
_{\mu}) (p_{\nu}-{\vec S}_q\cdot {\vec {\cal A}}_{\nu}) \Big) (q,m)\Big],
\nonumber \\
 &&{}\nonumber \\
 &&{\check \omega}^r\, {\buildrel \circ \over =}\, {{\partial H_{rel}}\over
 {\partial {\check S}^r}},\nonumber \\
 &&{\dot q}^{\mu}\, {\buildrel \circ \over =}\, {{\partial H_{rel}}\over {\partial
 p_{\mu}}}.
\label{b12}
\end{eqnarray}

\vfill\eject

\section{Some Formulas for the Case N=3.}

In this Appendix we justify Eq.(\ref{III25}).

To find the expression of ${\vec \pi}_{qa}\cdot \hat N$, ${\vec
\pi}_{qa}\cdot \hat \chi$, ${\vec \pi}_{qa}\cdot \hat N\times \hat \chi$
let us consider the following quantities

\begin{eqnarray}
{\vec S}_{q1}&=&{\vec \rho}_{q1}\times {\vec \pi}_{q1}=\rho_{q1} [\vec
N\times {\vec \pi}_{q1}+\vec \chi \times {\vec \pi}_{q1}],\nonumber \\
{\vec S}_{q2}&=&{\vec \rho}_{q2}\times {\vec \pi}_{q2}=\rho_{q2} [\vec
N\times {\vec \pi}_{q2}-\vec \chi \times {\vec \pi}_{q2}],\nonumber \\
&&{}\nonumber \\
 {\vec S}_q&=&{\vec S}_{q1}+{\vec S}_{q2}=\vec N\times
[\rho_{q1}{\vec
\pi}_{q1}+\rho_{q2}{\vec \pi}_{q2}]+\vec \chi \times [\rho_{q1}{\vec
\pi}_{q1}-\rho_{q2}{\vec \pi}_{q2}],\nonumber \\
{\vec W}_q&=&{\vec S}_{q1}-{\vec S}_{q2}=\vec N\times [\rho_{q1}{\vec
\pi}_{q1}-\rho_{q2}{\vec \pi}_{q2}]+\vec \chi \times [\rho_{q1}{\vec \pi}_{q1}
+\rho_{q2}{\vec \pi}_{q2}],\nonumber \\
 &&{}\nonumber \\
 {\vec S}_q\cdot \vec N &=& |\vec N| {\check S}^3_q=
 |\vec N| |\vec \chi | [\rho_{q1}{\vec
\pi}_{q1}-\rho_{q2}{\vec \pi}_{q2}]\cdot \hat N\times \hat \chi
=|\vec N| |\vec \chi | b_3,\nonumber \\
 {\vec S}_q\cdot \vec \chi &=& |\vec \chi | {\check S}^1_q=
 -|\vec N| |\vec \chi | [\rho_{q1}{\vec
\pi}_{q1}+\rho_{q2}{\vec \pi}_{q2}]\cdot \hat N\times \hat \chi
=-|\vec N| |\vec \chi | a_3,\nonumber \\
 {\vec S}_q\cdot \vec N\times \vec \chi &=&
 |\vec N \times \vec \chi | {\check S}^2_q=\nonumber \\
 &=&{\vec N}^2|\vec \chi |
[\rho_{q1}{\vec \pi}_{q1}+\rho_{q2}{\vec \pi}_{q2}]\cdot \hat \chi
-{\vec \chi}^2 |\vec N| [\rho_{q1}{\vec \pi}_{q1}-\rho_{q2}{\vec
\pi}_{q2}]\cdot \hat N=\nonumber \\
 &=&{\vec N}^2|\vec \chi | a_2-{\vec \chi}^2|\vec N| b_1,\nonumber \\
 &&{}\nonumber \\
{\vec W}_q\cdot \vec N&=&-{\vec S}_q\cdot \vec \chi ,
\nonumber \\
{\vec W}_q\cdot \vec \chi &=& -{\vec S}_q\cdot \vec N,\nonumber \\
{\vec W}_q\cdot \vec N\times \vec \chi &=& {\vec N}^2|\vec \chi
|[\rho_{q1}{\vec \pi}_{q1}-\rho_{q2}{\vec \pi}_{q2}]\cdot \hat \chi
-{\vec \chi}^2|\vec N| [\rho_{q1}{\vec \pi}_{q1}+\rho_{q2}{\vec
\pi}_{q2}]\cdot \hat N=\nonumber \\
 &=&{\vec N}^2|\vec \chi | b_2-{\vec \chi}^2|\vec N| a_1,\nonumber \\
&&{}\nonumber \\
\rho_{q1}{\tilde \pi}_{q1}&=&{\vec \rho}_{q1}\cdot {\vec
\pi}_{q1}=\rho_{q1}[{\vec \pi}_{q1}\cdot \vec N+{\vec \pi}_{q1}\cdot
\vec \chi ],\nonumber \\
 \rho_{q2}{\tilde \pi}_{q2}&=&
{\vec \rho}_{q2}\cdot {\vec \pi}_{q2}= \rho_{q2}[{\vec
\pi}_{q2}\cdot \vec N-{\vec \pi}_{q2}\cdot \vec \chi ],\nonumber \\
\rho_{q1}{\tilde \pi}_{q1}+\rho_{q2}{\tilde \pi}_{q2}
&=&{\vec \rho}_{q1}\cdot {\vec \pi}_{q1}+{\vec \rho}_{q2}\cdot {\vec
\pi}_{q2}=\nonumber \\
 &=&[\rho_{q1}{\vec \pi}_{q1}+\rho_{q2}{\vec \pi}_{q2}]\cdot \vec N+
[\rho_{q1}{\vec \pi}_{q1}-\rho_{q2}{\vec \pi}_{q2}]\cdot \vec
\chi =|\vec N| a_1+|\vec \chi | b_2,\nonumber \\
\rho_{q1}{\tilde \pi}_{q1}-\rho_{q2}{\tilde \pi}_{q2}
&=&{\vec \rho}_{q1}\cdot {\vec \pi}_{q1}-{\vec \rho}_{q2}\cdot {\vec
\pi}_{q2}=\nonumber \\
&=&[\rho_{q1}{\vec \pi}_{q1}-\rho_{q2}{\vec \pi}_{q2}]\cdot \vec N+
[\rho_{q1}{\vec \pi}_{q1}+\rho_{q2}{\vec \pi}_{q2}]\cdot \vec \chi
=|\vec N| b_1+|\vec \chi | a_2,
\label{cc1}
\end{eqnarray}

\noindent where

\begin{eqnarray}
a_1&=&[\rho_{q1}{\vec \pi}_{q1}+\rho_{q2}{\vec \pi}_{q2}] \cdot \hat
N,\nonumber \\
 a_2&=&[\rho_{q1}{\vec \pi}_{q1}+\rho_{q2}{\vec \pi}_{q2}] \cdot \hat
\chi ,\nonumber \\
 a_3&=&[\rho_{q1}{\vec \pi}_{q1}+\rho_{q2}{\vec \pi}_{q2}] \cdot \hat
 N\times \hat \chi ,\nonumber \\
 b_1&=&[\rho_{q1}{\vec \pi}_{q1}-\rho_{q2}{\vec \pi}_{q2}] \cdot \hat
 N,\nonumber \\
 b_2&=&[\rho_{q1}{\vec \pi}_{q1}-\rho_{q2}{\vec \pi}_{q2}]  \cdot \hat
 \chi ,\nonumber \\
 b_3&=&[\rho_{q1}{\vec \pi}_{q1}-\rho_{q2}{\vec \pi}_{q2}]  \cdot \hat
 N\times \hat \chi .
\label{cc2}
\end{eqnarray}

Eqs.(\ref{cc1}) imply

\begin{eqnarray}
a_1&=&\sqrt{ {{1+{\hat \rho}_{q1}\cdot {\hat \rho}_{q2}}\over 2}
}[\rho_{q1}{\tilde \pi}_{q1}+\rho_{q2}{\tilde \pi}_{q2}]- \sqrt{ {{1-
{\hat \rho}_{q1}\cdot {\hat \rho}_{q2}}\over 2} } {\vec W}_q\cdot \hat
N\times \hat \chi ,\nonumber \\
 a_2&=&\sqrt{ {{1-{\hat \rho}_{q1}\cdot {\hat \rho}_{q2}}\over 2}
}[\rho_{q1}{\tilde \pi}_{q1}-\rho_{q2}{\tilde \pi}_{q2}]+ \sqrt{ {{1+
{\hat \rho}_{q1}\cdot {\hat \rho}_{q2}}\over 2} } {\vec S}_q\cdot \hat
N\times \hat \chi ,\nonumber \\
 a_3&=&- {{ {\vec S}_q\cdot \hat \chi}\over {|\vec N|}},\nonumber \\
 b_1&=&\sqrt{ {{1+{\hat \rho}_{q1}\cdot {\hat \rho}_{q2}}\over 2}
}[\rho_{q1}{\tilde \pi}_{q1}-\rho_{q2}{\tilde \pi}_{q2}]- \sqrt{ {{1-
{\hat \rho}_{q1}\cdot {\hat \rho}_{q2}}\over 2} } {\vec S}_q\cdot \hat
N\times \hat \chi ,\nonumber \\
 b_2&=&\sqrt{ {{1-{\hat \rho}_{q1}\cdot {\hat \rho}_{q2}}\over 2}
}[\rho_{q1}{\tilde \pi}_{q1}+\rho_{q2}{\tilde \pi}_{q2}]+ \sqrt{ {{1+
{\hat \rho}_{q1}\cdot {\hat \rho}_{q2}}\over 2} } {\vec W}_q\cdot \hat
N\times \hat \chi ,\nonumber \\
 b_3&=& {{ {\vec S}_q\cdot \hat N}\over {|\vec \chi |}},
\label{cc3}
\end{eqnarray}

\noindent so that we get

\begin{eqnarray}
{\vec \pi}_{q1}\cdot \hat N&=&{{a_1+b_1}\over {2\rho_{q1}}}={\tilde
\pi}_{q1}\sqrt{ {{1+{\hat \rho}_{q1}\cdot {\hat \rho}_{q2}}\over 2} }
-{{ {\vec S}_{q1}\cdot \hat N\times \hat \chi}\over {\rho_{q1}}}
\sqrt{ {{1-{\hat \rho}_{q1}\cdot {\hat \rho}_{q2}}\over 2} },\nonumber \\
{\vec \pi}_{q2}\cdot \hat N&=&{{a_1-b_1}\over {2\rho_{q2}}}={\tilde
\pi}_{q2}\sqrt{ {{1+{\hat \rho}_{q1}\cdot {\hat \rho}_{q2}}\over 2} }
+{{ {\vec S}_{q2}\cdot \hat N\times \hat \chi}\over {\rho_{q2}}}
\sqrt{ {{1-{\hat \rho}_{q1}\cdot {\hat \rho}_{q2}}\over 2} },\nonumber \\
{\vec \pi}_{q1}\cdot \hat \chi &=&{{a_2+b_2}\over
{2\rho_{q1}}}={\tilde \pi}_{q1}\sqrt{ {{1- {\hat \rho}_{q1}\cdot {\hat
\rho}_{q2}}\over 2} }+{{ {\vec S}_{q1}\cdot \hat N\times \hat \chi}\over {\rho_{q1}}}
\sqrt{ {{1+{\hat \rho}_{q1}\cdot {\hat \rho}_{q2}}\over 2} },\nonumber \\
{\vec \pi}_{q2}\cdot \hat \chi &=&{{a_2-b_2}\over
{2\rho_{q2}}}=-{\tilde \pi}_{q2}\sqrt{ {{1- {\hat
\rho}_{q1}\cdot {\hat \rho}_{q2}}\over 2} } +{{ {\vec S}_{q2}\cdot \hat N\times
\hat \chi}\over {\rho_{q2}}}\sqrt{ {{1+
{\hat \rho}_{q1}\cdot {\hat \rho}_{q2}}\over 2} },\nonumber \\
 {\vec \pi}_{q1}\cdot \hat N\times \hat \chi &=&{{a_3+b_3}\over {2\rho_{q1}}}=
-{1\over {2\rho_{q1}}} {\vec S}_q\cdot [{{\hat \chi}\over {|\vec N|}}-{{\hat N}
\over {|\vec \chi |}}],\nonumber \\
{\vec \pi}_{q2}\cdot \hat N\times \hat \chi &=&{{a_3-b_3}\over
{2\rho_{q2}}}=-{1\over {2\rho_{q2}}} {\vec S}_q\cdot [{{\hat
\chi}\over {|\vec N|}} +{{\hat N}\over {|\vec \chi |}}].
\label{cc4}
\end{eqnarray}

This completes the study of the decomposition of the momenta ${\vec
\pi}_{qa}$ on the dynamical body frame.

Since ${\vec S}_q\cdot \hat N={\check S}^3_q$, ${\vec S}_q\cdot \hat
\chi =sin\, \psi cos\, \gamma ={\check S}^1_q$, ${\vec S}_q\cdot \hat N\times \hat \chi =
sin\, \psi sin\, \gamma ={\check S}^2_q$, ${\vec W}_q\cdot \hat
N\times \hat \chi =\xi \sqrt{1-{\vec N}^2}$, from Eq.(\ref{cc4}) we
get ${\vec \pi}_{qa}$ in terms of the final canonical variables

\begin{eqnarray}
{\vec \pi}_{q1}\cdot \hat N&=&{\tilde
\pi}_{q1} |\vec N|-{ {\sqrt{1-{\vec N}^2}}\over {2\rho_{q1}}}\Big[ {\check S}_q^2
 +\xi \sqrt{1-{\vec N}^2}\Big] ,\nonumber \\
 {\vec \pi}_{q2}\cdot \hat N&=&{\tilde \pi}_{q2} |\vec N|+
 { {\sqrt{1-{\vec N}^2}}\over {2\rho_{q2}}}\Big[  {\check S}_q^2
 -\xi \sqrt{1-{\vec N}^2}\Big] ,\nonumber \\
 {\vec \pi}_{q1}\cdot \hat \chi &=& {\tilde \pi}_{q1} \sqrt{1-{\vec N}^2}
+{ {|\vec N|}\over {2\rho_{q1}}}\Big[ {\check S}_q^2
 +\xi \sqrt{1-{\vec N}^2}\Big],\nonumber \\
 {\vec \pi}_{q2}\cdot \hat \chi &=&-{\tilde \pi}_{q2} \sqrt{1-{\vec N}^2}
+{ {|\vec N|}\over {2\rho_{q2}}}\Big[ {\check S}_q^2
  -\xi \sqrt{1-{\vec N}^2}\Big],\nonumber \\
 {\vec \pi}_{q1}\cdot \hat N\times \hat \chi &=&
{1\over {2\rho_{q1}}} \Big[ -{{ {\check S}_q^1 }\over {|\vec N|}}+
{{{\check S}^3_q}\over {\sqrt{1-{\vec N}^2}}}\Big] ,\nonumber \\
 {\vec \pi}_{q2}\cdot \hat N\times \hat \chi &=&-{1\over
{2\rho_{q2}}} \Big[ {{ {\check S}_q^1}\over {|\vec N|}}+ {{ {\check
S}^3_q}\over {\sqrt{1-{\vec N}^2}}}\Big].
 \label{cc5}
 \end{eqnarray}

For the spin quantities ${\vec W}_q$, ${\vec S}_{qa}$ we get

\begin{eqnarray}
{\vec W}_q&=&[-{{S_q}\over {|\vec N|\sqrt{1-{\vec N}^2}}}sin\, \psi
cos\, \psi cos\, \gamma +\xi \sqrt{1-{\vec N}^2} sin\, \psi sin\,
\gamma ]{\hat S}_q+\nonumber \\
 &+&[{{S_q}\over {|\vec N|}}\Big( {{ cos^2\, \psi}\over {\sqrt{1-{\vec N}^2}}}-
 \sqrt{1-{\vec N}^2}\Big) cos\, \gamma -\xi \sqrt{1-{\vec N}^2}
cos\, \psi sin\, \gamma ]\hat R -\nonumber \\
 &-&[ {{S_q|\vec N|} \over {\sqrt{1-{\vec N}^2}}}sin\, \gamma cos\, \psi
 +\xi \sqrt{1-{\vec N}^2} cos\, \gamma] {\hat S}_q \times \hat R
 =\nonumber \\
 &=& {\check W}^1_q \hat \chi + {\check W}^2_q \hat N \times \hat \chi +
 {\check W}^3_q \hat N = W^r_q {\hat f}_r =\nonumber \\
&=&{\vec W}_q[|\vec N|,\xi ;S_q,\alpha ;S^3_q,\beta ;{\check
 S}^3_q,\gamma ],\nonumber \\
  &&{}\nonumber \\
   W^1_q&=&-
  {{{\check S}^3_q \sqrt{(S_q)^2-({\check S}^3_q)^2}\sqrt{(S_q)^2-({S}^3_q)^2}}\over
  {|\vec N| \sqrt{1-{\vec N}^2} (S_q)^2}} cos\, \gamma cos\, \beta+\nonumber \\
  &+&{{\sqrt{(S_q)^2-({\check S}^3_q)^2}\sqrt{(S_q)^2-({S}^3_q)^2}}\over
  {(S_q)^2}} \sqrt{1-{\vec N}^2}  sin\, \gamma cos\, \beta \xi+\nonumber \\
  &+&\Big[ {{S_q}\over {|\vec N|}}\Big( {{({\check S}^3_q)^2}\over {(S_q)^2\sqrt{1-{\vec N}^2}}}
  -\sqrt{1-{\vec N}^2}\Big) cos\, \gamma -\nonumber \\
  &-&{{{\check S}^3_q}\over {S_q}}
  \sqrt{1-{\vec N}^2}sin\, \gamma \xi \, \Big] (sin\, \alpha sin\, \beta -
  {{S^3_q}\over {S_q}} cos\, \alpha cos\, \beta)-\nonumber \\
  &-& \Big[ {{|\vec N| {\check S}^3_q}\over {\sqrt{1-{\vec N}^2}}} +\sqrt{1-{\vec N}^2}
  \xi \Big] sin\, \gamma (cos\, \alpha sin\, \beta -{{S^3_q}\over {S_q}}
  sin\, \alpha cos\, \beta ),\nonumber \\
  W^2_q&=& -
  {{{\check S}^3_q \sqrt{(S_q)^2-({\check S}^3_q)^2}\sqrt{(S_q)^2-({S}^3_q)^2}}\over
  {|\vec N| \sqrt{1-{\vec N}^2} (S_q)^2}} cos\, \gamma sin\, \beta+\nonumber \\
  &+&{{\sqrt{(S_q)^2-({\check S}^3_q)^2}\sqrt{(S_q)^2-({S}^3_q)^2}}\over
  {(S_q)^2}} \sqrt{1-{\vec N}^2}  sin\, \gamma sin\, \beta \xi-\nonumber \\
  &-&\Big[ {{S_q}\over {|\vec N|}}\Big( {{({\check S}^3_q)^2}\over {(S_q)^2\sqrt{1-{\vec N}^2}}}
  -\sqrt{1-{\vec N}^2}\Big) cos\, \gamma -\nonumber \\
  &-&{{{\check S}^3_q}\over {S_q}}
  \sqrt{1-{\vec N}^2}sin\, \gamma \xi \Big] (sin\, \alpha cos\, \beta +
  {{S^3_q}\over {S_q}} cos\, \alpha sin\, \beta)+\nonumber \\
  &+& \Big[ {{|\vec N| {\check S}^3_q}\over {\sqrt{1-{\vec N}^2}}} +\sqrt{1-{\vec N}^2}
  \xi \Big] sin\, \gamma (cos\, \alpha cos\, \beta -{{S^3_q}\over {S_q}}
  sin\, \alpha sin\, \beta ),\nonumber \\
  W^3_q&=&-{{S^3_q{\check S}^3_q \sqrt{(S_q)^2-({\check S}^3_q)^2}}\over
  {|\vec N| \sqrt{1-{\vec N}^2} (S_q)^2}} cos\, \gamma +
  {{{\check S}^3_q{S}^3_q}\over {(S_q)^2}} \sqrt{1-{\vec N}^2}
  sin\, \gamma  \xi+\nonumber \\
  &+& {{\sqrt{(S_q)^2-(S^3_q)^2}}\over
  {|\vec N|}}\Big( {{({\check S}^3_q)^2}\over {(S_q)^2\sqrt{1-{\vec N}^2}}}
  -\sqrt{1-{\vec N}^2}\Big) cos\, \gamma cos\, \alpha -\nonumber \\
  &-&{{{\check S}^3_q\sqrt{(S_q)^2-(S^3_q)^2}}\over {(S_q)^2}}
  \sqrt{1-{\vec N}^2} cos\, \alpha sin\, \gamma \xi +\nonumber \\
  &+& \Big[ {{|\vec N| {\check S}^3_q\sqrt{(S_q)^2-(S^3_q)^2}}\over
  {\sqrt{1-{\vec N}^2} S_q}}sin\, \alpha sin\, \gamma +\sqrt{1-{\vec N}^2}
  {{\sqrt{(S_q)^2-(S^3_q)^2}}\over {S_q}}sin\, \alpha cos\, \gamma \xi ,\nonumber \\
  {\check W}^1_q &=& - {{|\vec N|}\over {\sqrt{1-{\vec N}^2}}} {\check S}^3_q,\nonumber \\
  {\check W}^2_q &=& \sqrt{1-{\vec N}^2} \xi ,\nonumber \\
  {\check W}^3_q &=& - {{\sqrt{1-{\vec N}^2}}\over {|\vec N|}} {\check S}^1_q,\nonumber \\
 &&{}\nonumber \\
\Rightarrow&& {\vec S}_{qa}={1\over 2}({\vec S}_q+(-)^{a+1} {\vec W}_q)
[|\vec N|,\xi ;S_q,\alpha ;S^3_q,\beta ;{\check S}^3_q,\gamma
],\nonumber \\
 &&{}\nonumber \\
 {\check S}^1_{qa} &=& {1\over 2} ( {\check S}^1_q - (-)^{a+1}
 {{|\vec N|}\over {\sqrt{1-{\vec N}^2}}} {\check S}^3_q),\nonumber \\
 {\check S}_{qa}^2 &=& {1\over 2} ( {\check S}^2_q +(-)^{a+1} \sqrt{1-{\vec N}^2}
 \xi ), \nonumber \\
 {\check S}^3_{qa} &=& {1\over 2} ( {\check S}^3_q -(-)^{a+1}
 {{\sqrt{1-{\vec N}^2}}\over {|\vec N|}} {\check S}^1_q).
\label{cc6}
\end{eqnarray}

We get also

\begin{eqnarray}
 {\vec \pi}^2_{q1}&=&{\tilde \pi}^2_{q1}+{1\over {4\rho^2_{q1}}} \Big[
 \xi^2(1-{\vec N}^2)+ ({\check S}_q^2)^2+{1\over { {\vec N}^2}} ({\check S}^1_q)^2
+\nonumber \\
 &+&{{({\check S}_q^3)^2}\over {1-{\vec N}^2}}+2 (\xi \sqrt{1-{\vec N}^2} {\check S}_q^2 -
 {{{\check S}_q^1{\check S}_q^3}\over {|\vec N| \sqrt{1-{\vec N}^2}}}  )
  \Big],\nonumber \\
 {\vec \pi}_{q2}^2&=&{\tilde \pi}^2_{q2}+{1\over {4\rho^2_{q2}}} \Big[
 \xi^2(1-{\vec N}^2)+ ({\check S}_q^2)^2+{1\over { {\vec N}^2}} ({\check S}^1_q)^2
+\nonumber \\
 &+&{{({\check S}_q^3)^2}\over {1-{\vec N}^2}}-2 (\xi \sqrt{1-{\vec N}^2} {\check S}_q^2 -
 {{{\check S}_q^1{\check S}_q^3}\over {|\vec N| \sqrt{1-{\vec N}^2}}}  )
  \Big],\nonumber \\
 {\vec \pi}_{q1}\cdot {\vec \pi}_{q2}&=& (2{\vec N}^2-1) {\tilde \pi}_{q1}{\tilde \pi}_{q2}
 +\nonumber \\
  &+&|\vec N| \sqrt{1-{\vec N}^2}\Big[ \Big( {{{\tilde \pi}_{q1}}\over
 {\rho_{q2}}}-{{{\tilde \pi}_{q2}}\over {\rho_{q1}}}\Big)
 {\check S}_q^2 - \Big({{{\tilde \pi}_{q1}}\over {\rho_{q2}}}+
 {{{\tilde \pi}_{q2}}\over {\rho_{q1}}}\Big) \xi \sqrt{1-{\vec N}^2}\Big]
+\nonumber \\
 &+&{1\over {4\rho_{q1}\rho_{q2}}} \Big[ (2{\vec N}^2-1)
 ({\check S}_q^2)^2+{1\over { {\vec N}^2}} ({\check S}^1_q)^2
-{{({\check S}_q^3)^2}\over {1-{\vec N}^2}}+\nonumber \\
&+& (1-{\vec N}^2)(2{\vec N}^2-1) \xi^2 \Big] ,
\label{cc7}
\end{eqnarray}

\vfill\eject

\section{Euler Angles.}

Let us denote by $\tilde \alpha$, $\tilde \beta$, $\tilde
\gamma$ the Euler angles chosen as orientation variables $\theta^{\alpha}$.

Let ${\hat f}_1=\hat i$, ${\hat f}_2=\hat j$, ${\hat f}_3=\hat k$ be
the unit 3-vectors along the axes of the space frame and ${\hat
e}_1=\hat \chi$, ${\hat e}_2=\hat N\times \hat \chi$, ${\hat e}_3=\hat
N$, the unit 3-vectors along the axes of a {\it body frame}. Then we
have

\begin{eqnarray}
{\vec S}_q&=& S^r_q {\hat f}_r =R^{rs}(\tilde \alpha, \tilde \beta,
\tilde \gamma ) {\check S}^s_q {\hat f}_r={\check S}^s_q {\hat e}_s,
\nonumber \\
 &&{\hat e}_s= (R^T)^{sr}(\tilde \alpha, \tilde \beta, \tilde \gamma )
 {\hat f}_r={\cal R}_s{}^r(\tilde \alpha, \tilde \beta,
\tilde \gamma ) {\hat f}_r.
\label{c1}
\end{eqnarray}

There are two main conventions for the definition of the Euler angles
$\tilde \alpha$, $\tilde \beta$, $\tilde \gamma$.

A) The {\it y-convention} (see Refs.\cite{c1} (Appendix B) and
\cite{c2}):\hfill\break
 i) perform a first rotation of an angle $\tilde \alpha$
around ${\hat f}_3$ [${\hat f}_1 \mapsto {\hat e}^{'}_1=cos\, \tilde
\alpha {\hat f}_1 +sin\, \tilde \alpha {\hat f}_2$, ${\hat f}_2
\mapsto {\hat e}^{'}_2=-sin\, \tilde \alpha {\hat f}_1+cos\, \tilde \alpha {\hat f}_2$,
${\hat f}_3 \mapsto {\hat e}^{'}_3={\hat f}_3$];\hfill\break
 ii) perform a second rotation of an angle $\tilde \beta$ around ${\hat e}^{'}_2$
 [${\hat e}^{'}_1 \mapsto {\hat e}^{"}_1=cos\, \tilde \beta {\hat e}^{'}_1
 -sin\, \tilde \beta {\hat e}^{'}_3$, ${\hat e}^{'}_2 \mapsto {\hat e}^{"}_2={\hat e}^{'}_2$,
 ${\hat e}^{'}_3 \mapsto {\hat e}^{"}_3=sin\, \tilde \beta {\hat e}^{'}_1+
 cos\, \tilde \beta {\hat e}^{'}_3$];\hfill\break
 iii) perform a third rotation of an angle $\tilde \gamma$ around ${\hat e}^{"}_3$
 [${\hat e}^{"}_1 \mapsto {\hat e}_1=cos\, \tilde \gamma {\hat e}^{"}_1+
 sin\, \tilde \gamma {\hat e}^{"}_2$, ${\hat e}^{"}_2 \mapsto {\hat e}_2=
 -sin\, \tilde \gamma {\hat e}^{"}_1+cos\, \tilde \gamma {\hat e}^{"}_2$].
 In this way we get

\begin{eqnarray}
&&\left( \begin{array}{c} \hat \chi \\ \hat N\times \hat \chi \\ \hat
N \end{array} \right) \equiv \left( \begin{array}{c} {\hat e}_1 \\
{\hat e}_2
\\ {\hat e}_3 \end{array} \right) ={\cal R}(\tilde \alpha ,\tilde
\beta ,\tilde \gamma ) \left( \begin{array}{c} {\hat f}_1 \\ {\hat f}_2 \\
{\hat f}_3 \end{array} \right) ,\nonumber \\
 &&{}\nonumber \\
 &&{\cal R}_r{}^s(\tilde \alpha ,\tilde \beta ,\tilde \gamma )=R^{T rs}(\tilde \alpha ,\tilde
\beta ,\tilde \gamma )=\nonumber \\
 &=&\left( \begin{array}{ccc} cos\, \tilde \gamma cos\, \tilde \beta
cos\, \tilde \alpha -sin\, \tilde \gamma sin\, \tilde \alpha & cos\,
\tilde \gamma cos\, \tilde \beta sin\, \tilde \alpha +sin\, \tilde \gamma cos\, \tilde \alpha &
-cos\, \tilde \gamma sin\, \tilde \beta \\
-(sin\, \tilde \gamma cos\, \tilde \beta cos\, \tilde \alpha +cos\, \tilde \gamma
sin\, \tilde \alpha & -sin\, \tilde \gamma cos\, \tilde \beta sin\,
\tilde \alpha +cos\, \tilde \gamma cos\, \tilde \alpha & sin\, \tilde \gamma
sin\, \tilde \beta \\ sin\, \tilde \beta cos\, \tilde \alpha & sin\,
\tilde \beta sin\, \tilde \alpha & cos\, \tilde \beta \end{array} \right) ,\nonumber \\
 &&{}\nonumber \\
 &&with\nonumber \\
 &&{}\nonumber \\
 tg\, \tilde \alpha &=& {{ {\hat N}^2}\over {{\hat N}^1}},\nonumber \\
 cos\, \tilde \beta &=& {\hat N}^3,\nonumber \\
 tg\, \tilde \gamma &=& - {{ {\hat \chi}^3}\over {(\hat N\times \hat \chi )^3}}.
\label{c2}
\end{eqnarray}

Since   $\hat N$ and $\hat \chi$ are functions of ${\vec \rho}_{qa}$
only, see Eq.(\ref{III18}), we get $\{ \tilde \alpha ,\tilde
\beta \} = \{ \tilde \beta ,\tilde \gamma \} = \{ \tilde \gamma , \tilde
\alpha \} =0$.

 B) The {\it x-convention} (see Refs.\cite{c3}, \cite{c1} (in the text) and
 \cite{pauri1}):
 the Euler angles $\theta$, $\varphi$ and $\psi$ are: i) $\theta =\tilde \beta$;
 ii) $cos\, \varphi =-sin\, \tilde \alpha$, $sin\, \varphi =cos\, \tilde \alpha$;
 iii) $cos\, \psi =sin\, \tilde \gamma$, $sin\, \psi =-cos\, \tilde \gamma$.

 We  use the {\it y-convention}. Following Ref.\cite{pauri1}, let us introduce
 the canonical momenta $p_{\tilde \alpha}$, $p_{\tilde \beta}$, $p_{\tilde
 \gamma}$ conjugated to $\tilde \alpha$, $\tilde \beta$, $\tilde \gamma$:
 $\{ \tilde \alpha ,p_{\tilde \alpha} \} = \{ \tilde \beta ,p_{\tilde
 \beta} \} = \{ \tilde \gamma ,p_{\tilde \gamma} \} =1$ [note that this Darboux chart
 does not exist globally]. Then, the results of Ref.\cite{pauri1} imply

 \begin{eqnarray}
S^1_q&=& -sin\, \tilde \alpha p_{\tilde \beta} +{{cos\, \tilde
\alpha}\over {sin\, \tilde \beta}} p_{\tilde \gamma} -cos\, \tilde \alpha ctg\,
\tilde \beta p_{\tilde \alpha},\nonumber \\
 S^2_q&=& cos\, \tilde \alpha p_{\tilde \beta} +{{sin\, \tilde \alpha}\over
 {sin\, \tilde \beta}} p_{\tilde \gamma} - sin\, \tilde \alpha
 ctg\, \tilde \beta p_{\tilde \alpha},\nonumber \\
 S^3_q&=& p_{\tilde \alpha},\nonumber \\
 &&{}\nonumber \\
 {\check S}^1_q&=& sin\, \tilde \gamma p_{\tilde \beta} -{{cos\, \tilde \gamma}\over
 {sin\, \tilde \beta}} p_{\tilde \alpha} +cos\, \tilde \gamma
 ctg\, \tilde \beta p_{\tilde \gamma},\nonumber \\
 {\check S}^2_q&=& cos\, \tilde \gamma p_{\tilde \beta} +
 {{sin\, \tilde \gamma}\over {sin\, \tilde \beta}} p_{\tilde \alpha}
 -sin\, \tilde \gamma ctg\, \tilde \beta p_{\tilde \gamma},\nonumber \\
 {\check S}^3_q&=& p_{\tilde \gamma},\nonumber \\
 &&\Downarrow \nonumber \\
 p_{\tilde \alpha} &=& S^3_q =-sin\, \tilde \beta cos\, \tilde \gamma {\check S}^1_q
 +sin\, \tilde \beta sin\, \tilde \gamma {\check S}^2_q +cos\, \tilde \beta
 {\check S}^3_q,\nonumber \\
 p_{\tilde \beta} &=& -sin\, \tilde \alpha S^1_q +cos\, \tilde \alpha S^2_q =
  sin\, \tilde \gamma {\check S}^1_q -cos\, \tilde \gamma {\check S}^2_q,\nonumber \\
  p_{\tilde \gamma}&=& {\check S}^3_q = cos\, \tilde \alpha sin\, \tilde \beta S^1_q
  +sin\, \tilde \alpha sin\, \tilde \beta S^2_q + cos\, \tilde \beta S^3_q.
 \label{c3}
 \end{eqnarray}

\vfill\eject

\section{The gauge potential in the xxzz gauge.}

From the Hamilton equations associated with the Hamiltonian
(\ref{III31}) the shape velocities are

\begin{eqnarray}
 {\dot \rho}_{q1}\, &{\buildrel \circ \over =}\,& {{\partial H_{rel}}\over
 {\partial {\tilde \pi}_{q1}}}=\nonumber \\
 &=&2k^{-1}{}_{11} {\tilde \pi}_{q1} +2k^{-1}{}_{12}
 \Big[ |\vec N|(1-{\vec N}^2) {{{\check S}^2_q}\over {\rho_{q2}}}
 +(2{\vec N}^2-1) {\tilde \pi}_{q2} -|\vec N| (1-{\vec N}^2) {{\xi}\over {\rho_{q2}}}\Big] ,
 \nonumber \\
 {\dot \rho}_{q2}\, &{\buildrel \circ \over =}\,& {{\partial H_{rel}}\over
 {\partial {\tilde \pi}_{q2}}}=\nonumber \\
 &=&2k^{-1}{}_{22} {\tilde \pi}_{q2} +2k^{-1}{}_{12}
 \Big[ -|\vec N|(1-{\vec N}^2) {{{\check S}^2_q}\over {\rho_{q1}}}
 +(2{\vec N}^2-1) {\tilde \pi}_{q1} -|\vec N| (1-{\vec N}^2) {{\xi}\over {\rho_{q1}}}\Big] ,
 \nonumber \\
 {\dot {|\vec N|}}\, &{\buildrel \circ \over =}\,& {{\partial H_{rel}}\over
 {\partial \xi}}=\nonumber \\
 &=&2 (1-{\vec N}^2)({{k^{-1}{}_{11}}\over {4\rho^2_{q1}}}
 +{{k^{-1}{}_{22}}\over {4\rho^2_{q2}}}) \xi+2\sqrt{1-{\vec N}^2}({{k^{-1}{}_{11}}\over
 {4\rho^2_{q1}}}-{{k^{-1}{}_{22}}\over {4\rho^2_{q2}}}){\check S}_q^2+\nonumber \\
 &+&2 k^{-1}{}_{12} (1-{\vec N}^2) \Big[ {{2{\vec N}^2-1}\over {4\rho_{q1}\rho_{q2}}} \xi
 -|\vec N|  ({{{\tilde \pi}_{q1}}\over {\rho_{q2}}}
 +{{{\tilde \pi}_{q2}}\over {\rho_{q1}}})\Big] .
 \label{d1}
 \eea

 Since we must have

 \bea
 {\tilde \pi}_{q1}{|}_{\dot q=0}&=& {\vec S}_q\cdot {\vec {\cal A}}_{q1}(q)=
 {\check S}^2_q {\check {\cal A}}^2_{q1}(\rho_{q1},\rho_{q2},|\vec N|),\nonumber \\
 {\tilde \pi}_{q2}{|}_{\dot q=0}&=& {\vec S}_q\cdot {\vec {\cal A}}_{q2}(q)=
 {\check S}^2_q {\check {\cal A}}^2_{q2}(\rho_{q1},\rho_{q2},|\vec N|),\nonumber \\
 \xi {|}_{\dot q=0}&=& {\vec S}_q\cdot {\vec {\cal A}}_{\xi}(q)=
 {\check S}^2_q {\check {\cal A}}^2_{\xi}(\rho_{q1},\rho_{q2},|\vec N|),
 \label{d2}
 \eea

 \noindent the gauge potential is

 \bea
 &&{\check {\cal A}}^1_{q1,q2,\xi}(q)={\check {\cal A}}^3_{q1,q2,\xi}(q)=0,
 \nonumber \\
 &&{}\nonumber \\
 &&{\check {\cal A}}_{\xi}^2(q)={{ \Big[ {{{\vec N}^2(1-{\vec N}^2)(k^{-1}_{12})^2}\over
 {k^{-1}_{11}k^{-1}_{22}-(2{\vec N}^2-1)^2(k^{-1}_{12})^2}}-{{\sqrt{1-{\vec N}^2}}\over
 4}\Big]\Big( {{k^{-1}_{11}}\over {\rho_1^2}}-{{k^{-1}_{22}}\over {\rho_2^2}}\Big)}\over
 {{1\over 4}\Big( {{k^{-1}_{11}}\over {\rho_1^2}}+{{k^{-1}_{22}}\over {\rho_2^2}}\Big)
 +{{(1-{\vec N}^2)(2{\vec N}^2-1)k^{-1}_{12}}\over {2\rho_1\rho_2}}-
 {{{\vec N}^2(k^{-1}_{12})^2}\over {k^{-1}_{11}k^{-1}_{22}-(2{\vec N}^2-1)(k^{-1}_{12})^2}}
 \Big( {{k^{-1}_{11}}\over {\rho_1^2}}+{{k^{-1}_{22}}\over {\rho_2^2}}-
 {{2(2{\vec N}^2-1)k^{-1}_{12}}\over {\rho_1\rho_2}}\Big] }} ,\nonumber \\
 &&{}\nonumber \\
 &&{\check {\cal A}}^2_{\rho_1}(q)=  -{{|\vec N| (1-{\vec N}^2) k^{-1}_{12}}\over
 {k^{-1}_{11}k^{-1}_{22}-(2{\vec N}^2-1)^2(k^{-1}_{12})^2}}\Big[ {{k^{-1}_{22}}\over {\rho_2}}
 +{{(2{\vec N}^2-1)k^{-1}_{12}}\over {\rho_1}}-\nonumber \\
 &-&({{k^{-1}_{22}}\over {\rho_2}}-
 {{(2{\vec N}^2-1)k^{-1}_{12}}\over {\rho_1}}) {\check {\cal A}}^2_{\xi}(q)\Big]
 ,\nonumber \\
 &&{}\nonumber \\
 &&{\check {\cal A}}^2_{\rho_2}(q)= {{|\vec N| (1-{\vec N}^2) k^{-1}_{12}}\over
 {k^{-1}_{11}k^{-1}_{22}-(2{\vec N}^2-1)^2(k^{-1}_{12})^2}}\Big[ {{k^{-1}_{11}}\over {\rho_1}}
 +{{(2{\vec N}^2-1)k^{-1}_{12}}\over {\rho_2}}+\nonumber \\
 &+&({{k^{-1}_{11}}\over {\rho_1}}-
 {{(2{\vec N}^2-1)k^{-1}_{12}}\over {\rho_2}}) {\check {\cal A}}^2_{\xi}(q)\Big] .\nonumber \\
 &&{}
 \label{d3}
 \eea

The term in the Hamiltonian quadratic in the shape momenta allows to
find the inverse metric

\beq
  {\tilde g}^{\mu\nu}(q)=\left( \begin{array}{lll}
  k^{-1}_{11} & (2{\vec N}^2-1)k^{-1}_{12} &
  -{{|\vec N| (1-{\vec N}^2)k^{-1}_{12}}\over {\rho_2}}\\
 (2{\vec N}^2-1)k^{-1}_{12} & k^{-1}_{22} &
 -{{\vec N| (1-{\vec N}^2)k^{-1}_{12}}\over {\rho_1}} \\
 -{{\vec N| (1-{\vec N}^2)k^{-1}_{12}}\over {\rho_2}} &
  -{{\vec N| (1-{\vec N}^2)k^{-1}_{12}}\over {\rho_1}} &
  {{1-{\vec N}^2}\over 4}\Big( {{k^{-1}_{11}}\over {\rho_1^2}}+
  {{k^{-1}_{22}}\over {\rho_2^2}}\Big) +
  {{(1-{\vec N}^2)(2{\vec N}^2-1)k^{-1}_{12}}\over {2\rho_1\rho_2}}
   \end{array} \right) .
\label{d4}
\eeq

The following equations for the angular velocity

\bea
{\check \omega}^r\, &{\buildrel \circ \over =}\,& {{\partial
H_{rel}}\over {\partial {\check S}_q^r}}={\check {\cal I}}^{-1\,
rs}(q){\check S}_q^s-{\check {\cal A}}^r_{\mu}(q) {\tilde
g}^{\mu\nu}(q) (p_{\nu}-{\vec S}_q\cdot {\vec {\cal
A}}_{\nu}(q)),\nonumber \\
 &&{}\nonumber \\
 {\check \omega}^1&=&{2\over {{\vec N}^2}}\Big( {{k^{-1}{}_{11}}\over {4\rho^2_{q1}}}+
 {{k^{-1}{}_{22}}\over {4\rho^2_{q2}}}+{{k^{-1}{}_{12}}\over {2\rho_{q1}\rho_{q2}}}\Big)
  {\check S}_q^1 -\nonumber \\
  &-&{2\over {|\vec N| \sqrt{1-{\vec N}^2}}}\Big( {{k^{-1}{}_{11}}\over {4\rho^2_{q1}}}
  -{{k^{-1}{}_{22}}\over {4\rho^2_{q2}}}\Big) {\check S}^3_q=
  {\check {\cal I}}^{-1 1s}(q){\check S}_q^s,\nonumber \\
  {\check \omega}^2&=&{2\over {{\vec N}^2}}\Big( {{k^{-1}{}_{11}}\over {4\rho^2_{q1}}}+
  {{k^{-1}{}_{22}}\over {4\rho^2_{q2}}}+{{k^{-1}{}_{12}(2{\vec N}^2-1)}\over {4\rho_{q1}\rho_{q2}}}
  \Big) {\check S}_q^2+\nonumber \\
  &+&\sqrt{1-{\vec N}^2}\Big[ 2\Big( {{k^{-1}{}_{11}}\over {4\rho^2_{q1}}}-
  {{k^{-1}{}_{22}}\over {4\rho^2_{q2}}}\Big) \xi +2k^{-1}{}_{12} |\vec N|
  \sqrt{1-{\vec N}^2}\Big( {{{\tilde \pi}_{q1}}\over {\rho_{q2}}}-
  {{{\tilde \pi}_{q2}}\over {\rho_{q1}}}\Big) \Big]=\nonumber \\
  &=&{\check {\cal I}}^{-1 2s}(q){\check S}_q^s-{\check {\cal A}}^2_{\mu}(q)
  {\tilde g}^{\mu\nu}(q)\Big( p_{\nu}-{\check S}^2_q {\check {\cal A}}_{\nu}^2(q)
  \Big),\nonumber \\
  {\check \omega}^3&=&{2\over {1-{\vec N}^2}}\Big( {{k^{-1}{}_{11}}\over
  {4\rho_{q1}^2}}+{{k^{-1}{}_{22}}\over {4\rho^2_{q2}}}-{{k^{-1}{}_{12}}\over
  {2\rho_{q1}\rho_{q2}}}\Big) {\check S}^3_q-\nonumber \\
  &-&{2\over {|\vec N| \sqrt{1-{\vec N}^2}}}\Big( {{k^{-1}{}_{11}}\over
  {4\rho^2_{q1}}}-{{k^{-1}{}_{22}}\over {4\rho^2_{q2}}}\Big) {\check S}^1_q=
  {\check {\cal I}}^{-1 3s}(q) {\check S}_q^s,\nonumber \\
  &&{}\nonumber \\
  &&{}\nonumber \\
  &&{\check \omega}^1{|}_{{\check {\vec S}}_q=0}={\check \omega}^3{|}_{{\check {\vec S}}_q=0}
  =0,\nonumber \\
  &&{\check \omega}^2{|}_{{\check {\vec S}}_q=0}=-{\check {\cal A}}^2_{\mu} \Big[
  {\tilde g}^{\mu q1} {\tilde \pi}_{q1}+{\tilde g}^{\mu q2} {\tilde \pi}_{q2}+
  {\tilde g}^{\mu \xi} \xi \Big], \nonumber \\
  &&{}\nonumber \\
  {\check \omega}^r =0 \,\, &\Rightarrow& \Big[ {\check {\cal I}}^{-1 rs}(q) +
  {\check {\cal A}}^r_{\mu}(q)\, {\tilde g}^{\mu\nu}(q)\, {\check {\cal A}}^s_{\nu}(q)\Big]
  \, {\check S}^s_q{|}_{{\check \omega}^t=0} = {\check {\cal A}}^r_{\mu}(q)
  {\tilde g}^{\mu\nu}(q)\, p_{\nu},\nonumber \\
  &&\Downarrow \nonumber \\
  {\check S}^r_q {|}_{{\check \omega}^s=0} &=& F^{r\nu}(q) p_{\nu},
  \label{d5}
  \eea

  \noindent allow to find the nonzero components of the inverse of the tensor of
  inertia in this canonical basis

  \bea
 {\check {\cal I}}^{-1, 11}(q)&=&
 {2\over {{\vec N}^2}}\Big( {{k^{-1}{}_{11}}\over {4\rho^2_{q1}}}+
 {{k^{-1}{}_{22}}\over {4\rho^2_{q2}}}+{{k^{-1}{}_{12}}\over {2\rho_{q1}\rho_{q2}}}\Big),
 \nonumber \\
 {\check {\cal I}}^{-1, 13}(q)&=& -
 {2\over {|\vec N| \sqrt{1-{\vec N}^2}}}\Big( {{k^{-1}{}_{11}}\over {4\rho^2_{q1}}}
  -{{k^{-1}{}_{22}}\over {4\rho^2_{q2}}}\Big) ,\nonumber \\
 {\check {\cal I}}^{-1, 22}(q)&=& {1\over {2{\vec N}^2}}\Big( {{k^{-1}_{11}}\over {\rho_1^2}}
 +{{k^{-1}_{22}}\over {\rho_2^2}}+{{(2{\vec N}^2-1)k^{-1}_{12}}\over {\rho_1\rho_2}}\Big)
 -\nonumber \\
 &-&{{|\vec N| (1-{\vec N}^2)k^{-1}_{12}}\over {\rho_1\rho_2}}
 \Big[ \rho_1 {\check {\cal A}}^2_{\rho_1}(q)-\rho_2 {\check {\cal A}}^2_{\rho_2}(q)\Big]
 -{{\sqrt{1-{\vec N}^2}}\over 2}\Big( {{k^{-1}_{11}}\over {\rho_1^2}}-{{k^{-1}_{22}}
 \over {\rho_2^2}}\Big) {\check {\cal A}}^2_{\xi}(q),\nonumber \\
 {\check {\cal I}}^{-1, 33}(q)&=&
 {2\over {1-{\vec N}^2}}\Big( {{k^{-1}{}_{11}}\over
  {4\rho_{q1}^2}}+{{k^{-1}{}_{22}}\over {4\rho^2_{q2}}}-{{k^{-1}{}_{12}}\over
  {2\rho_{q1}\rho_{q2}}}\Big).
 \label{d6}
 \end{eqnarray}

The other Hamilton equations besides Eqs.(\ref{d1}) are [we do not
give the explicit expression of the last three of them]

\bea
\dot \alpha \, &{\buildrel \circ \over =}\,& {{\partial H_{rel}}\over
{\partial S_q}}= S_q \Big[ {1\over 2} ({{cos^2\, \gamma}\over {{\vec
N}^2}}+sin^2\, \gamma)({{k^{-1}{}_{11}}\over
{\rho^2_{q1}}}+{{k^{-1}{}_{22}}\over {\rho^2_{q2}}})+\nonumber \\
 &+&({{cos^2\, \gamma}\over {{\vec N}^2}}+(2{\vec N}^2-1) sin^2\, \gamma )
 {{k^{-1}{}_{12}}\over {\rho_{q1}\rho_{q2}}}\Big] +\nonumber \\
 &+& {{S_q}\over {\sqrt{(S_q)^2-({\check S}^3_q)^2}}} \Big( sin\, \gamma
 \sqrt{1-{\vec N}^2} \Big[ {{\xi}\over 2} ({{k^{-1}{}_{11}}\over {\rho^2_{q1}}}-
 {{k^{-1}{}_{22}}\over {\rho^2_{q2}}})+\nonumber \\
 &+&2k^{-1}{}_{12} |\vec N| \sqrt{1-{\vec N}^2} ({{{\tilde \pi}_{q1}}\over {\rho_{q2}}}
 -{{{\tilde \pi}_{q2}}\over {\rho_{q1}}})\Big] -{{cos\, \gamma}\over
 {2 |\vec N| \sqrt{1-{\vec N}^2}}} ({{k^{-1}{}_{11}}\over {\rho^2_{q1}}}-
 {{k^{-1}{}_{22}}\over {\rho^2_{q2}}})\Big) ,\nonumber \\
\dot \beta \, &{\buildrel \circ \over =}\,& {{\partial H_{rel}}\over
{\partial S^3_q}}= 0,\nonumber \\
\dot \gamma \, &{\buildrel \circ \over =}\,& {{\partial H_{rel}}\over
{\partial {\check S}^3_q}}=-{\check S}^3_q \Big[ {1\over 2} ({{cos^2\,
\gamma}\over {{\vec N}^2}}+sin^2\, \gamma +{1\over {1-{\vec N}^2}})({{k^{-1}{}_{11}}\over
{\rho^2_{q1}}}+{{k^{-1}{}_{22}}\over {\rho^2_{q2}}})+\nonumber \\
 &+&({{cos^2\, \gamma}\over {{\vec N}^2}}+(2{\vec N}^2-1) sin^2\, \gamma -{1\over {1-{\vec N}^2}} )
 {{k^{-1}{}_{12}}\over {\rho_{q1}\rho_{q2}}}\Big] -\nonumber \\
 &-&{1\over {\sqrt{(S_q)^2-({\check S}^3_q)^2}}} \Big( sin\, \gamma
{\check S}^3_q
 \sqrt{1-{\vec N}^2} \Big[ {{\xi}\over 2} ({{k^{-1}{}_{11}}\over {\rho^2_{q1}}}-
 {{k^{-1}{}_{22}}\over {\rho^2_{q2}}})+\nonumber \\
 &+&2k^{-1}{}_{12} |\vec N| \sqrt{1-{\vec N}^2} ({{{\tilde \pi}_{q1}}\over {\rho_{q2}}}
 -{{{\tilde \pi}_{q2}}\over {\rho_{q1}}})\Big] +
 {{cos\, \gamma [(S_q)^2-2({\check S}^3_q)^2]}\over
 {2 |\vec N| \sqrt{1-{\vec N}^2}}} ({{k^{-1}{}_{11}}\over {\rho^2_{q1}}}-
 {{k^{-1}{}_{22}}\over {\rho^2_{q2}}})\Big) ,\nonumber \\
 &&{}\nonumber \\
 {\dot S}_q \, &{\buildrel \circ \over =}\,& -{{\partial H_{rel}}\over
{\partial \alpha}}= 0,\nonumber \\
 {\dot S}^3_q \, &{\buildrel \circ \over =}\,& -{{\partial H_{rel}}\over
{\partial \beta}}= 0,\nonumber \\
 {d\over {dt}} {\check S}^3_q \, &{\buildrel \circ \over =}\,& -{{\partial H_{rel}}\over
{\partial \gamma}}= sin\, \gamma cos\, \gamma [(S_q)^2-({\check
S}^3_q)^2] \Big[ {1\over 2}({1\over {{\vec
N}^2}}-1)({{k^{-1}{}_{11}}\over {\rho^2_{q1}}}+{{k^{-1}{}_{22}}\over
{\rho^2_{q2}}})+\nonumber \\
 &+&[{1\over {{\vec N}^2}}-(2{\vec
N}^2-1)]{{k^{-1}{}_{12}}\over {\rho_{q1}\rho_{q2}}}\Big] -
\sqrt{(S_q)^2-({\check S}^3_q)^2}\Big( cos\, \gamma \sqrt{1-{\vec
N}^2}\Big[ {{\xi}\over 2} ({{k^{-1}{}_{11}}\over {\rho^2_{q1}}}-
 {{k^{-1}{}_{22}}\over {\rho^2_{q2}}})+  \nonumber \\
 &+&2k^{-1}{}_{12} |\vec N| \sqrt{1-{\vec N}^2} ({{{\tilde \pi}_{q1}}\over {\rho_{q2}}}
 -{{{\tilde \pi}_{q2}}\over {\rho_{q1}}})\Big] +
  {{sin\, \gamma {\check S}^3_q}\over {2|\vec N| \sqrt{1-{\vec N}^2}}}
  ({{k^{-1}{}_{11}}\over {\rho^2_{q1}}}-
 {{k^{-1}{}_{22}}\over {\rho^2_{q2}}})  \Big),\nonumber \\
 &&{}\nonumber \\
 {\dot \pi}_{q1} \, &{\buildrel \circ \over =}\,& -{{\partial H_{rel}}\over
{\partial \rho_{q1}}},\nonumber \\
 {\dot \pi}_{q2} \, &{\buildrel \circ \over =}\,& -{{\partial H_{rel}}\over
{\partial \rho_{q2}}},\nonumber \\
 \dot \xi \, &{\buildrel \circ \over =}\,& -{{\partial H_{rel}}\over
{\partial |\vec N|}}.
\label{d7}
\eea

We recover the three non-Abelian constants of motion $S_q$, $\beta$,
$S^3_q$.

\vfill\eject

\section{4-Body Case}.

Let us give the main steps of  the construction of the spin basis in
the case  $N=4$. From Eqs.(\ref{III34}) we have

\bea
{\vec \rho}_{qa} &=& \rho_{qa} \Big[ |{\vec N}_{(ab)}| {\hat N}_{(ab)}
+\sqrt{1-{\vec N}_{(ab)}^2} {\hat \chi}_{(ab)}\Big] ,\nonumber \\
{\vec \rho}_{qb} &=& \rho_{qb} \Big[ |{\vec N}_{(ab)}| {\hat N}_{(ab)}
-\sqrt{1-{\vec N}_{(ab)}^2} {\hat \chi}_{(ab)}\Big] ,\nonumber \\
 {\vec \rho}_{qc} &=& \rho_{qc} {\hat R}_c.
\label{e1}
\eea

The definitions given after Eqs.(\ref{III34}) and Appendix D imply

\bea
\vec N \,&{\buildrel {def} \over =}\,& {\vec N}_{((ab)c)} =
{1\over 2} ({\hat R}_{(ab)} + {\hat R}_c),\nonumber \\
\vec \chi \,&{\buildrel {def} \over =}\,& {\vec \chi}_{((ab)c)} =
{1\over 2} ({\hat R}_{(ab)} - {\hat R}_c),\nonumber \\
 \vec N \times \vec \chi &=& - {1\over 2} {\hat R}_{(ab)} \times {\hat R}_c,\nonumber \\
 &&\Downarrow \nonumber \\
 {\vec \rho}_{qc} &=& \rho_{qc} \Big[ |\vec N| \hat N -\sqrt{1-{\vec N}^2} \hat \chi \Big],
 \nonumber \\
 \Rightarrow&& \rho^r_{qc} = {\cal R}^{rs}(\tilde \alpha ,\tilde \beta ,\tilde \gamma )
 {\check \rho}^s_{qc}(|\vec N|, \rho_{qc}).
\label{e2}
\eea

Let us remember that the dynamical shape variables of Eq.(\ref{III33})
are $q^{\mu}$ $= \{$ $|\vec N|,$$ \gamma_{(ab)}$, $|{\vec N}_{(ab)}|$,
$\rho_{qa}$, $\rho_{qb}$, $\rho_{qc} $$\}$. Then, from Eqs.
(\ref{III21}) and (\ref{III24}), for the subsystem $(ab)$ we get

\bea
{\hat R}_{ab} &=& \vec N + \vec \chi ,\qquad {\vec S}_{(ab)}\cdot
{\hat R}_{(ab)}=0,\nonumber \\
 {\vec S}_{(ab)}\, &{\buildrel {def} \over =}\,& {\vec S}_{qa} + {\vec S}_{qb}=\nonumber \\
 &=& S_{(ab)}\Big( sin\, \psi_{(ab)}\Big[ cos\, \gamma_{(ab)}{\hat \chi}_{(ab)}+sin\,
 \gamma_{(ab)}{\hat N}_{(ab)}\times {\hat \chi}_{(ab)}\Big] =cos\,
 \psi_{(ab)} {\hat N}_{(ab)}\Big),\nonumber \\
 &&{}\nonumber \\
 {\hat \chi}_{(ab)} &=& {1\over 2}({\hat R}_a - {\hat R}_b) =\nonumber \\
 &=& cos\, \gamma_{(ab)} \Big( sin\, \psi_{(ab)} {\hat S}_{(ab)} -
 cos\, \psi_{(ab)} {\hat R}_{(ab)}\Big) - sin\, \gamma_{(ab)}
 {\hat R}_{(ab)} \times {\hat S}_{(ab)},\nonumber \\
 {\hat N}_{(ab)} &=& cos\, \psi_{(ab)} {\hat S}_{(ab)} +
 sin\, \psi_{(ab)} {\hat R}_{(ab)},\nonumber \\
 {\hat N}_{(ab)} \times {\hat \chi}_{(ab)} &=& sin\, \gamma_{(ab)} \Big( sin\,
 \psi_{(ab)} {\hat S}_{(ab)} - cos\, \psi_{(ab)} {\hat R}_{(ab)}\Big) +
 cos\, \gamma_{(ab)} {\hat R}_{(ab)} \times {\hat S}_{(ab)},\nonumber \\
 &&{}\nonumber \\
 &&cos\, \psi_{(ab)}={{ \Omega_{(ab)}}\over {S_{(ab)}}},\qquad
 sin\, \psi_{(ab)} = \sqrt{1- ({{ \Omega_{(ab)}}\over {S_{(ab)}}})^2},\qquad
 \Omega_{(ab)} = {\vec S}_{(ab)}\cdot {\hat N}_{(ab)}.\nonumber \\
 &&
\label{e3}
\eea

We have moreover

\bea
 {\vec S}_{qc} &=& {1\over 2} ({\vec S}_q - {\vec W}_{q((ab)c)}),\nonumber \\
 &&{}\nonumber \\
 &&{\vec W}_{q((ab)c)}\cdot \vec N =-{\vec S}_q\cdot \vec \chi,\qquad
 {\vec W}_{q((ab)c)}\cdot \vec \chi =- {\vec S}_q\cdot \vec N,\nonumber \\
 &&{}\nonumber \\
 &&{\check W}^1_q=-{{|\vec N| {\check S}^3_q}\over {\sqrt{1-{\vec N}^2}}},\qquad
 {\check W}^2_q=\xi \sqrt{1-{\vec N}^2},\nonumber \\
 &&{\check W}^3_q=- {{\sqrt{1-{\vec N}^2}}\over {|\vec N|}} cos\, \gamma \sqrt{(S_q)^2-
 ({\check S}^3_q)^2},\nonumber \\
 &&{}\nonumber \\
 \gamma_{((ab)c)} &=& tg^{-1}\, {{{\vec S}_q\cdot (\hat N\times \hat \chi )}\over
 {{\vec S}_q\cdot \hat \chi}} = tg^{-1}\, {{{\check S}^2_q}\over {{\check S}^1_q}},
 \nonumber \\
 \xi_{((ab)c)} &=& {{{\vec W}_{q((ab)c)}\cdot (\hat N\times \hat \chi )}
 \over {\sqrt{1-{\vec N}^2}}},\nonumber \\
 &&{}\nonumber \\
 {\vec S}_{(ab)} &=& {1\over 2} ({\vec S}_q + {\vec W}_{q((ab)c)})=\nonumber \\
 &=& {\check S}^1_{(ab)} \hat \chi + {\check S}^2_{(ab)} \hat N \times
 \hat \chi + {\check S}^3_{(ab)} \hat N = \nonumber \\
 &=& {1\over 2}({\check S}^1_q +{\check W}^1_{q((ab)c)}) \hat \chi + {1\over 2}
 ({\check S}^2_q + {\check W}^2_{q((ab)c)}) \hat N \times \hat \chi +{1\over 2}
 ({\check S}^3_q + {\check W}^3_{q((ab)c)}) \hat N,\nonumber \\
 &&{}\nonumber \\
 {\vec S}_q&=&{\vec S}_{qa}+{\vec S}_{qb}+{\vec S}_{qc}={\vec S}_{(ab)}+{\vec S}_{qc}\,
 {\buildrel {def} \over =}\, {\vec S}_{q((ab)c)}=\nonumber \\
 &=&{\check S}^1_q {\hat \chi} +{\check S}^3_q \hat N +{\check S}^2_q
 \hat N\times \hat \chi ,\nonumber \\
 &&{}\nonumber \\
 &&{\check S}^1_{(ab)} ={1\over 2} \Big( {\check S}^1_q-{{|\vec N|}\over {\sqrt{1-{\vec N}^2}}}
 {\check S}^3_q\Big) ,\nonumber \\
 &&{\check S}^2_{(ab)} ={1\over 2} \Big( {\check S}^2_q +\xi_{((ab)c)} \sqrt{1-{\vec N}^2}
 \Big) ,\nonumber \\
 &&{\check S}^3_{(ab)} ={1\over 2} \Big( {\check S}^3_q -{{\sqrt{1-{\vec N}^2}}\over {|\vec N|}}
 {\check S}^1_q\Big) ,\nonumber \\
 &&{}\nonumber \\
 &&{\check S}^1_q= cos\, \gamma_{((ab)c)} \sqrt{(S_q)^2-({\check S}^3_q)^2},\qquad
 {\check S}^2_q= sin\, \gamma_{((ab)c)} \sqrt{(S_q)^2-({\check S}^3_q)^2},\nonumber \\
 &&{}\nonumber \\
 &&S_{(ab)} = \sqrt{ ({\check S}^1_{(ab)})^2 +({\check S}^2_{(ab)})^2+
 ({\check S}^3_{(ab)})^2 }=\nonumber \\
 &=&{1\over 2} \sqrt{ ({\check S}^2_q+\xi_{((ab)c)} \sqrt{1-{\vec N}^2})^2+
 ({{{\check S}^1_q}\over {|\vec N|}} -{{{\check S}^3_q}\over {\sqrt{1-{\vec N}^2}}})^2
 },\nonumber \\
 &&{}\nonumber \\
 {\hat S}_{(ab)} &=& {1\over {2S_{(ab)}}} \Big[ ({\check S}^2_q+\xi_{((ab)c)}
 \sqrt{1-{\vec N}^2}) \hat N\times \hat \chi +
 ({\check S}^1_q - {{|\vec N|}\over { \sqrt{1-{\vec N}^2} }}
 {\check S}^3_q) \hat \chi +\nonumber \\
   &+& ({\check S}^3_q-{ { \sqrt{1-{\vec N}^2} }\over {|\vec N|} } {\check S}^1_q)
 \hat N \Big],\nonumber \\
 {\hat R}_{(ab)} &=& |\vec N| \hat N +\sqrt{1-{\vec N}^2} \hat \chi ,\nonumber \\
 {\hat R}_{(ab)} \times {\hat S}_{(ab)} &=& {1\over {2S_{(ab)}}} \Big[
 ({\check S}^2_q +\xi_{((ab)c)} \sqrt{1-{\vec N}^2}) (\sqrt{1-{\vec N}^2}\hat N
 - |\vec N| \hat \chi ) +\nonumber \\
 &+& \Big( |\vec N| ({\check S}^1_q- {{|\vec N|}\over {\sqrt{1-{\vec N}^2}}} {\check
 S}^3_q) - \sqrt{1-{\vec N}^2} ({\check S}^3_q- {{\sqrt{1-{\vec N}^2}}\over {|\vec N|}}
 {\check S}^1_q) \Big) \hat N\times \hat \chi \Big].
\label{e4}
\eea

The final result is  [${\check S}^1_q={1\over
{S_q}}\sqrt{(S_q)^2-(S^3_q)^2} cos\, \beta_{((ab)c)}$, ${\check
S}^2_q={1\over {S_q}}\sqrt{(S_q)^2-(S^3_q)^2} sin\, \beta_{((ab)c)}$]

\bea
{\hat N}_{(ab)} &=& cos\, \psi_{(ab)} {\hat S}_{(ab)} + sin\,
\psi_{(ab)} {\hat R}_{(ab)} =\nonumber \\
 &=& \Big[ {{\Omega_{(ab)} ({\check S}^3_q- {{\sqrt{1-{\vec N}^2}}\over
 {|\vec N|}} {\check S}^1_q)}\over
 {2(S_{(ab)})^2}} + |\vec N| \sqrt{1-({{\Omega_{(ab)}}\over {S_{(ab)}}})^2}\Big] \hat N
 +\nonumber \\
 &+& \Big[ {{\Omega_{(ab)} ({\check S}^1_q- {{|\vec N|}\over {\sqrt{1-{\vec N}^2}}}
 {\check S}^3_q)}\over { 2(S_{(ab)})^2}} + \sqrt{1-{\vec N}^2} \sqrt{1-({{\Omega_{(ab)}}\over
  {S_{(ab)}}})^2} \Big] \hat \chi +\nonumber \\
 &+& {{\Omega_{(ab)} ({\check S}^2_q +\xi_{((ab)c)} \sqrt{1-{\vec N}^2})}\over
 {2 (S_{(ab)})^2}} \hat N \times \hat \chi ,\nonumber \\
 &&{}\nonumber \\
 {\hat \chi}_{(ab)} &=& cos\, \gamma_{(ab)}\Big( sin\, \psi_{(ab)} {\hat S}_{(ab)}-
 cos\, \psi_{(ab)} {\hat R}_{(ab)}\Big) -sin\, \gamma_{(ab)} {\hat R}_{(ab)}
 \times {\hat S}_{(ab)}=\nonumber \\
 &=&\Big( cos\, \gamma_{(ab)} \Big[ \sqrt{1-({{\Omega_{(ab)}}\over {S_{(ab)}}})^2}
 {{({\check S}^3_q- {{\sqrt{1-{\vec N}^2}}\over {|\vec N|}} {\check S}^1_q)}\over
 {2S_{(ab)}}} - |\vec N| {{\Omega_{(ab)}}\over {S_{(ab)}}}\Big] -\nonumber \\
 &-&{{\sqrt{1-{\vec N}^2}sin\, \gamma_{(ab)}}\over
 {2 S_{(ab)}}}({\check S}^2_q +\xi_{((ab)c)} \sqrt{1-{\vec N}^2})\Big) \hat N +\nonumber \\
 &+&\Big( cos\, \gamma_{(ab)} \Big[ \sqrt{1-({{\Omega_{(ab)}}\over {S_{(ab)}}})^2}
 {{({\check S}^1_q- {{|\vec N|}\over {\sqrt{1-{\vec N}^2}}}{\check S}^3_q)}\over
 {2S_{(ab)}}} -
 \sqrt{1-{\vec N}^2} {{\Omega_{(ab)}}\over {S_{(ab)}}}\Big] +\nonumber \\
 &+&{{|\vec N| sin\, \gamma_{(ab)}}\over
 {2 S_{(ab)}}}({\check S}^2_q +\xi_{((ab)c)} \sqrt{1-{\vec N}^2})\Big) \hat \chi +\nonumber \\
 &+& \Big( cos\, \gamma_{(ab)} \sqrt{1-({{\Omega_{(ab)}}\over {S_{(ab)}}})^2}
 {{({\check S}^2_q +\xi_{((ab)c)} \sqrt{1-{\vec N}^2})}\over
 {2S_{(ab)}}} -\nonumber \\
 &-& {{sin\, \gamma_{(ab)}}\over
 {2 S_{(ab)}}}\Big[ |\vec N| ({\check S}^1_q-{{|\vec N|}\over {\sqrt{1-{\vec N}^2}}}
 {\check S}^3_q)-\sqrt{1-{\vec N}^2}({\check S}^3_q- {{\sqrt{1-{\vec N}^2}}\over
 {|\vec N|}} {\check S}^1_q)\Big] \Big) \hat N \times \hat \chi .
\label{e5}
\eea

Then, Eqs.(\ref{e1}) and (\ref{e2}) imply

\bea
\rho^r_{qa} &=& {\cal R}^{rs}(\tilde \alpha ,\tilde \beta ,\tilde \gamma ) {\check \rho}^s_{qa}
[|\vec N|, \gamma_{(ab)}, |{\vec N}_{(ab)}|, \rho_{qa}; \xi ,
\Omega_{(ab)}; {\check S}^r_q],\nonumber \\
\rho^r_{qb} &=& {\cal R}^{rs}(\tilde \alpha ,\tilde \beta ,\tilde \gamma ) {\check \rho}^s_{qb}
[|\vec N|, \gamma_{(ab)}, |{\vec N}_{(ab)}|, \rho_{qb}; \xi ,
\Omega_{(ab)}; {\check S}^r_q] ,\nonumber \\
\rho^r_{qc} &=& {\cal R}^{rs}(\tilde \alpha ,\tilde \beta ,\tilde \gamma )
 {\check \rho}^s_{qc}[|\vec N|, \rho_{qc}].
\label{e6}
\eea

\noindent This means that ${\check \rho}^r_{qa\, or\, b}$ depend not only on
the dynamical shape variables $q^{\mu}=\{ |\vec N| = |{\vec
N}_{((ab)c)}|, \gamma_{(ab)}, |{\vec N}_{(ab)}|, \rho_{qa}, \rho_{qb},
\rho_{qc} \}$ but also on some of the conjugate  shape momenta $p_{\mu}
=\{ \xi_{((ab)c)}, \Omega_{(ab)}={\vec S}_{q(ab)}\cdot {\hat N}_{(ab)},
\xi_{(ab)}, {\tilde \pi}_{qa}, {\tilde \pi}_{qb}, {\tilde \pi}_{qc} \}$ and  the
dynamical body frame components ${\check S}^r_q$ of the spin.

Therefore, a potential $V({\vec \eta}_{ij}\cdot {\vec
\eta}_{hk})= V^{'}({\vec \rho}_{qA}\cdot {\vec
\rho}_{qB})=\tilde V(\rho^2_{qa}, \rho^2_{qb}, \rho^2_{qc}, {\vec \rho}_{qa}\cdot
{\vec \rho}_{qb}=
(2{\vec N}^2-1)\rho_{qa}\rho_{qb}, {\vec \rho}_{qa}\cdot {\vec
\rho}_{qc},{\vec \rho}_{qb}\cdot {\vec \rho}_{qc})$ becomes momentum dependent.
As a matter of fact the last two scalar products become

\bea
{\vec \rho}_{qa}\cdot {\vec \rho}_{qc} &=& \rho_{qa}\rho_{qc}
\Big[ |{\vec N}_{(ab)}| \Big( {{\Omega_{(ab)}}\over {(S_{(ab))^2}}}
(|\vec N| {\check S}^3_q -\sqrt{1-{\vec N}^2} {\check S}^1_q)-
 \sqrt{1-({{\Omega_{(ab)}}\over {S_{(ab)}}})^2}\Big)+ \nonumber \\
 &+& \sqrt{1-{\vec N}^2_{(ab)}} \Big( {{\Omega_{(ab)}}\over {S_{(ab)}}}+
 cos\, \gamma_{(ab)}  \sqrt{1-({{\Omega_{(ab)}}\over {S_{(ab)}}})^2}
 {{  (|\vec N| {\check S}^3_q -\sqrt{1-{\vec N}^2} {\check S}^1_q) }\over {S_{(ab)}}}
 \Big) \Big] ,\nonumber \\
{\vec \rho}_{qb}\cdot {\vec \rho}_{qc} &=& \rho_{qb}\rho_{qc}
\Big[ |{\vec N}_{(ab)}| \Big( {{\Omega_{(ab)}}\over {(S_{(ab))^2}}}
(|\vec N| {\check S}^3_q -\sqrt{1-{\vec N}^2} {\check S}^1_q)-
 \sqrt{1-({{\Omega_{(ab)}}\over {S_{(ab)}}})^2}\Big) - \nonumber \\
 &-& \sqrt{1-{\vec N}^2_{(ab)}} \Big( {{\Omega_{(ab)}}\over {S_{(ab)}}}+
 cos\, \gamma_{(ab)}  \sqrt{1-({{\Omega_{(ab)}}\over {S_{(ab)}}})^2}
 {{  (|\vec N| {\check S}^3_q -\sqrt{1-{\vec N}^2} {\check S}^1_q) }\over {S_{(ab)}}}
 \Big) \Big].
\label{e7}
\eea

\noindent This {\it spin basis} could be useful when the effective potential
has a negligible dependence on these two scalar products since ever
the main bonds (for instance chemical bonds in molecular dynamics) do
not involve them.

By using Eq.(\ref{cc5}), for the canonical momenta we get [$|\vec
N|=|{\vec N}_{((ab)c)}|$, $S_{(ab)}$ given by Eq.(\ref{e4})]

\bea
{\vec \pi}_{qa} &=&\Big[ {\tilde \pi}_{qa} |{\vec N}_{(ab)}|
-{{\sqrt{1-{\vec N}^2_{(ab)}}}\over {2\rho_{qa}}} (\sqrt{(S_{(ab)})^2-(\Omega_{(ab)})^2}
sin\, \gamma_{(ab)} +\xi_{(ab)} \sqrt{1-{\vec N}^2_{(ab)}})
\Big] {\hat N}_{(ab)}+\nonumber \\
 &+&\Big[ {\tilde \pi}_{qa} \sqrt{1-{\vec N}^2_{(ab)}} +{{|{\vec
N}_{(ab)}|}\over {2\rho_{qa}}} (\sqrt{(S_{(ab)})^2-(\Omega_{(ab)})^2}
sin\, \gamma_{(ab)} +\xi_{(ab)} \sqrt{1-{\vec N}^2_{(ab)}})
\Big] {\hat \chi}_{(ab)} -\nonumber \\
 &-&{1\over {2\rho_{qa}}}
 ( \sqrt{(S_{(ab)})^2-(\Omega_{(ab)})^2} {{cos\, \gamma_{(ab)}}\over {|{\vec
N}_{(ab)}|}}-{{\Omega_{(ab)}}\over {\sqrt{1-{\vec N}^2_{(ab)}}}})
 {\hat N}_{(ab)} \times {\hat \chi}_{(ab)},\nonumber \\
{\vec \pi}_{qb} &=&\Big[ {\tilde \pi}_{qb} |{\vec N}_{(ab)}|
+{{\sqrt{1-{\vec N}^2_{(ab)}}}\over {2\rho_{qb}}}
(\sqrt{(S_{(ab)})^2-(\Omega_{(ab)})^2} sin\, \gamma_{(ab)} -\xi_{(ab)}
\sqrt{1-{\vec N}^2_{(ab)}})\Big] {\hat N}_{(ab)}-\nonumber \\
&-&\Big[ {\tilde \pi}_{qb} \sqrt{1-{\vec N}^2_{(ab)}} -{{|{\vec
N}_{(ab)}|}\over {2\rho_{qb}}} (\sqrt{(S_{(ab)})^2-(\Omega_{(ab)})^2}
sin\, \gamma_{(ab)} -\xi_{(ab)} \sqrt{1-{\vec N}^2_{(ab)}})
\Big] {\hat \chi}_{(ab)} -\nonumber \\
&-&{1\over {2\rho_{qb}}}(\sqrt{(S_{(ab)})^2-(\Omega_{(ab)})^2} {{cos\,
\gamma_{(ab)}}\over {|{\vec N}_{(ab)}|}}+{{\Omega_{(ab)}}\over
{\sqrt{1-{\vec N}^2_{(ab)}}}}) {\hat N}_{(ab)}
\times {\hat \chi}_{(ab)},\nonumber \\ {\vec \pi}_{qc} &=& \Big[
{\tilde \pi}_{qc} |\vec N| +{{\sqrt{1-{\vec N}^2}}\over {2\rho_{qc}}}
({\check S}_q^2 - \xi_{((ab)c)}
\sqrt{1-{\vec N}^2})\Big] \hat N -\nonumber \\
 &-& \Big[ {\tilde \pi}_{qc}\sqrt{1-{\vec N}^2} - {{|\vec N|}\over {2\rho_{qc}}}
 ({\check S}_q^2 - \xi_{((ab)c)} \sqrt{1-{\vec N}^2})\Big] \hat \chi -\nonumber \\
  &-& {1\over {2\rho_{qc}}} ({{{\check S}_q^1}\over {|\vec N|}} +
  {{{\check S}^3_q}\over {\sqrt{1-{\vec N}^2}}}) \hat N \times \hat \chi ,\nonumber \\
  &&{}\nonumber \\
 {\vec \pi}^2_{qa} &=& {\tilde \pi}^2_{qa} +{1\over {4\rho^2_{qa}}} \Big[ \xi^2_{(ab)}
 (1-{\vec N}^2_{(ab)})+ \nonumber \\
 &+&\Big( sin^2\, \gamma_{(ab)}+{{cos^2\, \gamma_{(ab)}}\over
{{\vec N}^2_{(ab)}}}\Big) \Big( (S_{(ab)})^2-(\Omega_{(ab)})^2\Big) +
 {{(\Omega_{(ab)})^2}\over {1-{\vec N}^2_{(ab)}}}+ \nonumber \\
 &+&2 \sqrt{(S_{(ab)})^2-(\Omega_{(ab)})^2} (\xi_{(ab)} \sqrt{1-{\vec N}^2_{(ab)}}
  sin\, \gamma_{(ab)}- {{ \Omega_{(ab)} cos\, \gamma_{(ab)}}
  \over {|{\vec N}_{(ab)}| \sqrt{1-{\vec N}^2_{(ab)}}}} )\Big] ,\nonumber \\
 {\vec \pi}^2_{qb} &=& {\tilde \pi}^2_{qb} +{1\over {4\rho^2_{qb}}} \Big[ \xi^2_{(ab)}
 (1-{\vec N}^2_{(ab)})+\nonumber \\
 &+&\Big( sin^2\, \gamma_{(ab)}+ {{cos^2\,
\gamma_{(ab)}}\over {{\vec N}^2_{(ab)}}}\Big) \Big( (S_{(ab)})^2-(\Omega_{(ab)})^2\Big)
 + {{(\Omega_{(ab)})^2}\over {1-{\vec N}^2_{(ab)}}}-\nonumber \\
 &-& 2 \sqrt{(S_{(ab)})^2-(\Omega_{(ab)})^2}
 (\xi_{(ab)} \sqrt{1-{\vec N}^2_{(ab)}}
  sin\, \gamma_{(ab)}- {{ \Omega_{(ab)} cos\, \gamma_{(ab)}}
  \over {|{\vec N}_{(ab)}| \sqrt{1-{\vec N}^2_{(ab)}}}} )\Big] ,\nonumber \\
 {\vec \pi}^2_{qc} &=& {\tilde \pi}_{qc}^2 +{1\over {4\rho_{qc}^2}}
\Big[ \xi^2_{((ab)c)} (1-{\vec N}^2)+
({\check S}^2_q)^2+{{({\check S}^1_q)^2}\over {{\vec N}^2}}+
{{({\check S}^3_q)^2}\over {1-{\vec N}^2}}-\nonumber \\
 &-& 2 (\xi_{((ab)c)} \sqrt{1-{\vec N}^2} {\check S}^2_q -
 {{{\check S}^1_q{\check S}^3_q}\over {|\vec N| \sqrt{1-{\vec N}^2}}})\Big] .
\label{e8}
\eea

\noindent To evaluate ${\vec \pi}_{qc}$ one defines an auxiliary variable ${\vec \pi}_{(ab)}$
 such that ${\vec S}_{(ab)}={\hat R}_{(ab)}\times {\vec \pi}_{(ab)}$ [its component
along ${\hat R}_{(ab)}$ is arbitrary; remember that ${\hat
R}_{(ab)}\cdot {\vec S}_{(ab)}=0$]. Then we use the equations ${\vec
S}_q={\vec S}_{(ab)}+{\vec S}_{qc}$ and ${\vec W}_q={\vec
S}_{(ab)}-{\vec S}_{qc}$ (with ${\vec S}_{qc}={\vec
\rho}_{qc}\times {\vec \pi}_{qc}$)  to extract
the form of ${\vec \pi}_{qc}$ like in the 3-body case
[Eq.(\ref{cc5})].

Considering a set of Jacobi coordinates ${\vec \rho}_{qA}$, $A=1,2,3$,
with reduced masses $\mu_A$ we get the following expression for the
Hamiltonian to be compared with Eq.(\ref{II34}) [$|\vec N|=|{\vec
N}_{((ab)c)}|$]

\bea
H_{rel ((ab)c)} &=& {{{\vec \pi}^2_{qa}}\over {2\mu_a}} + {{{\vec
\pi}^2_{qb}}\over {2\mu_b}} + {{{\vec \pi}^2_{qc}}\over {2\mu_c}}=
{{{\tilde \pi}^2_{qa}}\over {2\mu_a}} + {{{\tilde
\pi}^2_{qb}}\over {2\mu_b}} + {{{\tilde \pi}^2_{qc}}\over {2\mu_c}}+
\nonumber \\
 &+&\Big( {1\over {8\mu_a\rho^2_{qa}}}+{1\over {8\mu_b\rho^2_{qb}}}\Big)
 \Big[ \xi^2_{(ab)} (1-{\vec N}^2_{(ab)})+\nonumber \\
 &+& {\cal R}\, (sin^2\, \gamma_{(ab)}+{{cos^2\, \gamma_{(ab)}}\over
 {{\vec N}^2_{(ab)}}}) + {{(\Omega_{(ab)})^2}\over {1-{\vec N}^2_{(ab)}}}\Big]
 +\nonumber \\
 &+& \Big( {1\over {8\mu_a\rho^2_{qa}}}-{1\over {8\mu_b\rho^2_{qb}}}\Big)
 \Big[ \xi_{(ab)} sin\, \gamma_{(ab)} \sqrt{1-{\vec N}^2_{(ab)}}-\nonumber \\
 &-& {{\Omega_{(ab)} cos\, \gamma_{(ab)}}\over {|{\vec N}_{(ab)}| \sqrt{1-{\vec N}_{(ab)}^2}}}
 \Big] \sqrt{{\cal R}} +\nonumber \\
 &+&{1\over {8\mu_c \rho^2_{qc}}} \Big[ \xi^2_{((ab)c)} (1-{\vec N}^2)+
 {{({\check S}^1_q)^2}\over {{\vec N}^2}}+({\check S}^2_q)^2+
 {{({\check S}^3_q)^2}\over {1-{\vec N}^2}}-\nonumber \\
 &-& 2 \Big( \xi_{((ab)c)} \sqrt{1-{\vec N}^2} {\check S}^2_q -{{{\check S}^1_q
 {\check S}^3_q}\over {|\vec N| \sqrt{1-{\vec N}^2}}}\Big) \Big] ,\nonumber \\
 &&{}\nonumber \\
 {\cal R} &=& {\cal R}(|\vec N|, \xi_{((ab)c)}, \Omega_{(ab)}; {\check S}^r_q)=
 (S_{(ab)})^2 - (\Omega_{(ab)})^2=\nonumber \\
  &=&{1\over 4} \Big[ ({\check S}^2_q + \xi_{((ab)c)}\sqrt{1-{\vec N}^2})^2 +
  \nonumber \\
  &+& ({{{\check S}^1_q}\over {|\vec N|}} - {{{\check S}^3_q}\over
  {\sqrt{1-{\vec N}^2}}})^2\Big] - (\Omega_{(ab)})^2.
\label{e9}
\eea

Due to the presence of the function ${\cal R}$, this Hamiltonian is
different from that of the static orientation-shape bundle approach
given in Eq.(\ref{II34}) with $N=4$: note that it is not a polynomial
of second order in the dynamical body frame components of the spin.
Yet, the rotational part of the Hamiltonian, determined by putting the
dynamical shape velocities equal to zero (${\dot q}^{\mu}=0$), can be
shown to have the standard form of the rigid body case, with a
non-standard {\it dynamical inertia-like tensor} depending only on the
shape variables.

To show this result, let us consider the Hamilton equations for the
velocities

\bea
{\dot \rho}_{qa} &{\buildrel \circ \over =}& {{\partial H_{rel
((ab)c)}}\over {\partial {\tilde \pi}_{qa}}}=
  {{{\tilde \pi}_{qa}}\over {\mu_a}},\nonumber \\
 {\dot \rho}_{qb} &{\buildrel \circ \over =}& {{\partial
H_{rel ((ab)c)}}\over {\partial {\tilde \pi}_{qb}}}=
 {{{\tilde \pi}_{qb}}\over {\mu_b}},\nonumber \\
 {\dot \rho}_{qc} &{\buildrel \circ \over =}& {{\partial
H_{rel ((ab)c)}}\over {\partial {\tilde \pi}_{qc}}}=
 {{{\tilde \pi}_{qc}}\over {\mu_c}},\nonumber \\
 {d\over {dt}} |{\vec N}_{((ab)c)}| &{\buildrel \circ \over =}&
 {{\partial H_{rel ((ab)c)}}\over {\partial \xi_{((ab)c)}}}=\nonumber \\
 &=& {1\over 2}\sqrt{1-{\vec N}^2} \Big[ ({\check S}^2_q +\xi_{((ab)c)} \sqrt{1-{\vec N}^2})
 \times \nonumber \\
 &&\Big( {1\over 8} ({1\over {\mu_a\rho^2_{qa}}} + {1\over {\mu_b\rho^2_{qb}}})
 ( sin^2\, \gamma_{(ab)}+ {{cos^2\, \gamma_{(ab)}}\over {{\vec N}^2_{(ab)}}} ) +\nonumber \\
 &+& {1\over {16}} ({1\over {\mu_a\rho^2_{qa}}} - {1\over {\mu_b\rho^2_{qb}}})
 ( \xi_{(ab)} \sqrt{1-{\vec N}^2_{(ab)}} sin\, \gamma_{(ab)} -
 {{\Omega_{(ab)} cos\, \gamma_{(ab)}}\over {|{\vec N}_{(ab)}| \sqrt{1-{\vec N}^2_{(ab)}} }}
 \Big) {1\over {\sqrt{{\cal R}}}}-\nonumber \\
 &-& {1\over {2\mu_c\rho^2_{qc}}} ({\check S}^2_q -\xi_{((ab)c)}
 \sqrt{1-{\vec N}^2})\Big],\nonumber \\
 {\dot \gamma}_{(ab)} &{\buildrel \circ \over =}& {{\partial
H_{rel ((ab)c)}}\over {\partial \Omega_{(ab)}}}=\nonumber \\
 &=& {1\over 4} ({1\over {\mu_a\rho^2_{qa}}} + {1\over {\mu_b\rho^2_{qb}}})
 \Omega_{(ab)}\nonumber \\
 && \Big[ {1\over {1-{\vec N}^2_{(ab)}}} -(sin^2\, \gamma_{(ab)}+{{cos^2\,
 \gamma_{(ab)}}\over {{\vec N}^2_{(ab)}}}) \Big] -\nonumber \\
 &-& {1\over 8} ({1\over {\mu_a\rho^2_{qa}}} - {1\over {\mu_b\rho^2_{qb}}})
 {{cos\, \gamma_{(ab)}}\over {|{\vec N}_{(ab)}| \sqrt{1-{\vec N}^2_{(ab)}}}}
 \Big[ \sqrt{{\cal R}} + {{(\Omega_{(ab)})^2}\over {\sqrt{{\cal R}}}}\Big],
 \nonumber \\
 {d\over {dt}} |{\vec N}_{(ab)}| &{\buildrel \circ \over =}&
 {{\partial H_{rel ((ab)c)}}\over {\partial \xi_{(ab)}}}=\nonumber \\
 &=& {1\over 4} ({1\over {\mu_a\rho^2_{qa}}} + {1\over {\mu_b\rho^2_{qb}}})
  \xi_{(ab)} (1-{\vec N}^2_{(ab)})+\nonumber \\
 &+& {1\over 8} ({1\over {\mu_a\rho^2_{qa}}} - {1\over {\mu_b\rho^2_{qb}}})
 sin\, \gamma_{(ab)} \sqrt{1-{\vec N}^2_{(ab)}} \sqrt{{\cal R}}.
\label{e10}
\eea

If we put all the  shape velocities equal to zero we get

\bea
i)&& {\dot \rho}_{qA}=0,\quad A=a,b,c,\nonumber \\
 &&\Downarrow\nonumber \\
 {\tilde \pi}_{qA}{|}_{\dot q=0} &=& 0,\quad A=a,b,c,
\label{e11}
\eea

\bea
ii)&& {d\over {dt}} | {\vec N}_{(ab)}| =0,\nonumber \\
 &&\Downarrow \nonumber \\
 \sqrt{{\cal R}{|}_{\dot q=0}} &=& \Big( {1\over 4}\Big[
 ({\check S}^2_q+\xi_{((ab)c)}{|}_{\dot q=0}\, \sqrt{1-{\vec N}^2})^2 +\nonumber \\
 &+&({{{\check S}^1_q}\over {|\vec N|}} - {{{\check S}^3_q}\over {\sqrt{1-{\vec N}^2} }})^2
 \Big] -(\Omega_{(ab)}{|}_{\dot q=0})^2 \Big)^{1/2} =\nonumber \\
 &&{}\nonumber \\
 &=&  -2 {{  {1\over {\mu_a\rho^2_{qa}}} + {1\over {\mu_b\rho^2_{qb}}} }\over
 { {1\over {\mu_a\rho^2_{qa}}} - {1\over {\mu_b\rho^2_{qb}}} }}
 {{ \sqrt{1-{\vec N}^2_{(ab)}}}\over {sin\, \gamma_{(ab)}}} \xi_{(ab)}{|}_{\dot q =0},
\label{e12}
\eea

\bea
iii)&& {\dot \gamma}_{(ab)} =0,\nonumber \\
 &&\Downarrow \nonumber \\
 {1\over {16}}&& \Big( {1\over {\mu_a\rho^2_{qa}}} + {1\over {\mu_b\rho^2_{qb}}}\Big)
 \xi_{(ab)}{|}_{\dot q=0}\, \Big[ \Big( {{{1\over {\mu_a\rho^2_{qa}}} +
 {1\over {\mu_b\rho^2_{qb}}}}\over { {1\over {\mu_a\rho^2_{qa}}} - {1\over {\mu_b\rho^2_{qb}}} }}
 \Big)^2 {{ sin\, \gamma_{(ab)} cos\, \gamma_{(ab)}}\over {|{\vec N}_{(ab)}|
 (1-{\vec N}^2_{(ab)})}} \Big( {{ \Omega_{(ab)}{|}_{\dot q=0}}\over
 {\xi_{(ab)}{|}_{\dot q=0}}} \Big)^2 +\nonumber \\
 &+& 4 \Big( sin^2\, \gamma_{(ab)} + {{cos^2\, \gamma_{(ab)}}\over {{\vec N}^2_{(ab)}}} -
 {1\over {1-{\vec N}^2_{(ab)}}}\Big) {{\Omega_{(ab)}{|}_{\dot q=0}}\over
 {\xi_{(ab)}{|}_{\dot q=0}}} - {{4 cos\, \gamma_{(ab)}}\over {|{\vec N}_{(ab)}|
 sin\, \gamma_{(ab)}}} \Big] =0,\nonumber \\
 &&\Downarrow \nonumber \\
 \Omega_{(ab)}{|}_{\dot q=0} &=& X_{\pm}(q)\, \xi_{(ab)}{|}_{\dot q=0},\qquad for\,
 \xi_{(ab)}{|}_{\dot q=0}\not= 0, \nonumber \\
 &&{}\nonumber \\
 X_{\pm}(q) &=& \Big[ \Big( {{{1\over {\mu_a\rho^2_{qa}}} - {1\over {\mu_b\rho^2_{qb}}}}\over
 {{1\over {\mu_a\rho^2_{qa}}} + {1\over {\mu_b\rho^2_{qb}}}}} \Big)^2\,
 {{sin\, \gamma_{(ab)} cos\, \gamma_{(ab)}}\over {|{\vec N}_{(ab)}|
 (1-{\vec N}^2_{(ab)})}} \Big]^{-1}\nonumber \\
 &&\Big[ -2 \Big( sin^2\, \gamma_{(ab)} + {{cos^2\, \gamma_{(ab)}}\over {|{\vec N}_{(ab)}|}}
 - {1\over {1-{\vec N}^2_{(ab)}}}\Big) \pm \nonumber \\
 &\pm& 2 \Big( [sin^2\, \gamma_{(ab)} +{{cos^2\, \gamma_{(ab)}}\over {|{\vec N}_{(ab)}|}}-
 {1\over {1-{\vec N}^2_{(ab)}}} ]^2 - \Big( {{
 {1\over {\mu_a\rho^2_{qa}}} - {1\over {\mu_b\rho^2_{qb}}} }\over
 { {1\over {\mu_a\rho^2_{qa}}} + {1\over {\mu_b\rho^2_{qb}}} }} \Big)^2
 {{cos^2\, \gamma_{(ab)}}\over { |{\vec N}_{(ab)}| (1-{\vec N}^2_{(ab)})}}
 \Big)^{1/2} \Big],  \nonumber \\
 &&
\label{e13}
\eea

\noindent and then

\bea
iv)&& {d\over {dt}} |\vec N| =0,\nonumber \\
 &&\Downarrow \nonumber \\
 {1\over 8}&& \Big( {1\over {\mu_a\rho^2_{qa}}} + {1\over {\mu_b\rho^2_{qb}}}\Big)
 \Big( {\check S}^2_q +\xi_{((ab)c)}{|}_{\dot q=0}\, \sqrt{1-{\vec N}^2}\Big)
 \Big[ sin^2\, \gamma_{(ab)} + {{cos^2\, \gamma_{(ab)}}\over { {\vec N}^2_{(ab)}}}+
 \nonumber \\
 &+& {1\over 4} \Big( {{{1\over {\mu_a\rho^2_{qa}}} - {1\over {\mu_b\rho^2_{qb}}} }\over
 {{1\over {\mu_a\rho^2_{qa}}} + {1\over {\mu_b\rho^2_{qb}}} }}\Big)^2 sin\, \gamma_{(ab)}
 \Big( {{cos\, \gamma_{(ab)}}\over {|{\vec N}_{(ab)}| \sqrt{1-{\vec N}^2_{(ab)}} }}
 {{ \Omega_{(ab)}{|}_{\dot q=0}}\over {\xi_{(ab)}{|}_{\dot q=0}}}-
 sin\, \gamma_{(ab)}\Big) \Big] -\nonumber \\
 &-& {1\over {2\mu_c \rho^2_{qc}}} \Big( {\check S}^2_q -\xi_{((ab)c)}{|}_{\dot q=0}\,
 \sqrt{1-{\vec N}^2}\Big) =0.
\label{e14}
\eea

Therefore we get

\bea
\xi_{((ab)c)}{|}_{\dot q=0} &=& g_{\xi}(q)\, {\check S}^2_q,\nonumber \\
 &&{}\nonumber \\
 g_{\xi}(q) &=& - {1\over { \sqrt{1-{\vec N}^2}}} \Big( sin^2\, \gamma_{(ab)} +
 {{cos^2\, \gamma_{(ab)}}\over {{\vec N}^2_{(ab)}}} - {1\over 4}
  \Big( {{{1\over {\mu_a\rho^2_{qa}}} - {1\over {\mu_b\rho^2_{qb}}}}\over
 {{1\over {\mu_a\rho^2_{qa}}} + {1\over {\mu_b\rho^2_{qb}}}}} \Big)^2\, sin\,
 \gamma_{(ab)} \Big[ sin\, \gamma_{(ab)} -\nonumber \\
 &-& {{cos\, \gamma_{(ab)} X_{\pm}(q)}\over {|{\vec N}_{(ab)}| \sqrt{1-
 {\vec N}^2_{(ab)}}}}\Big]  -4 {{ {1\over {\mu_c\rho^2_{qc}}} }\over
{{1\over {\mu_a\rho^2_{qa}}} + {1\over {\mu_b\rho^2_{qb}}}}} \Big)
\times \nonumber \\
 && \Big( sin^2\, \gamma_{(ab)} +
 {{cos^2\, \gamma_{(ab)}}\over {{\vec N}^2_{(ab)}}} - {1\over 4}
  \Big( {{{1\over {\mu_a\rho^2_{qa}}} - {1\over {\mu_b\rho^2_{qb}}}}\over
 {{1\over {\mu_a\rho^2_{qa}}} + {1\over {\mu_b\rho^2_{qb}}}}} \Big)^2\, sin\,
 \gamma_{(ab)} \Big[ sin\, \gamma_{(ab)} -\nonumber \\
 &-& {{cos\, \gamma_{(ab)} X_{\pm}(q)}\over {|{\vec N}_{(ab)}| \sqrt{1-
 {\vec N}^2_{(ab)}}}}\Big]  +4 {{ {1\over {\mu_c\rho^2_{qc}}} }\over
{{1\over {\mu_a\rho^2_{qa}}} + {1\over {\mu_b\rho^2_{qb}}}}}
\Big)^{-1}.
\label{e15}
\eea

Then, from ii) and iii) we get for $\xi_{(ab)}{|}_{\dot q=0}\not= 0$

\bea
\xi_{(ab)}{|}_{\dot q=0} &=& \pm {1\over 2} \Big( [(1+g_{\xi}(q)) \sqrt{1-{\vec N}^2}]
({\check S}^2_q)^2 + ( {{{\check S}^1_q}\over {|\vec N|}} -{{{\check
 S}^3_q}\over {\sqrt{1-{\vec N}^2}}} )^2 \Big)^{1/2} \times \nonumber \\
 && \Big( X^2_{\pm}(q) +4\Big( {{{1\over {\mu_a\rho^2_{qa}}} + {1\over {\mu_b\rho^2_{qb}}}}\over
 {{1\over {\mu_a\rho^2_{qa}}} - {1\over {\mu_b\rho^2_{qb}}}}}
 {{\sqrt{1-{\vec N}^2_{(ab)}}}\over {sin\, \gamma_{(ab)}}} \Big)^2
\Big)^{-1/2},\nonumber \\
 &&{}\nonumber \\
 \Omega_{(ab)}{|}_{\dot q=0} &=& X_{\pm}(q) \xi_{(ab)}{|}_{\dot q=0}.
\label{e16}
\eea

The non vanishing  shape momenta $p^{(o)}_{\mu} = p_{\mu}{|}_{\dot
q=0}$ do not have the form ${\check {\vec S}}_q\cdot {\check {\cal
A}}_{\mu}(q)$  of the static orientation-shape bundle approach.  A
discussion of the various solutions will be done elsewhere. Here we
are mainly interested  in determining the form of the rotational
kinetic energy. From Eq.(\ref{e16}) we get

\bea
(\xi_{(ab)}{|}_{\dot q=0})^2 &{\buildrel {def} \over =}& {\cal
I}^{rs}_{\xi (ab)}(q)\, {\check S}^r_q{\check S}^s_q,\nonumber \\
 (\Omega_{(ab)}{|}_{\dot q=0})^2 &{\buildrel {def} \over =}& {\cal I}^{rs}_{\Omega (ab)}(q)\,
 {\check S}^r_q{\check S}^s_q,\qquad {\cal I}^{rs}_{\Omega (ab)}(q)=
 X^2_{\pm}(q){\cal I}^{rs}_{\xi (ab)}(q),\nonumber \\
 &&{}\nonumber \\
 {\cal R}{|}_{\dot q=0} &{\buildrel {def} \over =}&
 {\cal I}^{rs}_R(q) {\check S}^r_q{\check S}^s_q,\nonumber \\
 &&{\cal I}^{rs}_R(q)=4\Big( {{{1\over {\mu_a\rho^2_{qa}}} + {1\over {\mu_b\rho^2_{qb}}}}\over
 {{1\over {\mu_a\rho^2_{qa}}} - {1\over {\mu_b\rho^2_{qb}}}}}
 {{\sqrt{1-{\vec N}^2_{(ab)}}}\over {sin\, \gamma_{(ab)}}} \Big)^2
 {\cal I}^{rs}_{\xi (ab)}(q),\nonumber \\
 &&{}\nonumber \\
 \sqrt{{\cal R}{|}_{\dot q=0}} &=& -2 {{{1\over {\mu_a\rho^2_{qa}}} +
 {1\over {\mu_b\rho^2_{qb}}}}\over
 {{1\over {\mu_a\rho^2_{qa}}} - {1\over {\mu_b\rho^2_{qb}}}}}
 {{\sqrt{1-{\vec N}^2_{(ab)}}}\over {sin\, \gamma_{(ab)}}}
 \xi_{(ab)}{|}_{\dot q=0},\nonumber \\
 &&{}\nonumber \\
 {1\over {8\mu_c\rho^2_{qc}}}&& \Big[ \xi^2_{((ab)c)}{|}_{\dot q=0}\, (1-{\vec N}^2) +
 {{({\check S}^1_q)^2}\over {{\vec N}^2}}+ ({\check S}^2_q)^2 +{{({\check S}^3_q)^2}\over
 {1-{\vec N}^2}}-\nonumber \\
 &&-2 \Big( \xi_{(ab)c)}{|}_{\dot q=0}\, \sqrt{1-{\vec N}^2} {\check S}^2_q -
 {{{\check S}^1_q{\check S}^3_q}\over {|\vec N| \sqrt{1-{\vec N}^2}}}\Big)
 \Big] \, {\buildrel {def} \over =}\, {\cal I}_c^{rs}(q) {\check S}^r_q{\check S}^s_q.
\label{e17}
\eea

Therefore,  from Eq.(\ref{e9}) we obtain

\bea
H^{(rot)}_{rel} &=& H_{rel}{|}_{\dot q=0} = {\cal I}^{rs}(q) {\check
S}^r_q{\check S}^s_q,\nonumber \\
 &&{}\nonumber \\
 {\cal I}^{rs}(q) &=& {\cal I}^{rs}_{(ab)}(q) + {\cal I}^{rs}_c(q),
 \nonumber \\
 {\cal I}^{rs}_{(ab)}(q) &=&  {1\over 8}
 \Big( {1\over {\mu_a\rho^2_{qa}}} + {1\over {\mu_b\rho^2_{qb}}}\Big)
 \Big[ (sin^2\, \gamma_{(ab)} +{{cos^2\, \gamma_{(ab)}}\over {{\vec N}^2_{(ab)}}})
 {\cal I}_R^{rs}(q)+\nonumber \\
 &+&\Big( {1\over {1-{\vec N}^2_{(ab)}}}+ {{2 cos\, \gamma_{(ab)} X_{\pm}(q)}\over
 {|{\vec N}_{(ab)}| sin\, \gamma_{(ab)}}}\Big) {\cal I}^{rs}_{\Omega (ab)}(q) -
 (1-{\vec N}^2_{(ab)}) {\cal I}^{rs}_{\xi (ab)}(q) \Big],
 \label{e18}
 \eea

 \noindent where ${\cal I}^{rs}(q)$ is an {\it inertia-like tensor}
 depending only on the dynamical shape variables and ${\cal I}_c^{rs}(q)$
 is defined in the last line of Eq.(\ref{e17}).

On the other hand, the {\it dynamical angular velocity} evaluated in
the dynamical body frame is

\bea
{\check \omega}^r_{((ab)c)} &=& {{\partial H_{rel ((ab)c)}}\over
{\partial {\check S_{qr}}}},\nonumber \\
 &&{}\nonumber \\
 {\check \omega}^1_{((ab)c)} &=& {1\over {16 |\vec N|}} \Big[ \Big(
 {1\over {\mu_a\rho^2_{qa}}} + {1\over {\mu_b\rho^2_{qb}}}\Big)
 \Big( sin^2\, \gamma_{(ab)} + {{cos^2\, \gamma_{(ab)}}\over {{\vec N}^2_{(ab)}}}\Big)+\nonumber \\
 &+& \Big( {1\over {\mu_a\rho^2_{qa}}} - {1\over {\mu_b\rho^2_{qb}}}\Big)
 {{ \xi_{(ab)} sin\, \gamma_{(ab)} \sqrt{1-{\vec N}^2_{(ab)}}-
 {{\Omega_{(ab)} cos\, \gamma_{(ab)}}\over {|{\vec N}_{(ab)}| \sqrt{1-{\vec N}^2_{(ab)}}}}
 }\over { 2\sqrt{{\cal R}} }} \Big] \Big( {{{\check S}^1_q}\over {|\vec N|}} -{{{\check
 S}^3_q}\over {\sqrt{1-{\vec N}^2}}} \Big) +\nonumber \\
  &+& {1\over {4\mu_c\rho^2_{qc} |\vec N|}} \Big( ( {{{\check S}^1_q}\over {|\vec N|}} +{{{\check
 S}^3_q}\over {\sqrt{1-{\vec N}^2}}} ) \Big),\nonumber \\
 {\check \omega}^2_{((ab)c)} &=&{1\over {8 |\vec N|}} \Big[ \Big(
 {1\over {\mu_a\rho^2_{qa}}} + {1\over {\mu_b\rho^2_{qb}}}\Big)
 \Big( sin^2\, \gamma_{(ab)} + {{cos^2\, \gamma_{(ab)}}\over {{\vec N}^2_{(ab)}}}\Big)+\nonumber \\
 &+& \Big( {1\over {\mu_a\rho^2_{qa}}} - {1\over {\mu_b\rho^2_{qb}}}\Big)
 {{ \xi_{(ab)} sin\, \gamma_{(ab)} \sqrt{1-{\vec N}^2_{(ab)}}-
 {{\Omega_{(ab)} cos\, \gamma_{(ab)}}\over {|{\vec N}_{(ab)}| \sqrt{1-{\vec N}^2_{(ab)}}}}
 }\over { 2\sqrt{{\cal R}} }} \Big] \times\nonumber \\
 && \Big( {\check S}^2_q+ \xi_{((ab)c)}\, \sqrt{1-{\vec N}^2}\Big)
 + {1\over {4\mu_c\rho^2_{qc}}}\Big( {\check S}^2_q- \xi_{((ab)c)}\,
 \sqrt{1-{\vec N}^2}\Big),\nonumber \\
 {\check \omega}^3_{((ab)c)} &=&   -{1\over {16 \sqrt{1-{\vec N}^2}}} \Big[ \Big(
 {1\over {\mu_a\rho^2_{qa}}} + {1\over {\mu_b\rho^2_{qb}}}\Big)
 \Big( sin^2\, \gamma_{(ab)} + {{cos^2\, \gamma_{(ab)}}\over {{\vec N}^2_{(ab)}}}\Big)+\nonumber \\
 &+& \Big( {1\over {\mu_a\rho^2_{qa}}} - {1\over {\mu_b\rho^2_{qb}}}\Big)
 {{ \xi_{(ab)} sin\, \gamma_{(ab)} \sqrt{1-{\vec N}^2_{(ab)}}-
 {{\Omega_{(ab)} cos\, \gamma_{(ab)}}\over {|{\vec N}_{(ab)}| \sqrt{1-{\vec N}^2_{(ab)}}}}
 }\over { 2\sqrt{{\cal R}} }} \Big]
 \Big( {{{\check S}^1_q}\over {|\vec N|}} -{{{\check
 S}^3_q}\over {\sqrt{1-{\vec N}^2}}} \Big) +\nonumber \\
  &+& {1\over {4\mu_c\rho^2_{qc} \sqrt{1-{\vec N}^2}}}
  \Big( ( {{{\check S}^1_q}\over {|\vec N|}} +{{{\check
 S}^3_q}\over {\sqrt{1-{\vec N}^2}}} ) \Big).
 \label{e19}
 \eea

We see therefore that for $N \geq 4$ in the spin bases the spin-
 angular velocity relation is different from the result (\ref{II33}) of the static
 orientation-shape bundle approach.

If we define  the {\it dynamical vibrations} as those corresponding to
the vanishing of the measurable angular velocity (there is no need of
connections like in the static orientation-shape bundle approach), we
get

\bea
&&{\check S}^1_q{|}_{{\check \omega}^s=0} = {\check S}^3_q{|}_{{\check
\omega}^s=0}=0,\nonumber \\
 && {\check S}^2_q{|}_{{\check \omega}^s=0} = f_1(q)\xi_{(ab)c)} + f_2(q)
 \xi_{(ab)} + f_3(q) \Omega_{(ab)}\  {\buildrel {def} \over =}\, f^{\mu}(q) p_{\mu},
\label{e20}
\eea

\noindent so that the vibrational Hamiltonian is

\bea
H^{(vib)}_{rel ((ab)c)} &=& H_{rel ((ab)c)}{|}_{{\check \omega}^r=0} =
 \sum_A^{a,b,c} {{ {\tilde \pi}^2_{qA}}\over {2\mu_A}}+ {1\over 8}
  \Big( {1\over {\mu_a\rho^2_{qa}}} + {1\over {\mu_b\rho^2_{qb}}}\Big)
  \Big[  \xi^2_{(ab)} (1-{\vec N}_{(ab)}^2)+\nonumber \\
  &+& {1\over 4} \Big( f^{\mu}(q) p_{\mu} +\xi_{((ab)c)} \sqrt{1-
  {\vec N}^2}\Big)^2 \Big( sin^2\, \gamma_{(ab)} +{{cos^2\, \gamma_{(ab)}}\over
  {{\vec N}^2_{(ab)}}}\Big) + {{(\Omega_{(ab)})^2}\over {1-{\vec N}^2_{(ab)}}}
  \Big] +\nonumber \\
  &+&{1\over {16}} \Big( {1\over {\mu_a\rho^2_{qa}}} - {1\over {\mu_b\rho^2_{qb}}}\Big)
  \Big[ \xi_{(ab)} sin\, \gamma_{(ab)} \sqrt{1-{\vec N}^2_{(ab)}} -
  \Omega_{(ab)} {{cos\, \gamma_{(ab)}}\over {|{\vec N}_{(ab)}| \sqrt{1-
  {\vec N}^2_{(ab)}}}} \Big] \nonumber \\
  &&\Big( f^{\mu}(q) p_{\mu}+
  \xi_{((ab)c)} \sqrt{1-{\vec N}^2}\Big) +\nonumber \\
  &+& {1\over {8\mu_c\rho^2_{qc}}} \Big( f^{\mu}(q) p_{\mu}-
  \xi_{((ab)c)} \sqrt{1-{\vec N}^2}\Big)^2.
\label{e21}
\eea

Let us stress that the Hamiltonian is not the sum of the rotational
and vibrational parts: $H_{rel ((ab)c)} \not= H^{(rot)}_{rel ((ab)c)}
+ H^{(vib)}_{rel ((ab)c)}$.

By calling $\theta^{\alpha}$ the Euler angles $\tilde \alpha$, $\tilde
\beta$, $\tilde \gamma$, the other Hamilton equations become
[${\check X}^{(R)\alpha}{}_{(\beta )}$ is the inverse of ${\check
\Lambda}^{(R)(\alpha )}{}_{\beta}$ after Eq.(\ref{II30}); see Appendix A for the
comparison with the static orientation-shape bundle approach]

\bea
{\dot \theta}^{\alpha} &{\buildrel \circ \over =}& {\check
X}^{(R)\alpha}{}_{(\beta =r)} {{\partial H_{rel ((ab)c)}}\over
{\partial {\check S}_{qr}}} = {\check X}^{(R)\alpha}{}_{(\beta =r)}
{\check \omega}^r_{((ab)c)},\nonumber \\
 &&{}\nonumber \\
 {d\over {dt}} {\check S}^r_q &{\buildrel \circ \over =}& -{{\partial H_{rel ((ab)c)}}\over
 {\partial \theta^{\alpha =r}}}- \epsilon^{sru} {\check S}_{qs}
{{\partial H_{rel ((ab)c)}}\over {\partial {\check S}_{qu}}}
=\nonumber \\
 &=& \epsilon^{rsu} {\check S}_{qs} {\check \omega}^u_{((ab)c)},\nonumber \\
 &&{}\nonumber \\
 {d\over {dt}} {\tilde \pi}_{qA} &{\buildrel \circ \over =}&   -
{{\partial H_{rel ((ab)c)}}\over {\partial \rho_{qA}}},\qquad
A=a,b,c,\nonumber \\
 {d\over {dt}} \xi_{((ab)c)} &{\buildrel \circ \over =}&   -
{{\partial H_{rel ((ab)c)}}\over {\partial |\vec N|}},\nonumber \\
{d\over {dt}} \xi_{(ab)} &{\buildrel \circ \over =}&  - {{\partial
H_{rel ((ab)c)}}\over {\partial |{\vec N}_{(ab)}|}},\nonumber \\
 {d\over {dt}} \Omega_{(ab)} &{\buildrel \circ \over =}& -
{{\partial H_{rel ((ab)c)}}\over {\partial \gamma_{(ab)}}}.
\label{e22}
\eea

For N=4 Ref.\cite{little1} gives the static shape coordinate basis
$w=\rho_{q1}^2+\rho_{q2}^2+\rho_{q3}^2$ ($\sqrt{w}$ is the
hyperradius), $w_1={{\sqrt{3}}\over 2} (\rho_{q1}^2-\rho_{q2}^2)$,
$w_2=\sqrt{3} {\vec \rho}_{q1}\cdot {\vec \rho}_{q2}$, $w_3=\sqrt{3}
{\vec \rho}_{q2}\cdot {\vec \rho}_{q3}$, $w_4=\sqrt{3} {\vec
\rho}_{q3}\cdot {\vec \rho}_{q1}$, $w_5={1\over 2} [-\rho_{q1}^2-
\rho_{q2}^2+2\rho_{q3}^2]$; $V={\vec \rho}_{q1}\cdot
({\vec \rho}_{q2}\times {\vec \rho}_{q3})$ must be added to solve the
problems of shapes connected by parity. By using Eqs.(\ref{e6}) these
variables can be reexpressed in our spin basis, in which they depend
on the dynamical shape variables, on the dynamical shape momenta and
also on the dynamical body frame components of the spin. Therefore, to
recover locally the above $N=4$ canonical basis of the static
orientation-shape bundle approach, a non-point canonical
transformation on all the variables is needed.

Note that there is in addition $w_6=\sqrt{3} \sqrt{ |{\vec
\rho}_{q1}\times {\vec
\rho}_{q2}|^2+|{\vec \rho}_{q2}\times {\vec \rho}_{q3}|^2+
|{\vec \rho}_{q3}\times {\vec \rho}_{q1}|^2} \geq 0$, with
$w^2=\sum_{i=1}^6w^2_i$. There are 3 democratic invariants $w$,
$a=\sum_{i=1}^5w_i^2$, and $b$ [see (4.25) of the paper]. Due to the
absence of the degeneracy submanifolds of planar configurations,
collinear configurations and 4-body scattering point, the physical
region has a complicated definition based on 3 eigenvales $\lambda_1
\geq \lambda_2 \geq \lambda_3 \geq 0$ and is topologically $B=R\times (S^5-P)$,
with $P\approx RP^2$ being the 2-dimensional projective plane.

\vfill\eject

\end{document}